\documentclass[twocolumn]{aastex62}

\usepackage{xcolor}
\usepackage{amsmath}
\usepackage[section]{placeins}
\usepackage{hyperref}

\graphicspath{{./}{figures/}}

\newcommand{\new}[1]{{#1}}

\begin{document}

\title{\textbf{A Combined Ground-based and JWST Atmospheric Retrieval Analysis: Both IGRINS and NIRSpec Agree The Atmosphere of WASP-77A b is Metal-Poor}}

\author{Peter C. B. Smith}
\affil{School of Earth and Space Exploration, Arizona State University, Tempe, AZ 85287}

\author{Michael R. Line}
\affil{School of Earth and Space Exploration, Arizona State University, Tempe, AZ 85287}

\author{Jacob L. Bean}
\affil{Department of Astronomy and Astrophysics, The University of Chicago, Chicago, IL 60637}

\author{Matteo Brogi}
\affil{Department of Physics, University of Turin, Via Pietro Giuria 1, I-10125, Turin, Italy}
\affil{INAF -- Osservatorio Astrofisico di Torino, Via Osservatorio 20, I-10025, Pino Torinese, Italy}

\author{Prune August}
\affil{Department of Space Research and Technology, Technical University of Denmark, Kongens Lyngby, Denmark}

\author{Luis Welbanks}
\affil{School of Earth and Space Exploration, Arizona State University, Tempe, AZ 85287}
\affil{NHFP Sagan Fellow}

\author{Jean-Michel Desert}
\affil{Anton Pannekoek Institute of Astronomy, University of Amsterdam, Amsterdam, Netherlands}

\author{Jonathan Lunine}
\affil{Department of Astronomy, Cornell University, Ithaca, NY 14850}

\author{Jorge Sanchez}
\affil{School of Earth and Space Exploration, Arizona State University, Tempe, AZ 85287}

\author{Megan Mansfield}
\affil{Steward Observatory, University of Arizona, Tucson, AZ 85719}
\affil{NHFP Sagan Fellow}

\author{Lorenzo Pino}
\affil{INAF -- Osservatorio Astrofisico di Arcetri, Largo Enrico Fermi 5, 50125 Firenze, Italy}

\author{Emily Rauscher}
\affil{Department of Astronomy, University of Michigan, Ann Arbor, MI 48109}

\author{Eliza Kempton}
\affil{Department of Astronomy, University of Maryland, College Park, MD 20742}

\author{Joseph Zalesky}
\affil{Department of Astronomy, University of Texas at Austin, Austin, TX 78712}

\author{Martin Fowler}

\begin{abstract}

    Ground-based, high-resolution and space-based, low-resolution spectroscopy are the two main avenues through which transiting exoplanet atmospheres are studied. Both methods provide unique strengths and shortcomings, and combining the two can be a powerful probe into an exoplanet's atmosphere. Within a joint atmospheric retrieval framework, we combined JWST NIRSpec/G395H secondary eclipse spectra and Gemini South/IGRINS pre- and post-eclipse thermal eclipse observations of the hot Jupiter WASP-77A b. Our inferences from the IGRINS and NIRSpec data sets are consistent with each other, and combining the two allows us to measure the gas abundances of H$_2$O and CO as well as the vertical thermal structure with higher precision than either data set provided individually. We confirm WASP-77A b's subsolar metallicty ([(C+O)/H]=-0.61$^{+0.10}_{-0.09}$) and solar C/O ratio (C/O = 0.57$^{+0.06}_{-0.06}$). The two types of data are complementary, and our abundance inferences are mostly driven by the IGRINS data while \new{inference of} the thermal structure is driven by the NIRSpec data. Our ability to draw inferences from the post-eclipse IGRINS data is highly sensitive to the number of singular values removed in the detrending process, potentially due to high and variable humidity. We also search for signatures for atmospheric dynamics in the IGRINS data and find that propagated ephemeris error can manifest as both an orbital eccentricity or a strong equatorial jet. Neither are detected when using more up-to-date ephemerides. However, we find moderate evidence of thermal inhomogeneity and measure a cooler nightside that presents itself in the later phases after secondary eclipse. 
    
\end{abstract}

\keywords{planets and satellites: atmospheres}

\section{Introduction}

A wealth of information is contained in the atmospheres of exoplanets. Measuring their compositions and thermal structures at a population level provides insight into a planet's formation pathways \citep{oberg2011, madhu2017} and atmospheric chemical and physical processes \citep{fortney2008, parmentier2018}. One of the major goals of exoplanet science is to synthesize this information and uncover population-level trends in compositional diagnostics such as metallicity and the carbon-to-oxygen ratio (e.g., \citealp{sing2016, welbanks2019, baxter2020, mansfield2021}). Spectroscopic campaigns with the \textit{Hubble} and \textit{Spitzer Space Telescopes} (hereafter HST and Spitzer) aimed to measure the molecules H$_2$O and CO, which make up a significant fraction of metals in hot Jupiters\new{, as well as CO$_2$, whose strong absorption can provide another indicator of atmospheric metallicity}. Applying Bayesian atmospheric parameter estimation tools (``retrievals") to these datasets enabled constraints on the water abundances in several planets \citep[e.g.,][]{kreidberg2014, haynes2015,welbanks2019}. However, due to limited wavelength coverage, similar measurements of CO \new{and CO$_2$} remained elusive, and estimates of metallicity and C/O were based largely on these H$_2$O constraints.

An alternative to space-based spectroscopy is High Resolution (R = $\lambda / \Delta \lambda > $20000) Cross-Correlation Spectroscopy (HRCCS) using ground-based telescopes. HRCCS leverages the planetary signal's Doppler shift due to its orbital motion to disentangle the planetary atmospheric emission or transmission lines from telluric and stellar lines. HRCCS has been used to detect several C- and O-bearing species \citep[e.g.,][]{dekok2013, birkby2013, hawker2018} in addition to numerous refractory metals \citep[e.g.,][]{hoeijmakers2019, gandhi2023, pelletier2023} in hot- and ultra-hot Jupiters. HRCCS also provides probes of atmospheric dynamics and has been used to measure wind and rotation speeds \citep{snellen2010, brogi2016, gandhi2022}. \cite{brogi2019} and \cite{gibson2020} demonstrated that absolute abundance and temperature profile constraints could be retrieved from these datasets, placing HRCCS observations on a similar footing as space-based transit spectoscopy for quantitative estimation. Using simulated HST/WFC3 and CRIRES data, \cite{brogi2019} also showed that that high- and low-resolution data can be combined in a joint retrieval process. The two types of data are complementary: low-resolution data contains continuum information and probes high pressures, and high-resolution data is sensitive to molecular line shapes and probes lower pressures where the strongest line cores are located. \cite{brogi2019} predicted that such combinations would enable more precise estimates of atmospheric properties than either data set could provide alone. Using real data this has indeed been the case in the literature, but there are only a handful of studies that have done such analysis \citep{gandhi2019, kasper2023, boucher2023}.

\cite{line2021} recently demonstrated the HRCCS capabilities of IGRINS (Immersion GRating INfrared Spectrometer, R$\sim$45,000, 1.45-2.6 micron) on Gemini-South by applying the HRCCS retrieval framework to pre-secondary-eclipse (0.325 $< \varphi <$ 0.47, $\varphi=$0.5 being secondary eclipse) data of the hot Jupiter WASP-77A b \citep[$T_\mathrm{eq} = 1700 \mathrm{K}, R_P = 1.21 R_J, P= 1.36$ day;][]{maxted2013}. Owing to the instrument's stability and large wavelength coverage, \cite{line2021} placed precise constraints on the planet's thermal structure and the abundances of H$_2$O and CO ($\pm$0.1-0.2 dex). From the latter, they inferred a sub-solar metallicity ([(C+O)/H] = -0.48$^{+0.15}_{-0.13}$) and near-solar C/O (0.59$\pm$0.08), indicative of more diverse formation pathways commonly predicted in the literature (such as \cite{madhu2014, mordasini2016, khorshid2021}).

\cite{mansfield2022} investigated the atmosphere of WASP-77A b at low resolution using HST/WFC3 (1.1-1.7 micron, R $\sim$ 70) and the Spitzer/IRAC channels centered at 3.6 and 4.5 microns. Using a similar free retrieval framework to \cite{line2021}, they were unable to obtain any informative composition constraints, only placing a subsolar lower limit on the metallicity broadly consistent with the IGRINS results. Again, WASP-77A b was recently observed in eclipse using NIRSpec/G395H on the \textit{James Webb Space Telescope} \citep[JWST,][]{august2023}. Bounded constraints could be placed on the abundances of H$_2$O and CO \new{and an upper limit was placed on the abundance of CO$_2$} when applying a free retrieval, showing a clear improvement over WFC3 and allowing the confirmation of the low metallicity measured with IGRINS by \cite{line2021}.

While previous combined high- and low-resolution retrievals have only used HST and Spitzer data \citep{gandhi2019, kasper2023, boucher2023}, no such analysis has been attempted with data taken with JWST. Both IGRINS and NIRSpec have been shown to provide stringent estimates on atmospheric composition and thermal structure, and combining the unique strengths of both instruments could provide powerful probes into transiting giant atmospheres. However, such a combination may also be challenged in ways the previous high- and low- resolution combinations were not due to the fact that the two instruments probe different altitudes and hence potentially different gas abundances. Thus, a test of its feasibility and effectiveness is needed. In this paper, we will perform a combined ground-based and JWST retrieval analysis using the IGRINS and NIRSpec data of WASP-77A b.

We first present two additional nights of IGRINS data covering the post-eclipse phases of WASP-77A b's orbit in Section \ref{sec:obs}. In the subsequent sections, we search for additional information that may be gained by incorporating these new data through molecular detection via cross-correlation in Section \ref{sec:cross correlation}, searching for signatures of atmospheric dynamics in Section \ref{sec:velocities}, and atmospheric retrieval in Section \ref{sec:retrieval}. The results of combining the IGRINS and NIRSpec data are also presented in Section \ref{sec:retrieval}. We place WASP-77A b's atmosphere into context and discuss the synergies between NIRSpec and IGRINS in Section \ref{sec:discussion} before concluding in Section \ref{sec:conclusions}.

\section{Observations \& Data Reduction}
\label{sec:obs}

\begin{table}[]
    \centering
    \begin{tabular}{c c}
        \hline
        \hline 
        \multicolumn{2}{c}{WASP-77A System Properties} \\
        \hline
        \hline
        \textit{Stellar Properties} \\
        Spec. Type & G8V$^{b}$ \\
        $R_\star$ & 0.955 \ $R_\odot^{a}$ \\
        $M_\star$ & $1.002 \pm 0.045$  \ $M_\odot^{b}$ \\
        $T_\mathrm{eff}$ & 5605 $K^{b}$\\
        log$g$ & $4.33 \pm$ \ 0.08$^{b}$\\
        $M_K$ & $8.405 \pm \ 0.031^{b}$ \\ 
        $\gamma$ & $1.6845 \pm$ 0.0004 km \ s$^{{-1}^{b}}$ \\
        $\mathrm{[Fe/H]}$ & $0.00 \pm \ 0.11^{b}$ \\
        \new{$\mathrm{[C/H]}$}  & \new{$-0.02 \pm 0.05^{d}$} \\
         & \new{$0.10 \pm 0.09^{e}$} \\
         & \new{$-0.04 \pm 0.06^{f}$} \\
        \new{$\mathrm{[O/H]}$} & \new{$0.06 \pm 0.07^{d}$ }\\
         & \new{$0.23 \pm 0.02^{e}$} \\
         & \new{$-0.14 \pm 0.06^{f}$} \\
         \new{$\mathrm{C/O}$}  &  \new{$0.46 \pm 0.09^{d}$} \\
          & \new{ $ {0.44^{+0.07}_{-0.08}}^{e}$  } \\
           & \new{$0.59 \pm 0.08^{f}$} \\
        
        \\
        \textit{Planet Properties} \\
        $R_P$ & $1.21 \pm \ 0.02$ \ $R_J^{b}$ \\
        $M_P$ & $1.76 \pm \ 0.06$ \ $M_J^{b}$ \\ 
        $T_{eq}$ & 1740 K$^{b}$ \\ 
        $K_P$ & 192 $\pm$ 4.5   km s$^{-1}$ \\
        $e$ & ${0.0074^{+0.007}_{-0.005}}^{c}$ \\
        $\omega_\star$ & ${-166^{+66}_{-75}}^{\circ,{c}}$ \\ 
        $T_C$ & $2455870.44977 \pm 0.00014$   $\mathrm{BJD}^{b}$\\
        & $2457420.88439^{+0.00080}_{-0.00085}$ $\ \mathrm{BJD}^{c}$ \\
        $P$ &  $1.3600309 \pm \ 0.0000020$ \ $\mathrm{day}^{b}$ \\
        & $1.36002854 \pm 0.00000062$ \ $\mathrm{day}^{c}$\\
        $a$ & 	$0.02405 \pm  \ 0.00036$ \ $\mathrm{AU}^{b}$
        \\
        \hline
        \hline 
    \end{tabular}
    \caption{Relevant stellar and planet parameters for WASP-77A b and its star. $K_P$ is calculated by assuming a circular orbit with the reported semi-major axis and period: $K_P = 2\pi a / P$. References: \textit{a}:  \citet{bonomo2017}; \textit{b} \citet{maxted2013};  \textit{c}: \citet{cortes}\new{; \textit{d}: \citet{polanski2022}; \textit{e}: \citet{reggiani2022}; \textit{f}: \citet{kolecki2022}}.}
    \label{tab:properties}
\end{table}

\begin{figure*}
    \centering
    \includegraphics[width=0.7\textwidth]{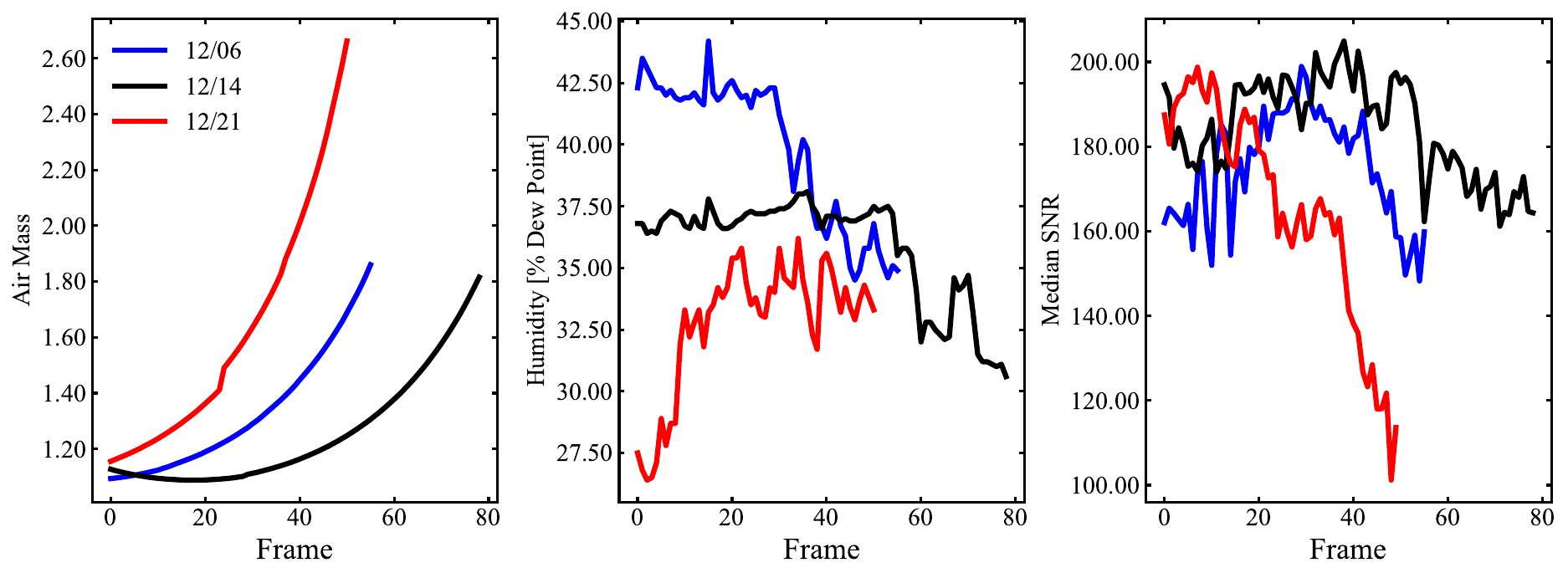}
    \caption{Observing conditions on each of the three nights data were taken.}
    \label{fig:observing conditions}
\end{figure*}

The high-resolution data were taken over the course of three half nights in December 2020 as part of observing program GS-2020B-Q-249 (P.I. J. Zalesky) using the IGRINS instrument on Gemini South \new{\citep{park2014,mace2016}}. The data taken on December 14, 2020, covering the pre-eclipse phases, were previously presented and analyzed in \cite{line2021}. Data taken on December 6 and December 21, each covering post-eclipse phases, are presented here for the first time\footnote{\new{The reduced spectral matrices and observing conditions for all three nights as well as the spectral templates used for the subsequent cross-correlation analysis are publicly available in a Zenodo repository \href{https://zenodo.org/records/10382053?token=eyJhbGciOiJIUzUxMiJ9.eyJpZCI6ImUwYzQ2OWY2LWUxMWItNGNmYy05MWNmLWNhN2YyYzc4YjEwMiIsImRhdGEiOnt9LCJyYW5kb20iOiI3MWIzOWZhNTI2MzAyOWIxODQ2NWEyOWFlNzQ0ZjVkMyJ9.2zZfrcXislbWzgsZg78Ra18eZzeOfALwPiUsSe4GQyApRu_IVNv34fe5cAhJmC26bkP93jzsK-kMZUOAmfA-JA}{linked here}.}}. For each night, we took a continuous sequence of 70s exposures using an AB-BA nodding pattern (140s per AB pair, hereafter referred to as ``frames"), consistently achieving high ($\sim$200) SNR per pixel per frame in most orders (Table \ref{tab:dates}). The primary star has a companion, WASP-77 B, 3.3" away ($\sim$ 10 slit widths) with a position angle 150$^\circ$ east of north. The slit was rotated from the default 90$^\circ$ to 60$^\circ$ to avoid contamination when nodding. Observing conditions on each night are summarized in Figure \ref{fig:observing conditions}.

\begin{table*}[]
    \centering
    \begin{tabular}{c|c|c|c|c|c|c|c}
    \hline
    \hline
       Date  & Orbital Phase & Start (BJD) & End (BJD) & \# Frames  & Med. SNR H & Med. SNR K & Air Mass \\
       \hline
       12/06/2020  & 0.535-0.605  & 2459189.61326 & 2459189.75917 & 39 & 195 & 185 & 1.09 $\rightarrow$1.86\\
       12/14/2020 & 0.325-0.47  & 2459197.53017 & 2459197.74095 & 79 & 205 & 190 & 1.12 $\rightarrow$ 1.09\\

        & & & & & & & \ $\rightarrow$ 1.82 \\
       
       12/21/2020 & 0.535-0.628  & 2459204.61064 & 2459204.74449 & 48 & 175 & 165 & 1.15 $\rightarrow$ 2.66 \\
       \hline
       \hline
    \end{tabular}
    \caption{Details for each of the three observing sequences taken with IGRINS.}
    \label{tab:dates}
\end{table*}

\begin{figure}
    \centering
    \includegraphics[width=\linewidth]{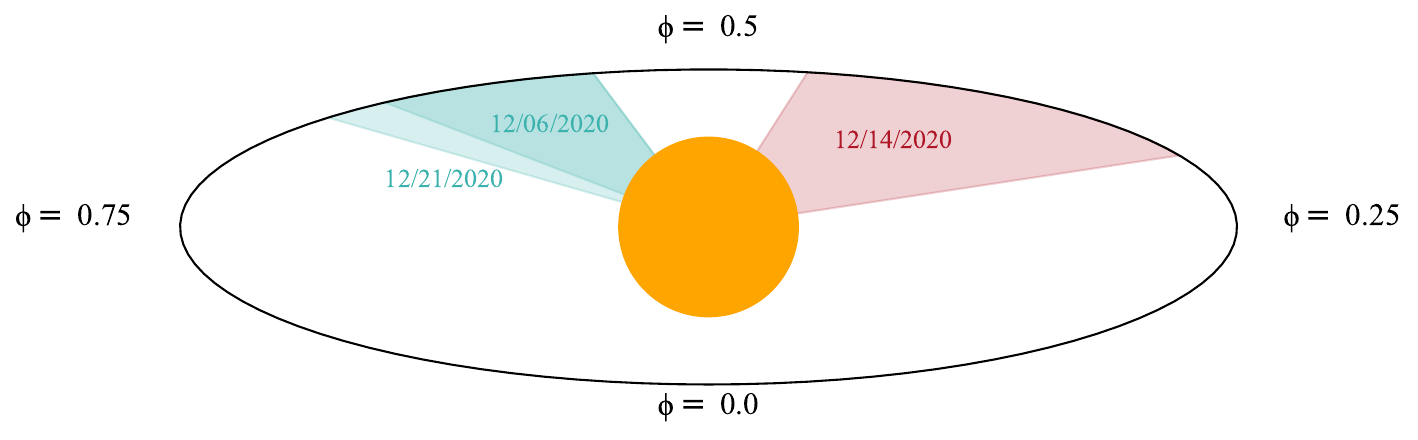}
    \caption{Observed phases. Transit is at $\phi$ = 0 and secondary eclipse is at $\phi$=0.5. The colors of the sectors indicate whether the planet signal is blue- or red-shifted during those phases. Not to scale.}
    \label{fig:phases}
\end{figure}

The raw 1D spectra were extracted by the IGRINS facility team using the IGRINS Pipeline Package \citep[PLP;][]{lee1016, mace2016}. While the PLP provides an initial wavelength solution, this solution can drift by up to 0.2 pixels (0.46 km s$^{-1}$) over the course of an observing sequence \citep[e.g., see][]{brogi2023}. To correct these small misalignments, for each sequence, we re-fit each frame to the last frame in the sequence to place the entire sequence on a common wavelength grid. We choose to align to the last frame in the sequence as this is the closest in time to when the wavelength calibrator is used by the PLP. We also normalize the counts in each frame, discard 8 orders due to low throughput and/or strong telluric contamination, and discard 200 pixels on the low throughput edges of each order. After this cleaning process, we place the data into data cubes of shape $N_\mathrm{orders} \times N_\mathrm{frames} \times N_\mathrm{pixels}$.

To detrend each raw spectral sequence, we apply a singular value decomposition (SVD) to each order from a data cube using \texttt{numpy.lingalg.svd} \citep[][]{dekok2013, brogi2019, line2021, brogi2023}. By visual inspection, there are no remaining telluric artifacts after the first 4 principal components are removed. Neither the cross-correlation nor the retrieval analyses change whether we choose 4, 6 or 8 principal components, and for our initial analysis we remove 4 for all three sequences in order to be consistent with the \cite{line2021} analysis of the pre-eclipse data. However, we find that removing only 3 is best for the post-eclipse nights (see Section \ref{subsec:poopy post}). For a given sequence, we save two matrices -- the data cube recomposed with the first 4 principal components removed and a scaling matrix of the data cube recomposed using only the first 4 -- as well as an array of the times of each frame in BJD. In the subsequent sections, we will refer to the sequence taken on 12/14/2020 as the pre-eclipse data. The sequences taken on 12/06/2020 and 12/21/2020 will be referred to together as the post-eclipse data, and in all analysis except the CCF trail, we sum their cross-correlation coefficients and log-likelihoods together and treat this sum as if from a single sequence.

\section{Molecular Detection via Cross-Correlation}
\label{sec:cross correlation}

\begin{figure*}[!ht]
    \centering
    \includegraphics[width=0.9\linewidth]{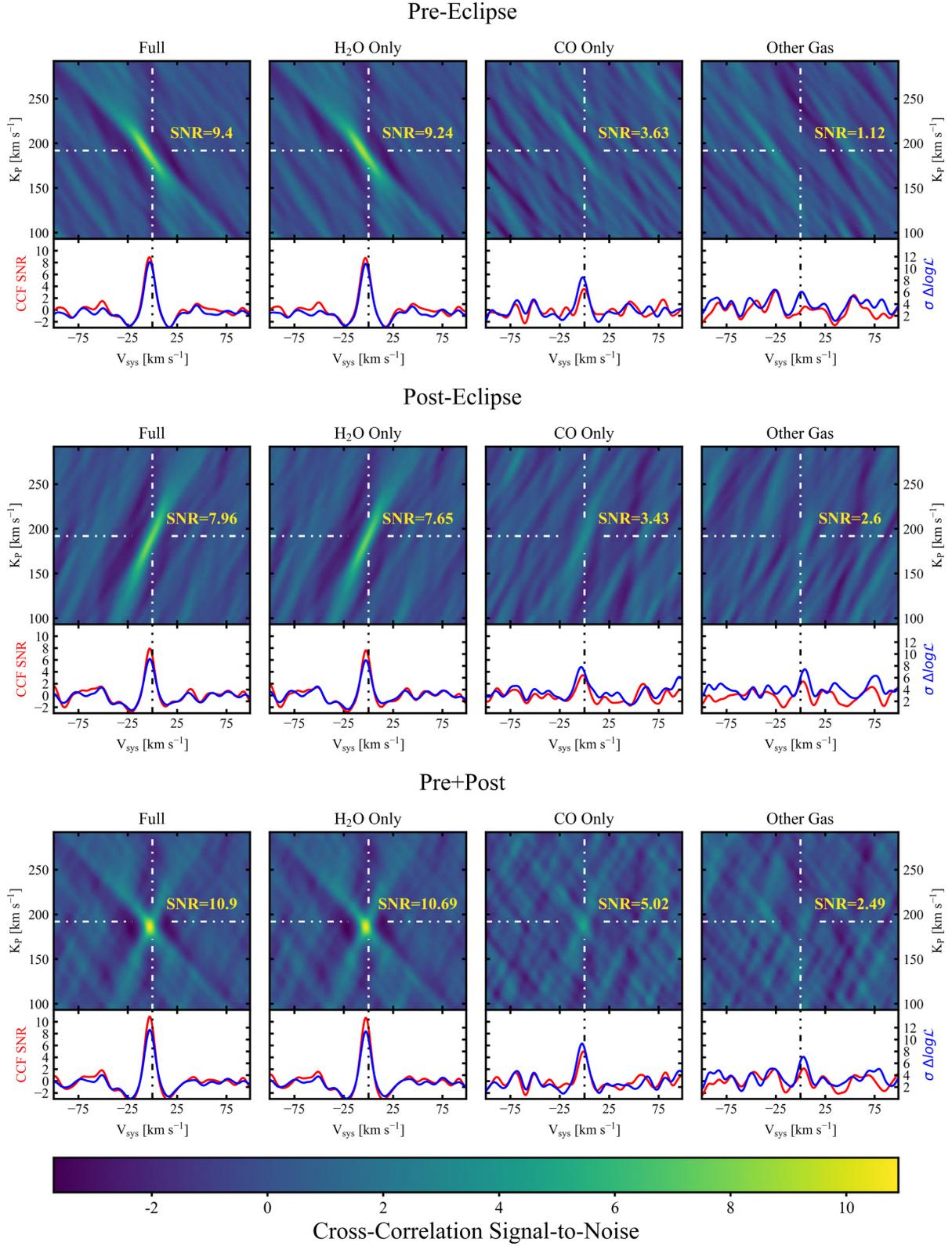}
    \caption{Cross-correlation maps for the pre- and post-eclipse sequences as well as from all 3 nights of IGRINS data combined. Each column is the resultant map from cross-correlating using a different atmospheric template. Below each map is a cross section of both the CCF SNR (red) and $\Delta$log$\mathcal{L}$ (divided by its standard deviation, blue) at the highest likelihood $K_P$.}
    \label{fig:all ccf}
\end{figure*}

As an initial check of the quality of the data and to determine the strength of the planetary signal, we cross-correlate a model spectrum with the post-SVD data. We used a solar composition ([X/H] = 0; C/O = 0.55) 1D radiative-convective-thermoequilibrium model (1D-RCTE) \new{using the ScCHIMERA modeling framework} as described in \cite{arcangeli2018, piskorz2018, mansfield2022}. \new{The ScCHIMERA model provides the dayside average 1D pressure-temperature (P-T) profile as well as gas abundance profiles.} For a hot Jupiter, C and O, in the form of H$_2$O and CO, will contain the majority of metals in the atmosphere. Therefore, H$_2$O and CO, both of which have many lines in the H and K bands, are of particular interest to detect.

A high-resolution (R=250,000) thermal emission model spectrum is computed \new{by passing} the 1D-RCTE atmospheric structure \new{through a GPU-accelerated version of the atmospheric forward modeling code CHIMERA \citep{line2013, brogi2019}}. The model spectrum is then convolved with the appropriate equatorial rotation kernel (vsin$i$=4.2 km s$^{-1}$) and a Gaussian intrumental profile at the IGRINS nominal resolution. Template spectra that include only H$_2$O and CO individually \new{(in addition to continuum opacities)} are also computed \new{using the same P-T and abundance profiles as output by the solar composition 1D-RCTE model but with all other gas abundances set to zero}. We include the most recent ExoMol \citep{tennyson2020}, HITEMP \citep{rothman2010}, and HITRAN opacities\footnote{H$_2$-H$_2$/He CIA cross-sections from \cite{karman2019}; H$_2$O line list from \cite{polyansky2018} and absorption cross-sections computed via the process described in \cite{gharib-nezhad2021}; the CO isotopologue cross-sections are from  \cite{li2015}; CH$_4$ from \cite{hargreaves2020}, H$_2$S from \cite{azzam2016}; NH$_3$ from \cite{coles2019}; and HCN from \cite{barber2014}.} from H$_2$-H$_2$ and H$_2$-He CIA, H$_2$O, $^{12}$CO, $^{13}$CO, CH$_4$, H$_2$S, NH$_3$, and HCN. 

We convert from $F_P$ to $F_P/F_\star$ by dividing by a smoothed (Gaussian over 200 pixels) PHOENIX stellar spectrum \citep{husser2013} at the appropriate $T_\mathrm{eff}$ and log$g$. Because the SVD process can modify and stretch the planet signal, before each cross-correlation or likelihood evaluation we follow the model injection procedure outlined in \cite{brogi2019}, \cite{line2021}, and \cite{brogi2023} to similarly modify the model spectrum. 

The planetary lines Doppler shift at each orbital phase as the planet orbits the star, and \new{t}he total line-of-sight velocity at a given time $t$ is:

\begin{equation}
    V_\mathrm{LOS}(t) = \gamma  + V_\mathrm{bary}(t) + V_P(t) + V_\mathrm{sys}
    \label{eqn:velocity}
\end{equation}
where $\gamma$ is the star-planet system's radial velocity, $V_\mathrm{bary}(t)$ is the Solar System barycentric radial velocity in the observatory's frame (via \texttt{radial\_velocity\_correction} from \texttt{astropy}), $V_P(t)$ is the planet's velocity in the star-planet system's barycentric frame, and $V_\mathrm{sys}$ is an additive term to account for any systematic offset. Assuming a circular orbit (however, see Section \ref{sec:velocities}) $V_P(t)$ becomes $K_P \sin[2 \pi \times \varphi(t)]$, where $K_P$ is the planet's radial velocity semi-amplitude and $\varphi$ is its orbital phase. As is typical in the literature \citep[e.g.,][]{birkby2018, brogi2019, line2021}, we Doppler shift and cross-correlate the model spectrum on a 2D grid of possible $K_P$ and $V_\mathrm{sys}$ values, creating a 2D map of correlation coefficients. We then median subtract this map, find the standard deviation of a 3-sigma-clipped copy, and divide the map by this standard deviation to determine the detection signal-to-noise ratio of planetary absorption or emission lines \citep{kasper2021, kasper2023}. In addition to calculating correlation coefficients, we use the likelihood function from \cite{brogi2019} to calculate a log-likelihood based detection map. Figure \ref{fig:all ccf} shows the cross-correlation (CCF) detection maps for the atmospheric signal from the above mentioned model.

Similar to \cite{line2021}, we achieve a strong detection of absorption lines from the planet's thermal emission with a signal-to-noise-ratio (SNR) of 9.4 when cross-correlating the full atmospheric model with the pre-eclipse data alone (Figure \ref{fig:all ccf}, top left). We note that this is different from \cite{line2021}, who report a SNR of 12.3 due to their different method of normalizing the CCF map. This and the many other different normalizing methods in the literature demonstrates that the CCF SNR alone is not necessarily a robust metric for planet signal strength. Using more statistically robust methods, such as a Welch t-test \citep{brogi2012} or the log-likelihood approach, may be more appropriate. Nonetheless, to keep in line with the literature, we list here the CCF SNR values, but for each model-data combination we also show the detection significances via the log-likelihood maps in Figure \ref{fig:all ccf}, all of which are comparable to the CCF SNR.

Searching for the individual gases of interest, we detect H$_2$O with a SNR of 9.2 and CO with a SNR of 3.6. We also use a spectral template without H$_2$O and CO to assess the presence of additional species still in the model like HCN, CH$_4$, and NH$_3$. We are unable to detect these species in the CCF alone (Fig. \ref{fig:all ccf}, top right), which is unsurprising because they have significantly lower abundances in the model compared to H$_2$O and CO.

Absorption features are also detected in both post-eclipse data sets. Summing over both post-eclipse nights, the full-atmospheric SNR is 8 while H$_2$O and CO are detected with SNRs of 7.7 and 3.4, respectively (Figure \ref{fig:all ccf}, middle row). Combining all three nights increases the full atmosphere SNR to 10.9. H$_2$O, and CO have SNRs of 10.7 and 5.0, respectively (Figure \ref{fig:all ccf}, bottom row). We note that at this relatively high SNR, many of the structures in the CCF maps are aliases of the planet signal or wings of the central CCF peak. Therefore, it is difficult to truly estimate the CCF noise, and the actual detection significance is likely underestimated. When cross-correlating the spectral template with H$_2$O and CO removed for both the post-eclipse data and all nights combined, the remaining gases (CH$_4$, NH$_3$, HCN, $^{13}$CO, and H$_2$S) are not detected to any significance.

\begin{figure}
    \centering
    \includegraphics[width=\linewidth]{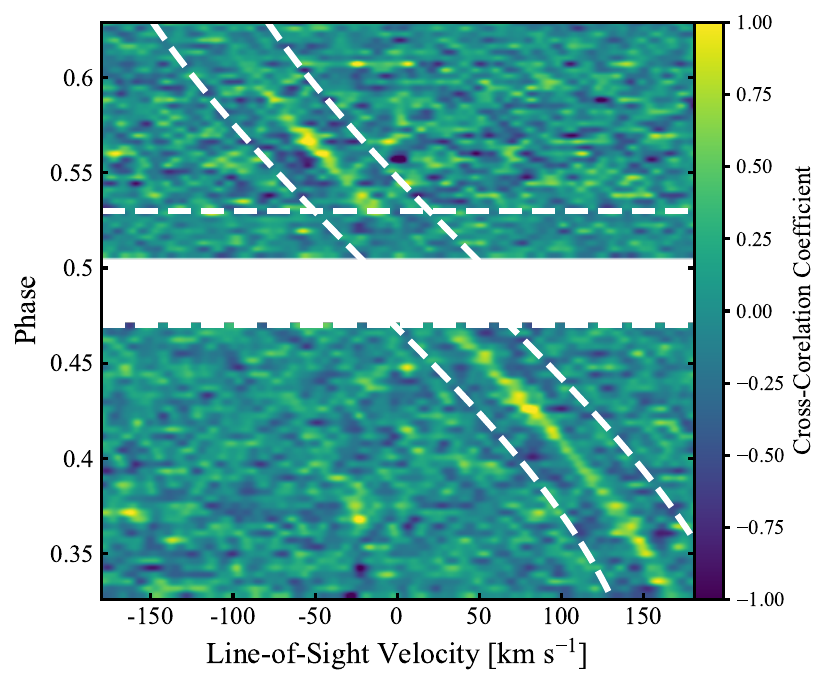}
    \caption{Cross correlation trails tracking the planet signal throughout the phases captured in this program. The same 1D equilibrium model used to create the 2D CCF maps in Figure \ref{fig:all ccf} was cross correlated with each individual frame at a range of line-of-sight velocities (horizontal axis). For visual clarity, the two post-eclipse sequences have been combined and pairs of frames have been binned together to boost the visible CCF. The dashed lines are the expected planet radial velocity per the best-fit values of K$_P$ and V$_\mathrm{sys}$ from Section \ref{sec:velocities}, offset by 35 km s$^{-1}$ on either side of the actual CCF trail for clarity. The horizontal dashed lines indicate secondary eclipse, when the planet is occulted by the star. The white space indicates phases at which no data were taken.}
    \label{fig:orbit trail}
\end{figure}

We also note that the correlated streaks in the CCF maps in Figure \ref{fig:all ccf} lean in opposite directions between pre- and post-eclipse phases, indicative that we are indeed seeing the planet signal as it moves away and then back toward the line-of-sight. We also search for the planetary signal's line-of-sight Doppler trail with orbital phase (Figure \ref{fig:orbit trail}). If the planetary signal is truly present, the CCF should trace out the predicted radial velocity with orbital phase given the literature reported $K_P$ and $V_\mathrm{sys}$. Figure \ref{fig:orbit trail} shows this is indeed the case. The fact that all correlation coefficients along the planet's path are positive also indicates that we are detecting absorption features, not emission features, as would be expected for a hot Jupiter with no thermal inversion in the infrared photosphere.

\section{Searching for Signatures of Atmospheric Dynamics}
\label{sec:velocities}

\begin{figure*}
    \centering
    \includegraphics[width=\linewidth]{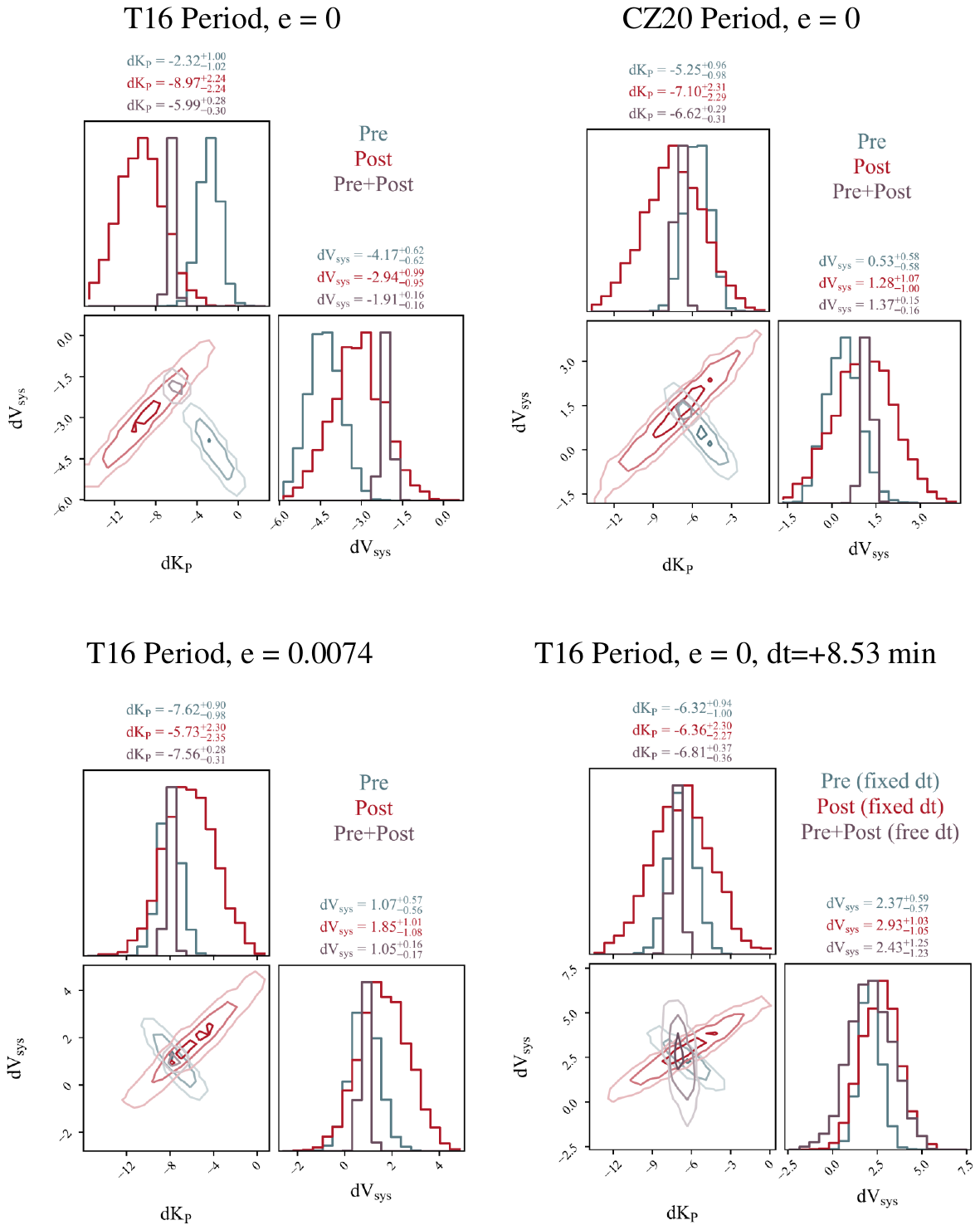}
    \caption{Posterior distributions for the radial velocity semi-amplitude, $K_P$, and systemic offset velocity, $V_\mathrm{sys}$, with different assumptions about WASP-77A b's orbit as described in Section \ref{sec:velocities}. Each label states the period reference and eccentricity. T16 stands for \cite{turner2016} and CZ20 stands for \cite{cortes}. }
    \label{fig:velocity summary}
\end{figure*}

The sub-km s$^{-1}$ velocity precision from current high resolution spectroscopy can enable sensitivity to the main atmospheric dynamical features in hot- or ultra-hot Jupiters such as day-night winds or equatorial jets \citep{flowers2019,beltz2021}. Phase-resolved CCFs can provide a powerful tool to probe these dynamics by tracking the planet signal's velocity in time \citep{ehrenreich2020, gandhi2022, pino2022}. In order to test for velocity consistency between the pre- and post-eclipse sequences and search for possible signatures of dynamics, we use the log-likelihood formalism with the PyMultinest sampler \citep{feroz2009, buchner2016} to obtain quantitative constraints on $K_P$ and V$_\mathrm{sys}$ for a given atmospheric template spectrum. We also include a multiplicative scale factor, $a$, that can stretch the model spectrum to account for any line amplitude mis-matches either due to a model inaccuracy or the SVD process. If multiple nights are considered at once, we include a separate $a$ for each individual night. We fit for $K_P$, $V_\mathrm{sys}$, and log$_{10}(a)$ using the full solar composition, H$_2$O only, and CO only models from Section \ref{sec:cross correlation} to search for any velocity offsets between gases as were found in \cite{brogi2023} or between pre- and post-eclipse sequences as were found in \cite{pino2022}.

\begin{figure*}
    \centering
    \includegraphics[width=\textwidth]{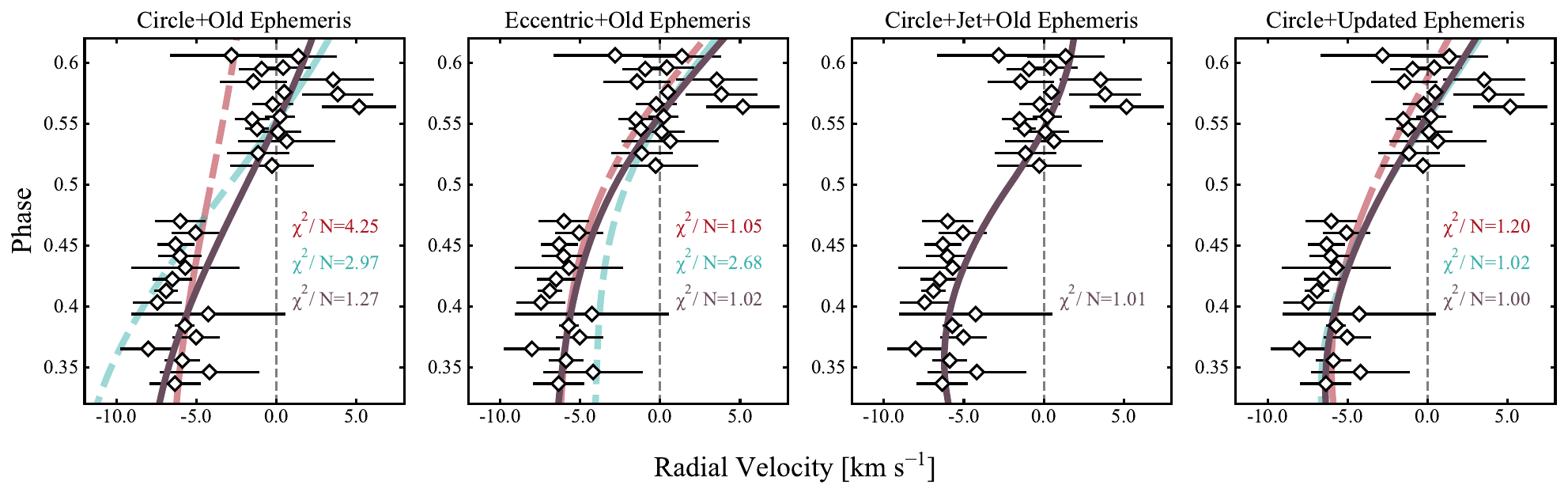}
    \caption{Measured velocity from the CCF signal in the planet literature rest frame (diamonds) and best fits from various assumptions about WASP-77A b's orbit. The purple lines are the estimated velocity trail using both IGRINS sequences, and the red and blue lines are estimates from each sequence alone. Pre-eclipse is red and post-eclipse is blue. The reduced $\chi^2$ from each best fit is also listed in its corresponding color. Only a combined pre+post fit appears for the jet model because after that initial analysis it was clear the jet was not a viable path of inquiry. Note that some of the data points appear to be at the same phase because the CCF frames were binned in BJD time and have been phase folded for visualization.}
    \label{fig:RV trails}
\end{figure*}

The planet's time-resolved velocity depends on its orbital phase, typically calculated by dividing the time elapsed since a measured transit midpoint by the period: $\varphi(t) = (t - T_C)/P$. Using up-to-date ephemerides is crucial -- initially we used the midtransit time and period from WASP-77A b's entry in the Transiting Exoplanet Catalogue (TEPCat, \citealp{southworth2011}, midtransit time from \citealp{bonomo2017} and period from \citealp{turner2016}) and found the pre- and post-eclipse $K_P$'s to be inconsistent by 3$\sigma$ (Figure \ref{fig:velocity summary}, top left). To assess goodness-of-fit provided by each subset of the IGRINS data, we fit Gaussian profiles to each frame of the CCF orbit trail to measure the velocity at each frame time. To build signal, we took the average of bins of 5 frames weighted by the individual CCF amplitudes and treated the standard error of that average as the 1$\sigma$ uncertainty on the velocity in that time bin. As can be seen in Figure \ref{fig:RV trails}, the best fit $K_P$ and $V_\mathrm{sys}$ from using both sequences provides a somewhat adequate fit ($\chi^2/N$=1.27, N=32), but the individual sequences fail to predict each other (pre-eclipse and post-eclipse giving $\chi^2/N$'s of 4.25 and 2.97, respectively). 

Such velocity asymmetries can arise from an orbital eccentricity or atmospheric dynamics, so we expand our velocity model to include either an eccentricity or an equatorial jet (the details of both are described in Appendix \ref{appendix:eccentricty model}). While the best fit eccentric model achieves consistency between pre- and post-eclipse and provides a better fit ($\chi^2/N$=1.02 using both sequences, Figure \ref{fig:RV trails}, second panel), we can only place an upper limit on the eccentricity and there is insufficient evidence to favor a freely eccentric orbit over a circular one. The jet model provides an even better fit ($\chi^2/N$=1.01, Figure \ref{fig:RV trails}, third panel) but ``detects" a supersonic westward jet at 3$\sigma$ (by comparing Bayes factors), which is physically implausible.

A more recent ephemeris analysis of WASP-77A b lists the eccentricity as $e=0.0074^{+0.007}_{-0.005}$. Fixing the eccentricity to this value once again achieves consistency between pre- and post-eclipse phases and is favored over the circular orbit by 4.2$\sigma$ (Figure \ref{fig:velocity summary}, lower left). However, when using the more up-to-date midtransit time from \cite{cortes}, the $K_P$ values become inconsistent again and this eccentricity is \textit{disfavored} over a circular orbit, indicating 0.0074 is too large of an eccentricity. Again, there is insufficient evidence for a\new{n} eccentricity measured on our own, and a circular orbit perfectly fits the data ($\chi^2$/N=1.00, Figure \ref{fig:RV trails}) and yields consistent $K_P$ values between sequences (Figure \ref{fig:velocity summary}, upper right). Fitting for a jet speed with the updated ephemerides is also disfavored over a simply circular orbit by 2.5$\sigma$.

This points to the velocity asymmetries arising from propagated ephemeris error. Indeed, if we add a time correction parameter to our initial circular orbit analysis with the TEPCat ephemerides, we are able to measure that WASP-77A b appears 8.53$^{+2.41}_{-2.39}$ minutes behind in phase, which translates to an initial measured period error of 0.21$\pm$0.06 seconds. Including this offset achieves consistent velocities between pre- and post-eclipse and is favored over the initial circular orbit fit by 3.6$\sigma$ (Figure \ref{fig:velocity summary}, bottom right). Fitting for such a correction with the \cite{cortes} midtransit time and period yields a value of 0 minutes to one part in a thousand. Therefore, we can conclude that WASP-77A b's orbit is effectively circular, but propagated ephemeris error from the TEPCat values induced an effective eccentricity. It should be noted that had the true eccentricity been as ``large" as 0.0074, we would have been sensitive to its effects.

In the future, the impact of propagated ephemeris error can be mitigated by using the most recent measured midtransit time before a given observation when calculating orbital phase. This may seem obvious, but many planets only have a few published ephemerides. In the likely scenario that published ephemerides measured recently before a given HRCCS observation don't exist, we recommend \new{taking advantage of the several campaigns monitoring known exoplanet host stars including with the Transiting Exoplanet Survey Satellite (e.g., \citealp[][]{wong2020}) and citizen science projects like ExoClock \citep{kokori2022} and Exoplanet Watch \citep{zellum2020}.} Searching for entries in the American Association of Variable Star Observers (AAVSO) database\footnote{https://www.aavso.org/}, we discovered that a transit of WASP-77A b was observed with the 6-inch MicroObservatory and subsequently analyzed through the Exoplanet Watch citizen science program in late October of 2020. Estimating once again WASP-77A b's orbital velocity using the \cite{turner2016} period with this midtransit time, the pre- and post-eclipse $K_P$'s are consistent.

Besides line positions, line \textit{shapes} are also affected by atmospheric dynamics. Assuming a circular orbit, we added an average line full width at half maximum (FWHM) parameter. This is the FWHM in pixels of a Gaussian kernel we convolved the 1D-RCTE model spectrum with instead of the nominal instrument profile and rotation kernels. For the pre- and post-eclipse sequences we were able to measure this FWHM to be 9.34$^{+0.59}_{-0.49}$ and 8.39$^{+0.76}_{-0.68}$ pixels, respectively, which is consistent with what we would expect from a combination of the instrument profile (4.5 pixels) and tidally locked planetary rotation (4.52 km s$^{-1}$, $\sim$7 pixels). There is nothing to suggest that dynamics or anything else beyond these two sources are affecting the line shape.

Finally, we find no significant velocity offsets between different gases. Therefore, we detect no signatures of atmospheric dynamics that may bias a 1D atmospheric retrieval analysis. Such analysis is presented in the following section.

\section{Estimating Gas Abundances and Vertical Thermal Structure}
\label{sec:retrieval}

To estimate WASP-77A b's atmospheric composition and vertical thermal structure, we apply the same retrieval framework described in \cite{line2021} and \cite{brogi2023}. Using PyMultinest paired with a GPU-accelerated version of CHIMERA \citep{line2013, line2021}, we estimate the constant-with-altitude abundances of the same gases mentioned in Section \ref{sec:cross correlation}, the 6-parameter analytic pressure-temperature (P-T) profile described in \cite{madhu2009}, and the velocities and scale factor estimated in Section \ref{sec:velocities}. \new{As in \cite{line2021}, we indirectly estimate the $^{13}$CO/$^{12}$CO isotopologue ratio by estimating directly [$^{13}$CO/$^{12}$CO] = log($^{13}$CO/$^{12}$CO)$_\mathrm{Planet}$ - log($^{13}$CO/$^{12}$CO)$_\mathrm{Earth}$, where ($^{13}$CO/$^{12}$CO)$_\mathrm{Earth}$ = 1/89 \citep{meibom2007}. For example, a retrieved [$^{13}$CO/$^{12}$CO] of 0.5 corresponds to a $^{13}$CO/$^{12}$CO ratio of 1/(89$^{-0.5}$) = 1/28.} Table \ref{tab:results} lists each parameter and their uniform prior bounds \new{as well as derived quantities such as the C/O ratio}.

\begin{figure*}
    \centering
    \includegraphics[width=0.7\textwidth]{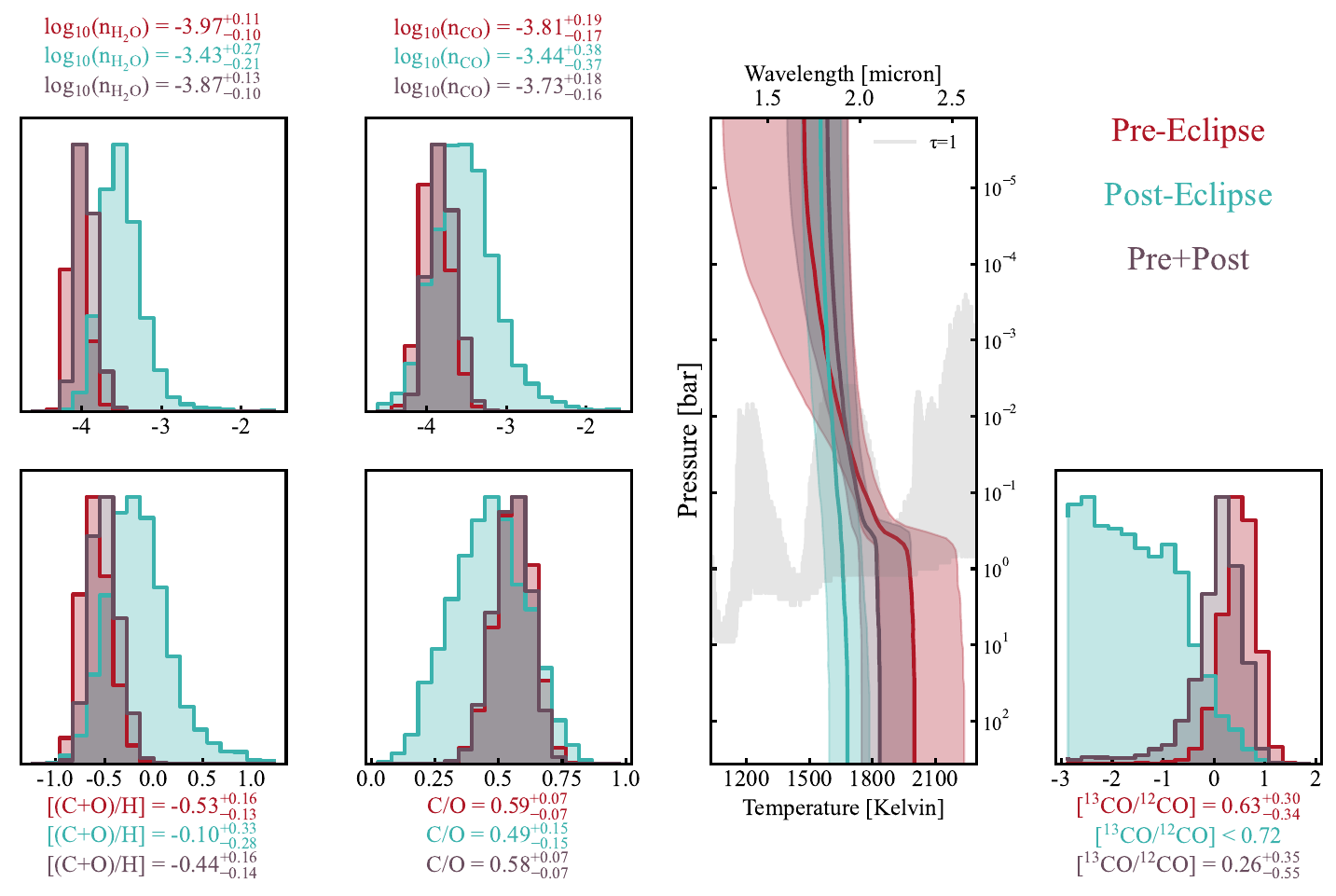}
    \caption{Marginalized posterior distributions of relevant gas abundances as well as the P-T profile from the three different subsets of IGRINS data: pre-eclipse (red), post-eclipse (blue), and all three nights (purple). The vertical P-T panel shows the median posterior profile (solid lines) and the shaded regions are the 1$\sigma$ confidence intervals. Also plotted is the $\tau$=1 spectrum assuming the best fit parameters from the pre+post retrieval.}
    \label{fig:pre post panel}
\end{figure*}

For retrievals on the NIRSpec/G395H data, the model spectra are computed at R=100,000 over those wavelengths ($\sim$2-5 micron). We then assume a top hat profile to bin the spectra down to the same wavelength bin widths as the data, which has an average R$\sim$250. We perform retrievals with the following data combinations: IGRINS pre-eclipse only, IGRINS post-eclipse nights combined, all IGRINS data combined, NIRSpec eclipse only, and IGRINS pre-eclipse and NIRSpec eclipse combined.  Table \ref{tab:results} lists each parameter, its prior, and its posterior median value for each retrieval. When considering both IGRINS and NIRSpec data simultaneously, we add the log-likelihood of the low-resolution data ($-\frac{1}{2} \chi^2$) to the IGRINS log-likelihood following \cite{brogi2019}. For the retrievals using only IGRINS data we use 500 live points, and we use 2500 for any retrieval including the NIRSpec data. This is because in our initial, exploratory retrieval analyses, the NIRSpec retrieval results varied with model resolution and number of live points. This variance asymptotically decreased as we increased both to the values adopted here. This behavior was not observed for the IGRINS-only retrievals. In the following subsections, we summarize the dependence of gas abundance and P-T profile estimates on the inclusion of different nights of IGRINS data (Figure \ref{fig:pre post panel}) and the NIRSpec data (Figure \ref{fig:high low panel}).

\begin{table*}[]
\centering
\begin{tabular}{l|l|l|l|l|l|l}
\hline
\hline
Parameter                                      & Prior                   & Pre-Eclipse             & Post-Eclipse            & Pre+Post                & NIRSpec                 & Pre+NIRSpec                                    \\
\hline
$\log_{10}(n_\mathrm{H_2O})$                   & $\mathcal{U}$(-12, 0)   & -3.97$^{+0.11}_{-0.10}$ & \textgreater -3.89      & -3.80$^{+0.16}_{-0.12}$ & -3.80$^{+0.34}_{-0.28}$ & -4.02$^{+0.07}_{-0.06}$                       \\
$\log_{10}(n_\mathrm{CO})$                     & $\mathcal{U}$(-12, 0)   & -3.81$^{+0.19}_{-0.17}$ & \textgreater -4.28      & -3.60$^{+0.22}_{-0.20}$ & -3.73$^{+0.55}_{-0.52}$ & -3.91$^{+0.13}_{-0.13}$                        \\
$\log_{10}(n_\mathrm{CH_4})$                   & $\mathcal{U}$(-12, 0)   & \textless -5.97         & \textless -1.77         & \textless -5.57         & -                       & \textless -6.93                                \\
$\log_{10}(n_\mathrm{H_2S})$                   & $\mathcal{U}$(-12, 0)   & \textless -4.45         & \textless -1.50         & \textless -4.36         & -                       & \textless -5.07                                \\
$\log_{10}(n_\mathrm{NH_3})$                   & $\mathcal{U}$(-12, 0)   & \textless -5.88         & \textless -3.72         & \textless -6.01         & -                       & \textless -6.03                                \\
$\log_{10}(n_\mathrm{HCN})$                    & $\mathcal{U}$(-12, 0)   & \textless -5.69         & \textless -2.43         & \textless -5.46         & -                       & \textless -.612                               \\
$[^{13} \mathrm{CO}/^{12}\mathrm{CO}]_\oplus$ & $\mathcal{U}$(-5, 5)    & 0.63$^{+0.30}_{-0.34}$  & \textless 0.70          & 0.26$^{+0.33}_{-0.43}$  & \textless 1.78          & 0.33$^{+0.35}_{-0.71}$                         \\
$\log_{10}(n_\mathrm{CO_2})$                   & $\mathcal{U}$(-12, 0)   & -                       & -                       & -                       & \textless -6.34         & \textless -7.25                                \\
$T_0$ {[}K{]}                                  & $\mathcal{U}$(500,2500) & 1470$^{+210}_{-370}$    & 1480$^{+110}_{-110}$    & 1460$^{+180}_{-280}$    & 1300$^{+80}_{-50}$      & 1400$^{+30}_{-40}$                             \\
$\log_{10} P_1$ {[}$\log_{10}$ bar{]}          & $\mathcal{U}$(-5.5,2.5) & \textgreater -4.25      & -1.78$^{+0.81}_{-0.61}$ & \textgreater -2.14      & \textgreater -1.37      & 0.69$^{+0.22}_{-0.24}$                         \\
$\log_{10} P_2$ {[}$\log_{10}$ bar{]}          & $\mathcal{U}$(-5.5,2.5) & Unc.                    & Unc.                    & Unc.                    & Unc.                    & Unc.                                          \\
$\log_{10} P_3$ {[}$\log_{10}$ bar{]}          & $\mathcal{U}$(-2,2)     & \textgreater{}-0.83     & Unc.                    & -0.59$^{+0.24}_{-0.18}$ & \textless -1.16         & \textgreater -0.51                             \\
$\alpha_1$                                     & $\mathcal{U}$(0.02,2)   & \textgreater 0.31       & \textgreater 0.29       & \textgreater 0.29       & 0.48$^{+0.04}_{-0.03}$  & 0.60$^{+0.04}_{-0.05}$                         \\
$\alpha_2$                                     & $\mathcal{U}$(0.02,2)   & Unc.                    & Unc.                    & Unc.                    & Unc.                    & \textgreater 0.03                              \\
$dK_P$ {[}km/s{]}                              & $\mathcal{U}$(-20,20)   & -1.26$^{+0.85}_{-0.85}$ & -9.62$^{+2.02}_{-2.07}$ & -5.75$^{+0.27}_{-0.27}$ & -                       & -1.29$^{+0.75}_{-0.78}$                        \\
$dV_\mathrm{sys}$ {[}km/s{]}                   & $\mathcal{U}$(-20,20)   & -5.27$^{+0.51}_{-0.53}$  & -3.84$^{+0.86}_{-0.86}$ & -2.46$^{+0.15}_{-0.16}$ & -                       & -5.27$^{+0.48}_{-0.47}$                        \\
$\log(a_1)$                                    & $\mathcal{U}$(-2,2)     & -                       & 0.80$^{+0.37}_{-0.44}$  & 0.04$^{+0.53}_{-0.33}$  & -                       & -                                              \\
$\log(a_2)$                                    & $\mathcal{U}$(-2,2)     & 0.15$^{+0.52}_{-0.53}$  & 0.73$^{+0.38}_{-0.43}$  & 0.14$^{+0.54}_{-0.32}$  & -                       & 0.03$^{+0.04}_{-0.04}$   \\
$\log(a_3)$                                    & $\mathcal{U}$(-2,2)     & -                       & -                       & -0.05$^{+0.55}_{-0.33}$ & -                       & -                                              \\
$\log(a_4)$                                    & $\mathcal{U}$(-2,2)     & -                       & -                       & -                       & 0.01$^{+0.02}_{-0.03}$  & 0.03$^{+0.01}_{-0.02}$                         \\
\hline
C/O                                            &                         & 0.59$^{+0.07}_{-0..07}$ & 0.58$^{+0.16}_{-0.19}$  & 0.61$^{+0.08}_{-0.07}$  & 0.54$^{+0.15}_{-0.16}$  & 0.57$^{+0.06}_{-0.06}$                         \\
{[}(C+O)/H{]}$_\odot$                          &                         & -0.53$^{+0.16}_{-0.13}$ & \textgreater -0.68      & -0.32$^{+0.20}_{-0.17}$ & -0.42$^{+0.49}_{-0.42}$ & -0.61$^{+0.11}_{-0.10}$                        \\
{[}O/H{]}                                      &                         & -0.54$^{+0.15}_{-0.12}$ & \textgreater -0.62      & -0.34$^{+0.19}_{-0.16}$ & -0.41$^{+0.45}_{-0.38}$ & -0.61$^{+0.10}_{-0.09}$                        \\
{[}C/H{]}                                      &                         & -0.51$^{+0.15}_{-0.12}$ & \textgreater{}-0.97     & -0.29$^{+0.22}_{-0.20}$ & -0.43$^{+0.55}_{-0.52}$ & -0.60$^{+0.13}_{-0.13}$                      \\ 
\new{$^{13}\mathrm{CO} / ^{12}\mathrm{CO}$} &    &  \new{1/21$^{+25}_{-11}$}  & \new{\textless 1/18}         &  \new{1/50$^{+127}_{-27}$}  & \new{\textless 1/1.48}          & \new{1/31$^{+92}_{-18}$}  \\
\hline
\hline
\end{tabular}
\caption{Free parameters and their retrieved values for each retrieval. Values listed with uncertainties are bounded constraints, while values with a $>$ or $<$ are 3-$\sigma$ lower or upper limits, respectively. Entries with ``Unc." are unconstrained, and those with ``\new{-}" were not included as free parameters for that specific retrieval. The parameters below the horizontal line are derived and not retrieved for directly. The scale factors, $a_i$, correspond to each specific night/data set. 1, 2, 3, and 4 are IGRINS nights 12/6, 12/14, 12/21, and the low resolution data, in that order.}
\label{tab:results}
\end{table*}

\subsection{Pre-Eclipse Data Only}
\label{subec:pre}

Constraints from the pre-eclipse data alone, shown in red in Figure \ref{fig:pre post panel}, are identical within the statistical noise of the nested sampler to those presented in \cite{line2021}. We place bounded constraints\footnote{We refer to marginalized posterior distributions that are not against the prior bounds as ``bounded" and report the median value and 1$\sigma$ quantiles on either side of this median. For posterior distributions against the upper (lower) prior, we report 3$\sigma$ lower (upper) limits. Posteriors against both prior bounds are ``unconstrained".} on the the gas abundance of both H$_2$O and CO to within 0.2 dex as well as a bounded constraint on the $^{13}$CO\new{/}$^{12}$CO \new{isotopologue ratio at 1/22$^{+14}_{-10}$}. This ratio is enriched compared to the terrestrial standard \citep[1\new{/}89;][]{meibom2007} and local ISM \citep[1\new{/}70;][]{wilson1994}. We can only place upper limits on the other gases in the model, all of which are expected to be much less abundant than H$_2$O and CO following expectations from equilibrium chemistry. The median retrieved P-T profile is non-inverted and temperature is monotonically increasing with pressure. These gas abundances are physically plausible under assumptions of both equilibrium and disequilibrium chemistry, and the P-T profile is consistent with an atmosphere with efficient day-to-night heat transport and/or day-night cold trapping \citep{line2021}. All bounded values are within 1$\sigma$ of their reported values in \cite{line2021}. For comparison to NIRSpec, we also performed a retrieval on the pre-eclipse data with the abundance of CO$_2$ as an added parameter while leaving out the rest of the trace gases besides H$_2$O and CO. Unsurprisingly, only an upper limit is placed on the CO$_2$ abundance.

\subsection{Including the Post-Eclipse Data}
\label{subsec:post}

The retrieval constraints from the post-eclipse data alone are summarized in blue in Figure \ref{fig:pre post panel}. Initially, when removing the first 4 principal components, we struggled to make any informative inferences and could only place lower limits on the abundances of H$_2$O and CO. Adjusting to remove only the first 3 principal components makes a drastic improvement in our inference capabilities and enable us to place bounded constraints on both H$_2$O and CO consistent with the pre-eclipse values albeit with slightly lower precision. We can only place an upper limit on the $^{13}$CO\new{/}$^{12}$CO ratio, but it is consistent with the retrieved pre-eclipse value. The top-of-atmosphere temperature is well constrained, but the rest of the retrieved P-T profile is poorly constrained and near-isothermal. The slightly less stringent constraints compared to the pre-eclipse retrieval may be due to the detrending process but also due to a changing P-T profile with phase that the 1D model struggles to capture. Both are discussed further in Section \ref{sec:discussion}.

When we combine all three nights of IGRINS data, shown in purple in Figure \ref{fig:pre post panel}, the constraints on H$_2$O and CO are consistent with both the pre- and post-eclipse sequences individually. The median retrieved values are between those from the pre- and post-eclipse sequences, weighted more in favor of the pre-eclipse sequence, which appears to be driving these inferences. These constraints are not more precise than those provided by the pre-eclipse data, suggesting that the post-eclipse data contributed little new compositional information.

\subsection{NIRSpec}
\label{subsec:nirspec}

\begin{figure*}
    \centering
    \includegraphics[width=0.7\textwidth]{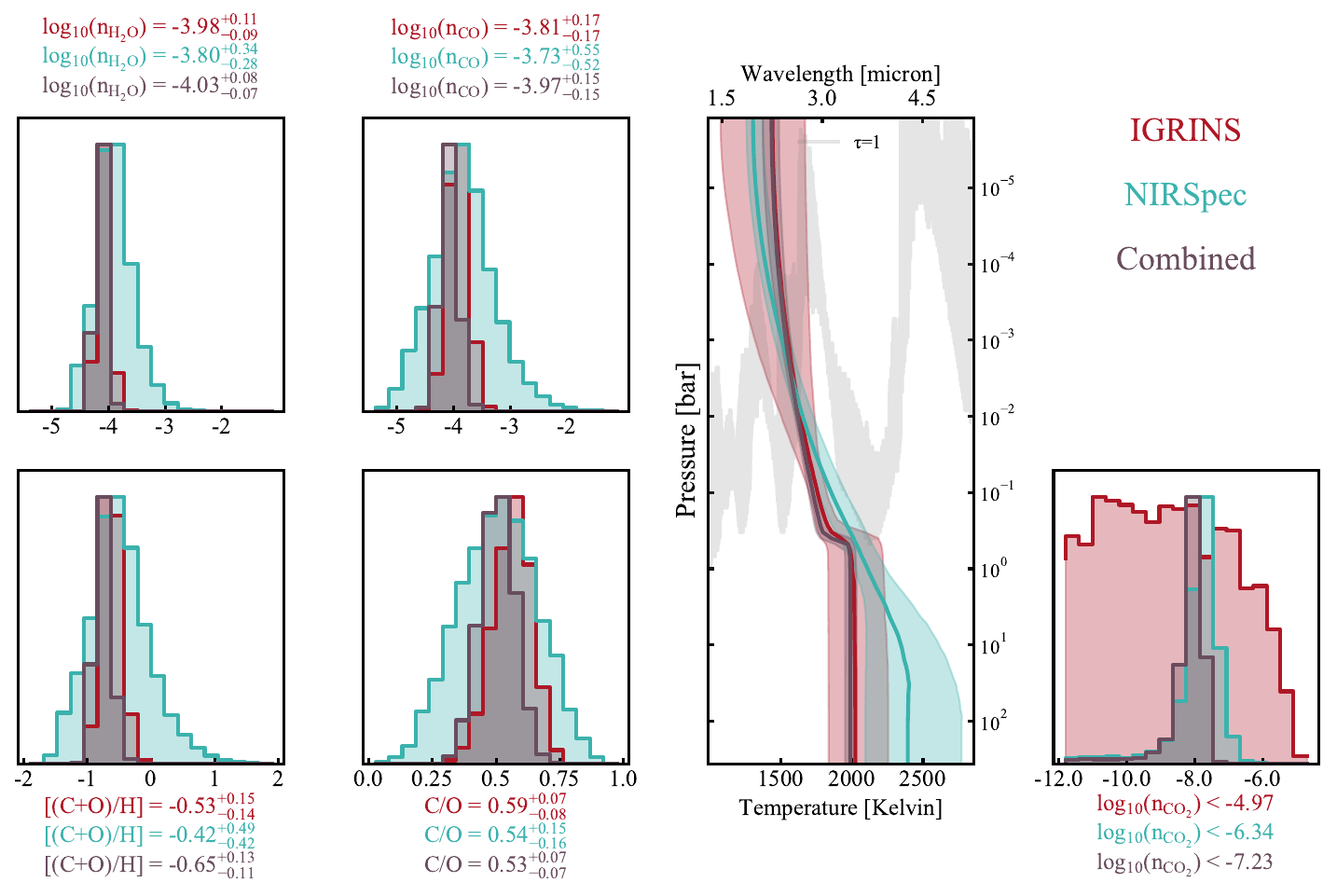}
    \caption{Similar to Figure \ref{fig:pre post panel}, marginalized posterior distributions of relevant gas abundances as well as the P-T profile from the pre-eclipse IGRINS data alone (red), NIRSpec alone (blue), and both combined (purple). From the confidence intervals of these parameters, it is evident that in the combined retrieval, the composition inferences are driven by the high resolution data while the P-T inferences are driven by the low resolution data.}
    \label{fig:high low panel}
\end{figure*}

We apply the same retrieval framework to the NIRSpec/G395H secondary eclipse data presented in \cite{august2023}. For initial exploratory retrievals with lower model spectral resolution and fewer live points, we include all of the gases used in the high-resolution retrievals in addition to CO$_2$ and SO$_2$, prominent absorbers over the 4-5 micron region. Just as with the IGRINS retrievals, we could only place bounded constraints on H$_2$O and CO and upper limits on all the other gases. The upper limits on the other gases did not get more stringent than the IGRINS limits, and for the final, fiducial retrieval with 2500 live points and R=100,000, we only include H$_2$O, $^{12}$CO, $^{13}$CO, and CO$_2$ in addition to the continuum opacities.

The results of this retrieval are summarized in blue in Figure \ref{fig:high low panel}. The constraints both on the composition and vertical thermal structure are consistent with IGRINS. The constraints on the H$_2$O and CO abundances are slightly less precise ($\sim 3\times$) than what we achieved with IGRINS. We can only place an upper limit on the $^{13}$CO\new{/}$^{12}$CO ratio, but it is consistent with the measured IGRINS value. The P-T profile is more precisely constrained in the upper atmosphere due to the lower pressures probed at these longer wavelengths. Only an upper limit can be placed on the abundance of CO$_2$, and its inclusion as a model parameter is only favored by 1.8$\sigma$ (Bayes factor 2). \new{Within the context of theoretical predictions that the abundance of CO$_2$ is highly sensitive to atmospheric metallicity (e.g., \citealp{lodders2002, zahnle2009, moses2013}),} the absence of a clear CO$_2$ absorption feature around 4.5 microns further qualitatively confirms the sub-solar metallicity of WASP-77A b's atmosphere as previously found with the IGRINS data (\new{in contrast to} published JWST results showing CO$_2$, e.g., \citealp{ers2023, bean2023}).

\subsection{IGRINS+NIRSpec}
\label{subsec:igrins+nirspec}

\begin{figure*}
    \centering
    \includegraphics[width=\linewidth]{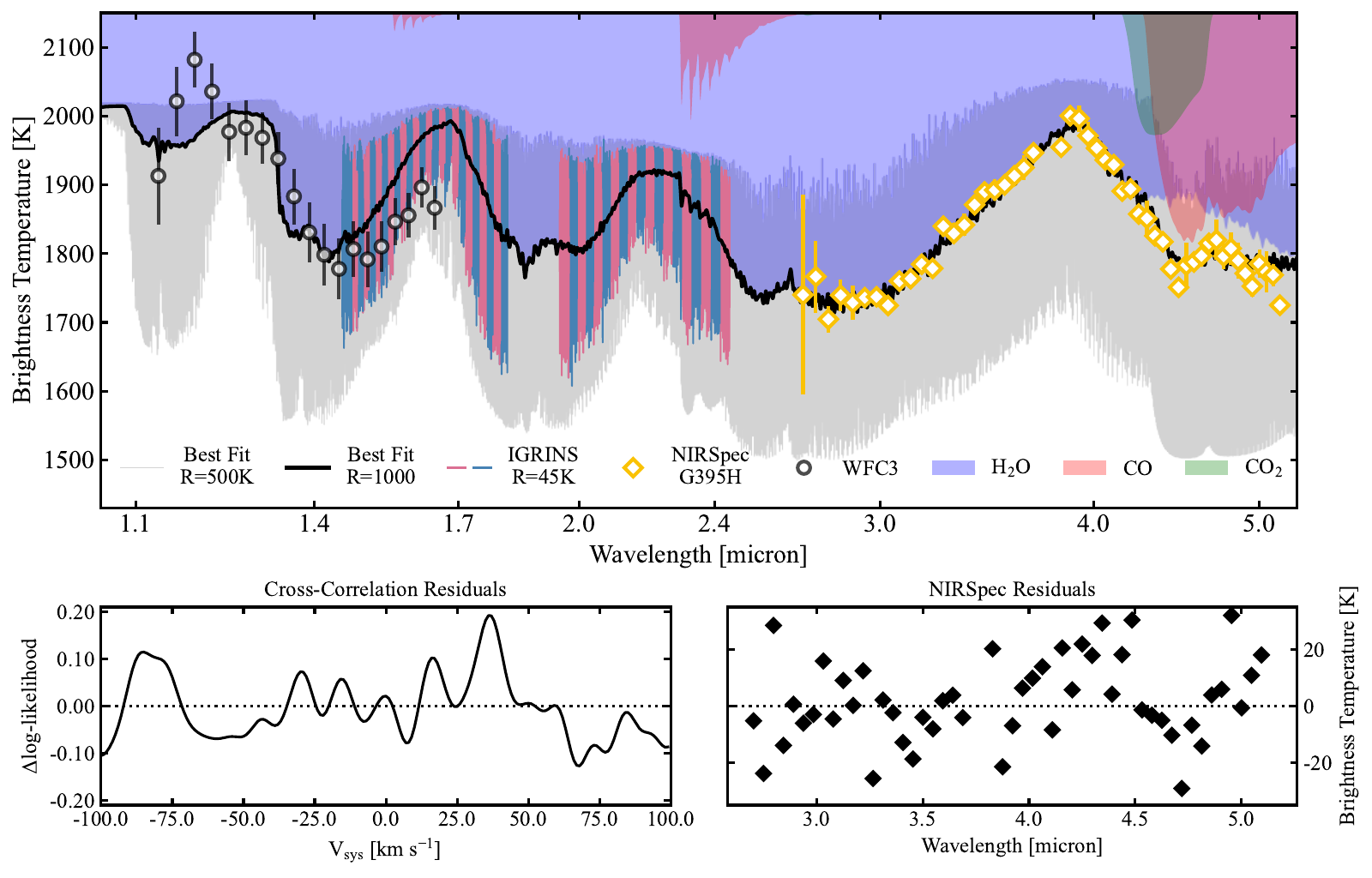}
    \caption{\textit{Top:} Best fit model spectrum from the combined IGRINS+NIRSpec retrieval at R=500,000 (gray), converted to brightness temperature and smoothed to R=1000 (black) for visual clarity. The shaded regions illustrate contributions and light absorbed from H$_2$O, CO, and CO$_2$ individually. The NIRSpec data is shown as gold diamonds and has been further binned down to R$\sim$85 for visual clarity. While the planet spectrum is not visible in the actual IGRINS data, to give a sense of its wavelength coverage and spectral resolution, the best fit has been further convolved to the IGRINS resolution (R=45,000) and  interpolated onto the instrument's wavelength grid, alternating colors with echelle order. Goodness of fit with both data sets is discussed in detail in Section \ref{subsec:predictive}. \new{Also shown for comparison are the previously published HST/WFC3 eclipse measurements presented in \cite{mansfield2022} but not considered in this study (circles).} \textit{Bottom Left:} Cross-correlation residuals with the IGRINS data after the best-fit spectrum has been divided out. A 2D detection map was calculated as in Section \ref{sec:cross correlation}, and shown here is a horizontal slice at the planet's $K_P$. The lack of a signal at rest velocity indicates the best fit spectrum is an adequate fit to the true planet signal. This process is described more in Section \ref{subsec:predictive}. \textit{Bottom Right:} Residuals from subtracting the binned NIRSpec data from the best fit binned onto the same wavelength grid (black squares in top panel).}
    \label{fig:best fit fig}
\end{figure*}

Here, we combine the IGRINS and NIRSpec data into a single retrieval. Because the post-eclipse data did not provide new information, we only use the pre-eclipse data in the interest of computational efficiency (high resolution retrievals take days to weeks to complete). The results of this combined retrieval are shown in purple in Figure \ref{fig:high low panel}. The inferences of the planet's atmosphere are consistent with both the IGRINS and NIRSpec individual analyses, and the constraints on both the composition and vertical thermal structure are more precise than either data set provided alone (composition by about 30\% compared to pre-eclipse only and P-T profile by about 50\% compared to NIRSpec alone). It is apparent that the composition constraints are driven by the IGRINS data, whereas the NIRSpec observations provide more precise contraints on the P-T profile. This is not surprising as the low resolution data retains continuum information while the high resolution data is more sensitive to gas abundances via individual molecular line shapes and ratios. Similar to the NIRSpec-only retrieval, we can only place an upper limit on the abundance of CO$_2$, but a bounded constraint is placed on the $^{13}$CO\new{/}$^{12}$CO ratio consistent with the IGRINS-only retrieval. The best fit model spectrum is shown in Figure \ref{fig:best fit fig}, showing remarkable agreement with both data sets across this wide wavelength range (1.5-5 micron).

\section{Discussion}
\label{sec:discussion}

\subsection{Bulk Composition and Placing WASP-77A b into Context}
\label{subsec:context}

A major goal of exoplanet science is to tie back the composition of planetary atmospheres to formation pathways via diagnostics like bulk metallicity and the C/O ratio. These diagnostics can be derived from our inferred gas abundances like so:

\begin{equation}
    \begin{aligned}
     [\mathrm{(C+O)/H}]_\odot &= \\
     &\mathrm{log}_{10} \Bigg[ \frac{n_\mathrm{H2O} + 2n_\mathrm{CO} + n_\mathrm{CH4}  + n_\mathrm{HCN} + 3n_\mathrm{CO_2}}{2n_\mathrm{H2} [(n_\mathrm{O} + n_\mathrm{C})/n_\mathrm{H}]_\odot} \Bigg]
     \end{aligned}
\end{equation}

and

\begin{equation}
    \mathrm{C/O} = \frac{n_\mathrm{C}}{n_\mathrm{O}} = \frac{n_\mathrm{CO} + n_\mathrm{CH4} + n_\mathrm{HCN} + n_\mathrm{CO2}}{n_\mathrm{H2O} + n_\mathrm{CO} + 2n_\mathrm{CO_2}}
\end{equation}
where $n_i$ is the abundance of gas $i$. We assume the solar values from \cite{asplund2009}. The median values of these derived quantities from each retrieval set up are listed in Table \ref{tab:results}. For the combined IGRINS+NIRSpec analysis, these are [(C+O)/H] = -0.61$^{+0.11}_{-0.10}$, C/O = 0.57$^{+0.06}_{-0.06}$, [O/H] = -0.61$^{+0.10}_{-0.09}$, and [C/H] = -0.60$^{+0.13}_{-0.13}$. These values are consistent with \cite{august2023}'s values for [M/H] (-0.91$^{+0.24}_{-0.16}$) and C/O  (0.36$^{+0.10}_{-0.09}$) within $\sim$2$\sigma$. The slight differences can be attributed to the many differences between the two analyses including modeling assumptions, model resolution, number of live points, and the modeling code itself. Slight differences in interpretations between modeling codes is a known occurrence for other JWST data sets (\citealp[e.g.,][]{taylor2023}), and model synthesis in the context of JWST exoplanet retrievals is an ongoing effort beyond the scope of this paper.

\cite{line2021} interpreted WASP-77A b's metallicity to be sub-stellar as previous studies had measured WASP-77A's metallicity (via [Fe/H]) to be consistent with solar \citep[0.00 $\pm$ 0.11, -0.10$^{+0.10}_{-0.11}$][respectively]{maxted2013, cortes}. More recent studies measuring the stellar [O/H] and [C/H] have since placed WASP-77A's [(C+O)/H] at slightly super-solar \citep[0.32 $\pm$0.04, 0.33$\pm$0.09;][]{polanski2022, reggiani2022}, and the qualitative interpretation of the planet's sub-stellar metallicity does not change. \cite{polanski2022} and \cite{reggiani2022} also both measured WASP-77A's C/O ratio to be slightly sub-solar (0.46 $\pm$0.09 and 0.44$^{+0.07}_{-0.08}$, respectively). While our median retrieved C/O ratio for the planet could be interpreted then as super-solar, it is consistent with these new stellar values within about 2$\sigma$. \new{On the other hand, \cite{kolecki2022} measure WASP-77A's C/O ratio to be $0.59 \pm 0.08$, in which case the the planet's C/O ratio would be almost exactly stellar}. Furthermore, if we account for partial sequestration of the total O inventory due to rainout of refractory condensates \citep{burrows1999}\footnote{$n_{O, true} = n_{O, observed} + 3.28 \times n_{Si}$. For $n_{Si}$ we assumed the Si/O ratio is the same as the star using [O/H] and [Si/H] from \cite{polanski2022}.}, which is plausible as discussed in Section \ref{subsec:TCGs}, the C/O ratio drops to 0.52$\pm$0.06, which is consistent with the stellar values \new{from \cite{polanski2022} and \cite{reggiani2022}}.

\new{While we present here extremely precise composition estimates, linking a planet's composition to its formation history is not trivial and depends heavily on planet and disk modeling assumptions}. As stated in \cite{line2021}, the combination of a substellar metallicity and stellar C/O ratio is not a common prediction from \new{many} planet formation theories (\citealp[e.g.,][]{oberg2016,madhu2017, khorshid2021}), but it is not implausible. One possible pathway under this scenario is formation via pebble accretion interior to the H$_2$O ice line from C-depleted gas with little-to-no planetesimal pollution or core erosion \citep{madhu2017}. \new{Formation beyond the CO$_2$ ice line is also possible if the planet migrated after disk-dissipation regardless of if the planet's final C/O is stellar \citep{madhu2014, schneider2021} or super-stellar \citep{reggiani2022}. \cite{schneider2021} further predict that low metal content is more common for planets that formed at large distances in the presence of pebble evaporation. However, low metallicities combined with stellar C/O ratios are rare in these models and depend on a low disk viscosity. \cite{mousis2019} construct a model in which devolatilization of amorphous ice from pebbles enriches the gas in Jupiter’s feeding zone, which enhances the C/O ratio while keeping the overall metallicity modest, but this model is tuned to the specific problem of Jupiter’s composition.}

\new{These formation pathways are supported by more recent, targeted studies -- \cite{bitsch2022} and \cite{khorshid2023} run formation simulations specifically for WASP-77A b, and both infer that the planet most likely formed beyond the CO$_2$ ice line and migrated late. The enrichment of $^{13}$CO in WASP-77A b's atmosphere may also be an indicator of formation at large distances. \cite{zhang2021} suggest a similar enrichment in the atmosphere of the planetary mass companion YSES-1 b may be due to a combination of less efficient $^{12}$CO self shielding and more efficient isotope exchange reactions at large orbital separations in the protoplanetary disk. However, this is speculative and no standard model linking isotope ratios with exoplanet formation has been developed in the literature. Ultimately, these measurements of WASP-77A b's composition will be more meaningful in the future in the context of a larger sample size of similar measurements with which we can test the wide range of planet formation theories.}

\begin{figure}
    \centering
    \includegraphics[width=\linewidth]{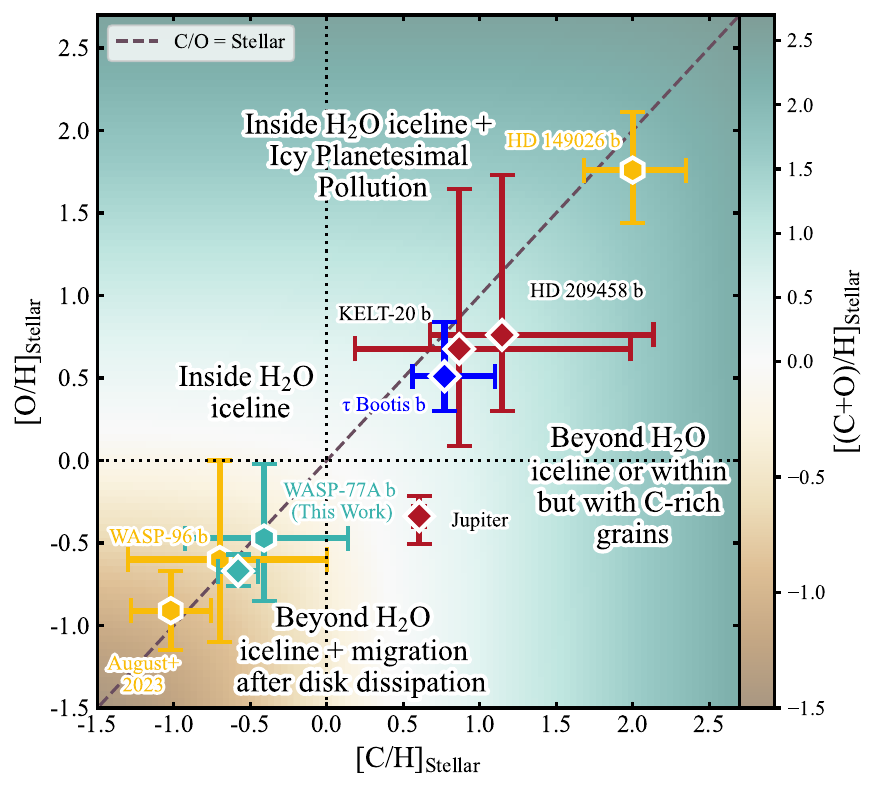}
    \caption{WASP-77A b in context of other hot- and ultra-hot Jupiters with both measured [O/H] and [C/H] values. These include HD 209458 b (\citealp{gandhi2019}, VLT/CRIRES and HST/WFC3), \new{$\tau$ Bootis b (\citealp{pelletier2021}, CFHT/SPIRou),} KELT-20 b (\citealp{kasper2023}, GN/MAROON-X and HST/WFC3), HD 149026 b (\citealp{bean2023}, JWST/NIRCam), \new{and WASP-96 b (\citealp{taylor2023}, JWST/NIRISS)}. Note the first two studies also combined high- and low-resolution data to obtain these measurements. The values and uncertainties here for WASP-77A b are from the combined IGRINS+NIRSpec retrieval \new{(turquoise diamond)}, which were mostly driven by the IGRINS data, but the NIRSpec-only constraints are also shown for comparison \new{(denoted with a hexagon)}. Also included is Jupiter \citep{atreya2016} to illustrate the high precision achieved on WASP-77A b's composition with the combined IGRINS and NIRSpec analysis, as well as regions in this parameter space associated with several broad predicted formation pathways as summarized in \cite{reggiani2022}.}
    \label{fig:context}
\end{figure}

Such inferences about a planet's formation history were more difficult in the era of the \textit{Hubble Space Telescope}. WFC3 was mainly sensitive to H$_2$O and not CO \new{or CO$_2$}, so attempting to measure the complete O or C inventory in a hot Jupiter was impossible. With the advent of high resolution retrievals \citep{brogi2019, gibson2020}, the abundances of both H$_2$O and CO, and indirectly [O/H] and [C/H], have been measured in a handful of hot- and ultra-hot Jupiters in the past few years \citep{gandhi2019, line2021,pelletier2021, kasper2023}. The launch of JWST has increased our capabilities to do so and already the sample of such measurements has grown \citep{bean2023, august2023, taylor2023}. Nonetheless, the current dearth of planets in which these two have been reliably estimated is stark (Figure \ref{fig:context}), highlighting the long path ahead for characterizing the transiting giant planet population as a whole. 

In the landscape of other transiting giants for which [O/H] and [C/H] have both been measured in some capacity  (i.e., \textit{constrained}), IGRINS has provided some of the most precise measurements to date, on par with measurements of [O/H] and [C/H] for planets in the solar system. However, as we demonstrated here and as apparent in the case of WASP-18 b \citep{brogi2023, coulombe2023}, JWST has a better grasp on vertical thermal structure. Measurements of both composition and climate go hand-in-hand and are necessary to interpret each other -- thus there is high value in combining high- and low-resolution data as the capabilities of both are expanded.

\subsection{Comparison to Mansfield et al. 2022}
\label{subsec: mansfield}

\new{WASP-77A b was previously observed in eclipse by \cite{mansfield2022} using HST/WFC3 and Spitzer's IRAC channels centered at 3.6 and 4.5 microns. Using a similar ``free" chemistry prescription and the same analytic P-T profile parameterization as used in this paper, only a rough lower limit could be placed on the planet's atmospheric metallicity. However, through a 1D-RCTE model grid search, a moderately super-solar metallicity was inferred, apparently in tension with the sub-solar metalicity preferred by the IGRINS data. A frequentist $\chi^2$ statistic could not reliably place a clear preference for either case, with the low metallicity IGRINS best-fit from \cite{line2021} giving a $\chi^2_\nu$ of 1.32 and the super-solar metallicity best grid fit yielding 1.24. Qualitatively, the WFC3 data fall well within the \cite{line2021}'s posterior distribution of model spectra, indicating that while the two data sets are consistent, the WFC3 data may be insufficient, either in quality or wavelength coverage, to reliably place precise constraints on gas abundances.}

\new{When incorporating the WFC3 data in their NIRSpec analysis, \cite{august2023} inferred a higher, solar metallicity than from the NIRSpec data alone. However, again frequentist metrics could not place a clear preference between the solar and sub-solar metallicity best fit models, and those authors suggest the WFC3 data may be unreliable. In initial exploratory retrieval analyses, we tested combining the WFC3 data with the IGRINS data and found negligible differences in the inferences or precision on the atmospheric composition. Because the data did not appear to contribute new information, we chose not to include the WFC3 and Spitzer data in the combined IGRINS and NIRSpec analyses. In regards to pressures probed and estimating the vertical thermal structure, NIRSpec has access to the same pressures as WFC3 (the role this plays in inferences of the P-T profile is discussed further in Section \ref{subsec:predictive}), so little vertical thermal information was lost by excluding the WFC3 data.}

\new{Our combined IGRINS+NIRSpec best fit spectrum is broadly consistent with the WFC3 data (Figure \ref{fig:best fit fig}). However, the spectral slope starts to diverge in the reddest data points and the reduced chi-square statistic with the WFC3 data is $\chi^2/N$ = 1.92. Investigating whether any specific points are driving inferences toward a higher or lower metallicity, such as with a leave-one-out cross validation analysis \citep{welbanks2023}, may elucidate why both IGRINS and NIRSpec are yielding different results than WFC3. However, such an analysis is beyond the scope of this paper and the application of leave-one-out cross validation has not yet been validated for use on high-resolution data. That NIRSpec has the ability to give meaningful constraints on individual gas abundances when WFC3 struggled to do so for the same planet speaks to the exceptional quality of JWST data and the dramatic improvements in the capabilities of space-based exoplanet spectroscopy in little over a year's time.}

\subsection{Comparing the Predictive Power of IGRINS and NIRSpec}
\label{subsec:predictive}

The joint IGRINS and NIRSpec retrieval presented in Section \ref{subsec:igrins+nirspec} shows that these two datasets are a powerful combination for atmospheric model parameter inference. Estimates from this retrieval are more precise than either data set provided alone, and it appears that the composition inferences were primarily driven by IGRINS while the P-T inferences were driven by NIRSpec. In this subsection, we will identity in what specific ways the two datasets are complementary and how they support the shortcomings of each other.

\begin{figure*}
    \centering
    \includegraphics[width=\linewidth]{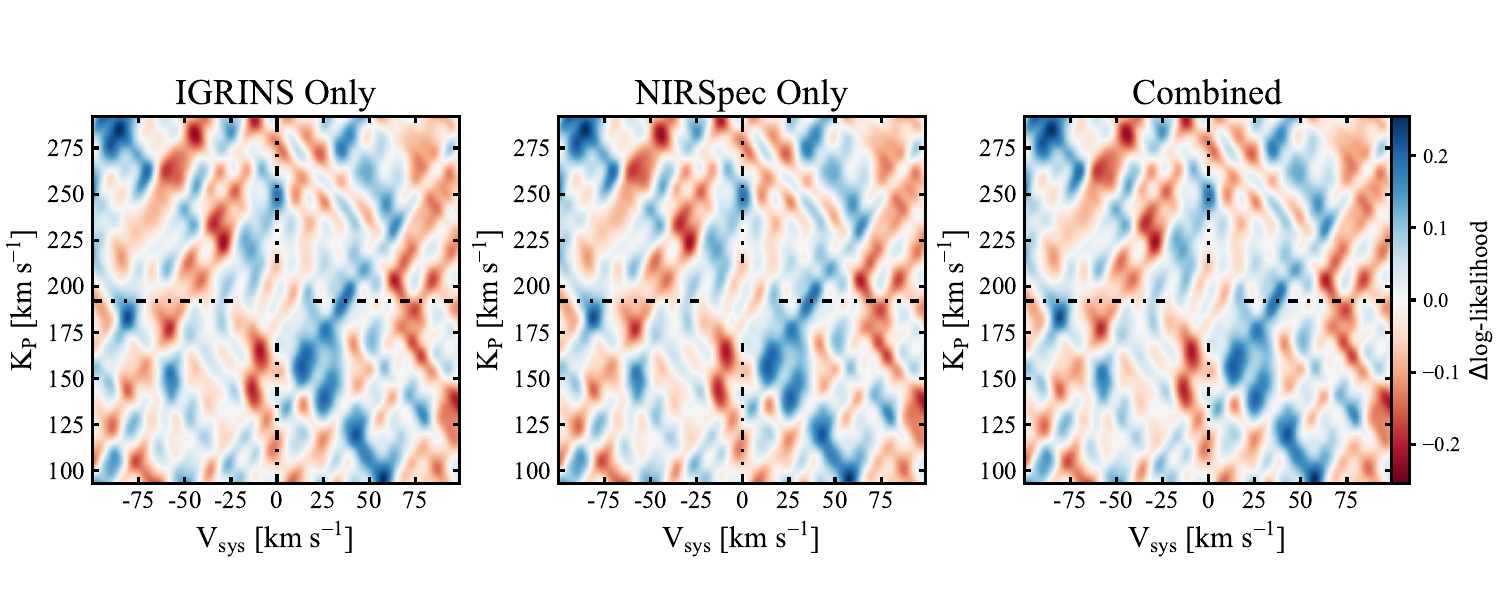}
    \caption{Residual $\Delta$log-likelihood maps from cross-correlating the solar composition 1D-RCTE template with the IGRINS data after the best fit model spectra from each of the three main retrievals were divided out. The maps have been median subtracted, and in contrast to the CCF maps in Figure \ref{fig:all ccf}, there are no peaks near the expected planet velocity, indicating the best fit spectra matched the true underlying planet signal within the precision of the IGRINS data.}
    \label{fig:predictive ccfs}
\end{figure*}

To compare how well each retrieval predicts the IGRINS data, we divide out the best fit model spectrum from the IGRINS, NIRSpec, and IGRINS+NIRSpec retrievals and then calculate cross-correlation maps using the same solar composition model as in Section \ref{sec:cross correlation}. If a given model is a good fit, there should be no significant peak at the planet $K_P$ and V$_\mathrm{sys}$, and the maps will only show the cross-correlation signal of residual noise. This is indeed the case for all three models (Figure \ref{fig:predictive ccfs}). There is no significant difference between the residual maps themselves, indicating that each of the three best fit spectra predicted the IGRINS data equally well to within the quality of the IGRINS data.

\begin{figure*}
    \centering
    \includegraphics[width=\textwidth]{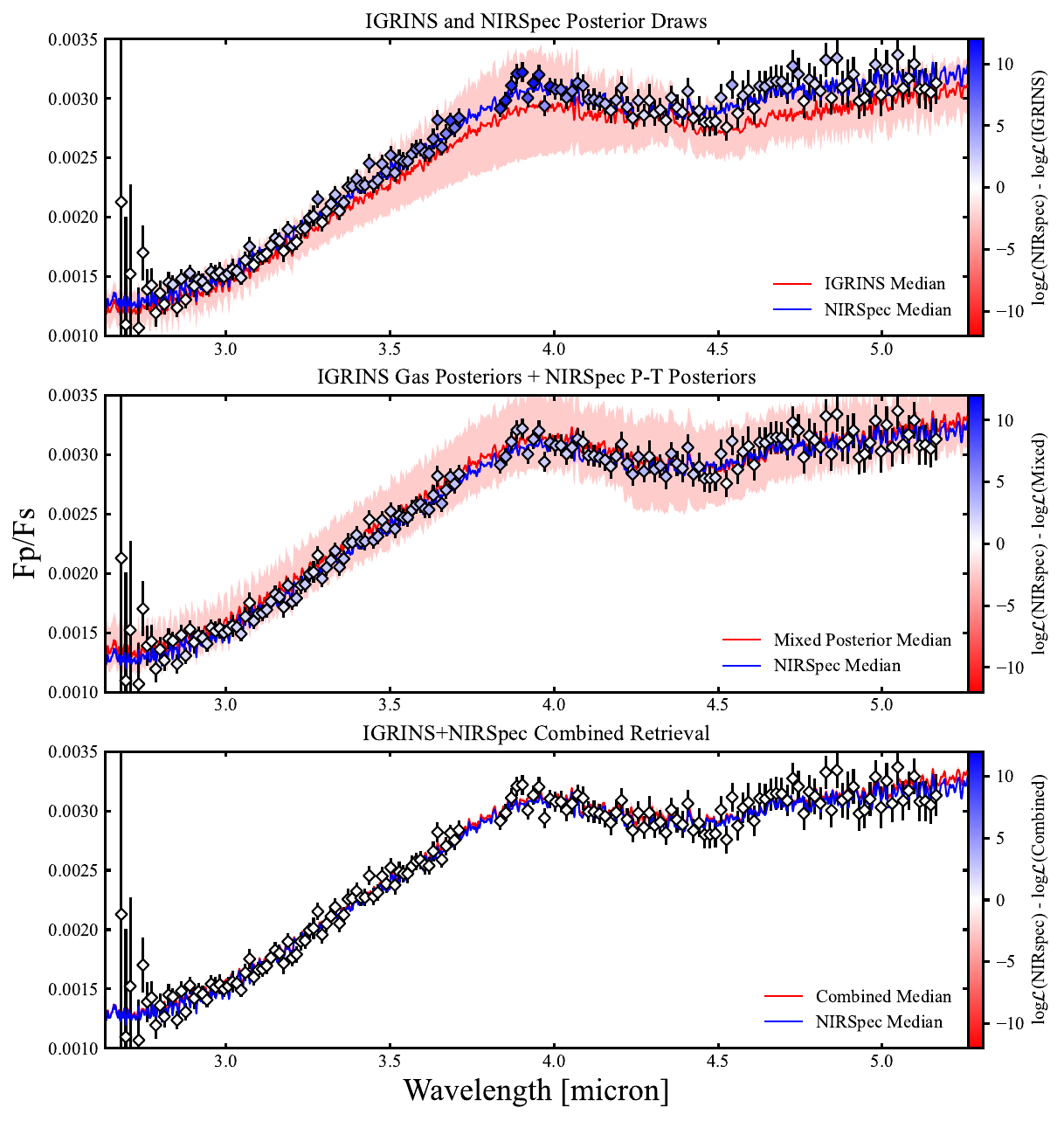}
    \caption{\textit{Top:} The median spectrum from 2000 random draws from the IGRINS pre-eclipse retrieval (red line) and 1$\sigma$ distribution about the median (red shaded region). The median spectrum from 2000 draws from the NIRSpec retrieval is in blue, and the 1$\sigma$ distribution is so tightly around this median it is not visible in this figure. The NIRSpec data is plotted in diamonds, with the color indicating the difference in the median log-likelihood While the predictions from the IGRINS retrieval are largely consistent with the NIRSpec data, it struggles to predict the data around 4 microns the most. \textit{Middle:} Similar to the top panel, but now in red is the median spectrum resulting from combining random draws from the IGRINS gas posteriors but the NIRSpec P-T posteriors. \textit{Bottom:} Draws from the combined IGRINS+NIRSpec retrieval. Similar to the NIRSpec retrieval, the 1$\sigma$ contours are not visible.}
    \label{fig:draws}
\end{figure*}

To compare how well the IGRINS and NIRSpec retrievals predict the NIRSpec data, we post-process 2000 random draws from each posterior distribution to the NIRSpec wavelengths and calculate the by-point log-likelihood for each draw. Unsurprisingly, the NIRSpec posteriors better predict the data with a median $\chi^2/N=0.77 \pm 0.02 (N=160)$ and a best fit $\chi^2/N=0.72$. The best fit spectrum from the IGRINS-only retrieval adequately fits the data ($\chi^2/N=1.72$), and the distribution of the draws is largely consistent with the data as well, albeit with a much wider range of $\chi^2/N$ ( median $\chi^2/N=7.56\pm9.93$). Taking the difference in the medians of the by-point log-likelihood distributions, we can see that the points IGRINS struggles the most to predict are around 4 microns (dark blue points in the top panel of Figure \ref{fig:draws}), at which it tends to underpredict the flux.

This region is outside the CO$_2$ feature at 4.5 microns as well as wavelength regions in which NIRSpec probes higher altitudes than IGRINS. Plotting the joint IGRINS+NIRSpec best fit photosphere against the data, we can see that these wavelengths probe $\sim$10$^0$-10$^{-2}$ bar. Comparing the median retrieved P-T profiles from IGRINS and NIRSpec in Figure \ref{fig:high low panel}, we can see that NIRSpec estimates the atmosphere to be hotter, and therefore brighter, than IGRINS does at these pressures. When we again post-process draws from the IGRINS posterior but replace the P-T draws with those from the NIRSpec posterior, the fits to the data improve. The median $\chi^2$/N reduces to 5.2 and the by-point differences in log$\mathcal{L}$ reduce as well (middle panel of Figure \ref{fig:draws}). Therefore, we can conclude that NIRSpec's ability to probe deep into the atmosphere and more accurately measure the temperature at those pressures is what gave it the edge over IGRINS in this particular wavelength region.

The joint retrieval combining the IGRINS and NIRSpec data predicts the NIRSpec data just as well as the NIRSpec data alone without a significant improvement. The median $\chi^2/N$ is 0.78$\pm$0.03, and the best fit $\chi^2/N$ is 0.75. Additionally, the by-point differences in log$\mathcal{L}$ tend to be negligible (bottom panel of Figure \ref{fig:draws}). Between the NIRSpec-only and combined retrievals, the precision on the gas abundances increased by a factor of $\sim$4-5, while the P-T constraints remained largely the same. We can then conclude that, at least in this low metallicity case with weak molecular features, fits to the NIRSpec data are more sensitive to the P-T profile than the composition, and additional information about the gas abundances did not improve the fit. This further highlights the greater sensitivity to composition of IGRINS for the molecules it is sensitive to, which does not include CO$_2$, which NIRSpec was able to place a more stringent upper limit on.

The slight improvements to the precision on the gas abundances between the IGRINS-only and combined retrievals can be attributed to the decreased uncertainty on the P-T profile at pressures less than $\sim$ 1 mbar. The gas abundances are slightly anti-correlated with the P-T parameters $T_0$, log$P_1$, and $\alpha_1$, and between the IGRINS-only and combined retrieval the confidence intervals on all three decreased significantly. As mentioned above, the constraints on the P-T profile are mostly driven by the NIRSpec data, and the certainty on the P-T profile it provided propagated back to decreased uncertainty on the gas abundances themselves.

Ultimately, the combined analysis did not qualitatively change the previous interpretations of WASP-77A b's atmosphere by \cite{line2021} and \cite{august2023}. However, the excellent agreement between IGRINS and NIRspec greatly increases the confidence in the accuracy of these interpretations as well as those from the combined analysis itself. The most significant improvement is the more precise constraints on the P-T profile. This can enable comparisons to predictions from global circulation models to interrogate assumptions about the distribution of heat in WASP-77A b's atmosphere, but such comparisons are beyond the scope of this paper.

\subsection{The Effect of Ephemeris Error on Atmospheric Inferences}
\label{subsec:velocity discussion}

As shown in Section \ref{sec:velocities}, propagated ephemeris error can have a non-negligible impact on measurements of an exoplanet's velocity. As the field of high resolution exoplanet spectroscopy moves toward the measurement of atmospheric dynamics through the measurement of wind speeds \citep{gandhi2022, pino2022} and velocity offsets between gases \citep{brogi2023}, it is crucial to avoid false positives from small eccentricities or ephemeris error. While our ``detection" of a superrotating westward jet is physically implausible and certainly not reality, it is conceivable that a similar error in ephemeris could lead to the false detection of a more believable dynamical result for another hot- or ultra-hot Jupiter atmosphere.

The degeneracy between eccentricity and dynamics has been discussed before by \cite{pino2022}, who also found asymmetries in their retrieved $K_P$ values between pre- and post-eclipse phases for the ultra-hot Jupiter KELT-9 b. When fitting for an eccentricity, they obtain a value much higher than the upper limit derived from previous photometric studies and conclude that the orbit is only \textit{effectively} non-circular. Instead, \cite{pino2022} suggest the anomalous Doppler shifts in the planet signal are indeed from rotation and winds. As they note, the degeneracy between eccentricity and atmospheric dynamics is difficult to break. In the two cases of WASP-77A b and KELT-9 b, this degeneracy was able to be broken using physical intuition and prior information, but more ambiguous cases are bound to occur in the future. For a simple 1D analysis of an atmosphere like this paper, attempting to break the degeneracy between ephemeris error, winds, and orbital eccentricity may be beyond the scope for such a study. If one measures a velocity asymmetry between pre- and post-eclispe phases, it may be adequate to simply give each sequence separate $K_P$ and $V_\mathrm{sys}$ values.

We test whether the effective deviation from a circular orbit due to propagated ephemeris error is enough to affect the planet signal-to-noise ratio, and we repeat the calculation of CCF and $\Delta$log$\mathcal{L}$ maps as described in Section \ref{sec:cross correlation}. Assuming either a circular or eccentric orbit and using the midtransit time and period values reported by \cite{cortes} to determine the phase, we find no significant difference in the CCF SNR and $\Delta$log$\mathcal{L}$ values compared to when we assumed a circular orbit and the \cite{bonomo2017} midtransit time. We also repeat the three-night IGRINS retrieval performed in Section \ref{subsec:post} with the \cite{cortes} period and midtransit time in order to test whether the decrease in precision on the gas abundances compared to the pre-eclipse retrieval was due to the planet signal being imperfectly summed along its velocity. The inferences and precision were unaffected.

\subsection{Sensitivity of the Post-Eclipse Data to the Number of Principal Components Removed}
\label{subsec:poopy post}

\begin{figure*}
    \centering
    \includegraphics[width=0.7\linewidth]{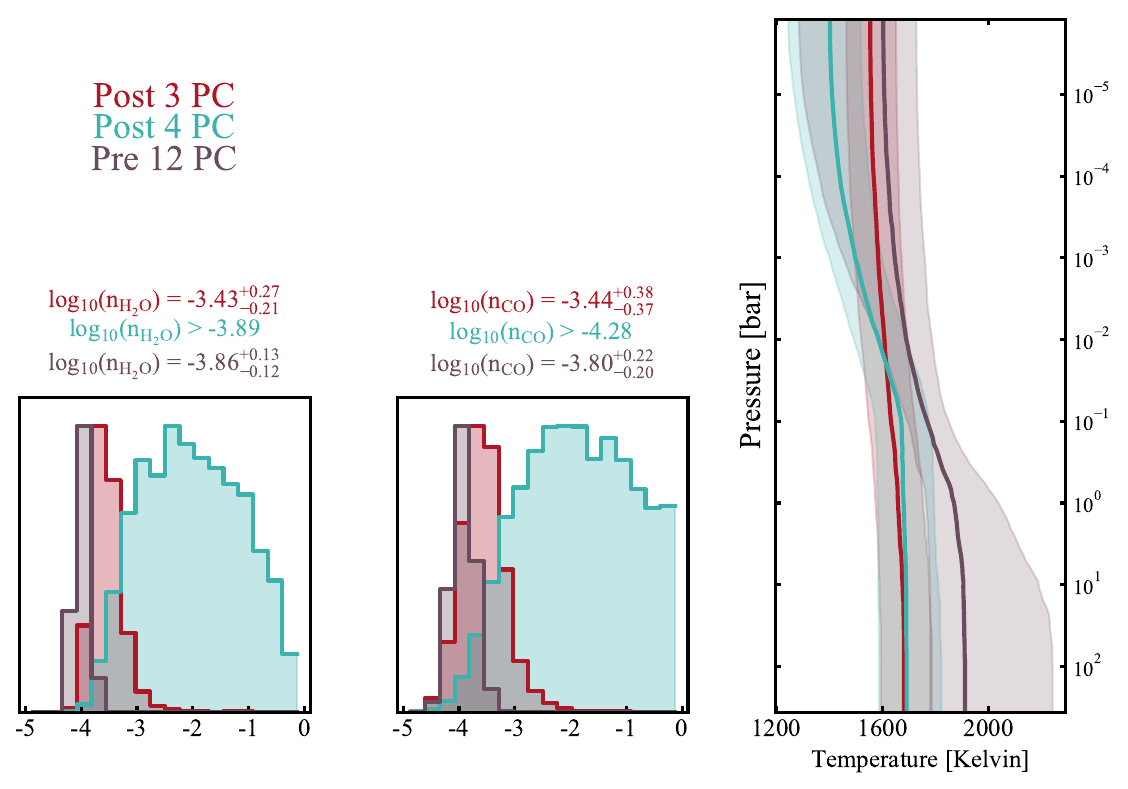}
    \caption{Marginalized posterior distributions for the abundances of H$_2$O and CO as well as posterior draws of the P-T profile for when removing 3 principal components (red) and 4 principal components (blue) from the post-eclipse data. For comparison, the same marginalized posteriors for the pre-eclipse data when 12 principal components are removed.}
    \label{fig:post exploration}
\end{figure*}

\begin{figure}
    \centering
    \includegraphics[width=\linewidth]{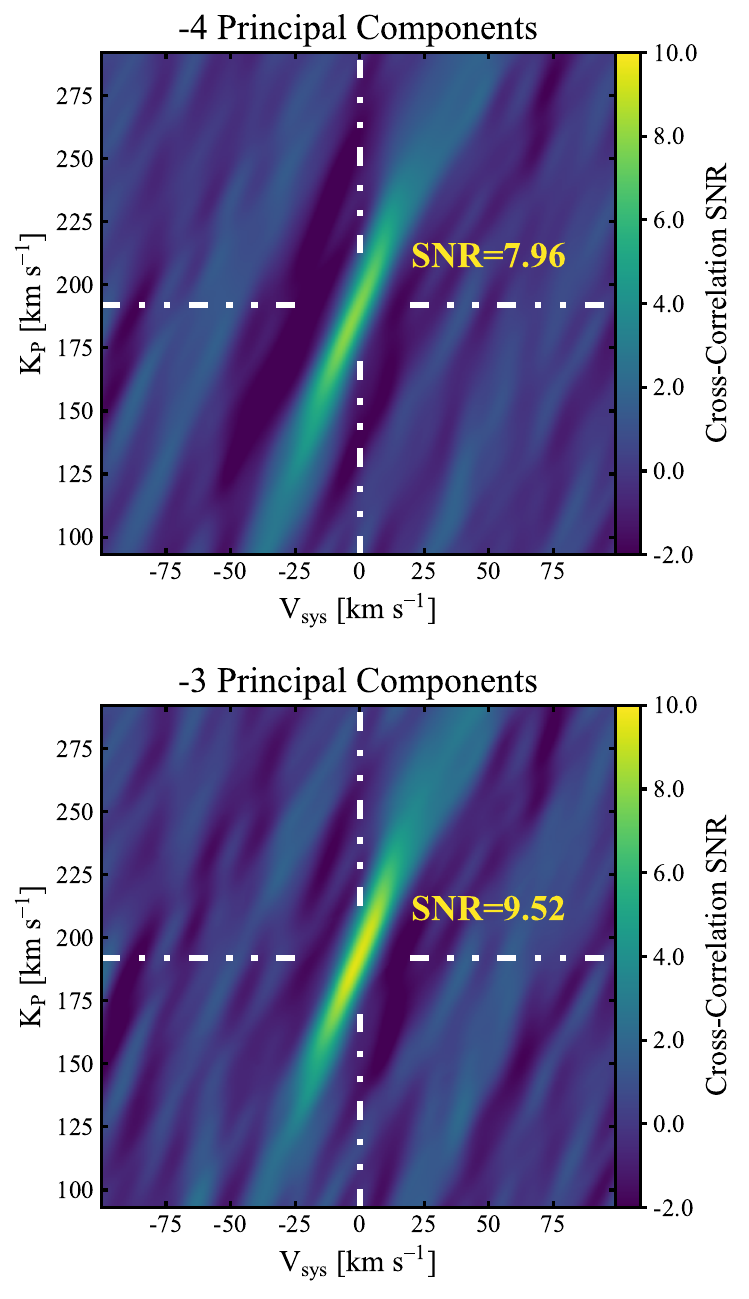}
    \caption{\textit{Top}: Cross-correlation map for the post eclipse data calculated as described in Section \ref{sec:cross correlation} when removing the first 4 principal components from the data. \textit{Bottom}: Similar cross-correlation map but when removing only the first 3 principal components.}
    \label{fig:3 vs 4}
\end{figure}

As noted in Section \ref{subsec:post}, the post-eclipse data are highly sensitive to the number of principal components (PCs) removed in the detrending process, and we chose to remove 3 for our main reported retrieval results on this sequence. The CCF SNR is stronger after only removing 3 PCs (9.5, compared to 8.0 when removing 4 PCs and 9.4 for pre-eclipse, Figure \ref{fig:3 vs 4}), but we discounted removing so few principal components in our initial analysis because in some orders telluric artifacts were still visible by eye in the post-SVD matrices. However, no obvious spurious telluric peaks appear in the CCF map. The choice to initially remove 4 PCs was also motivated as an attempt to be consistent with the previous analysis by \cite{line2021} on the pre-eclipse data. However, as can be seen in Figure \ref{fig:post exploration}, which compares the retrieval results from the 3 PC and 4 PC cases in red and blue, respectively, removing no more than 3 PCs is required to get any informative constraints from the post-eclipse data. While this conforms to intuition that removing more principal components removes more of the planet signal, this sensitivity is concerning, especially considering there was no indication of such a dramatic shift in quality from the small improvement in the CCF SNR, and repeating retrievals several times to test the number of principal components to remove is a time intensive process that many studies are likely to forgo.

\begin{figure}
    \centering
    \includegraphics[width=0.8\linewidth]{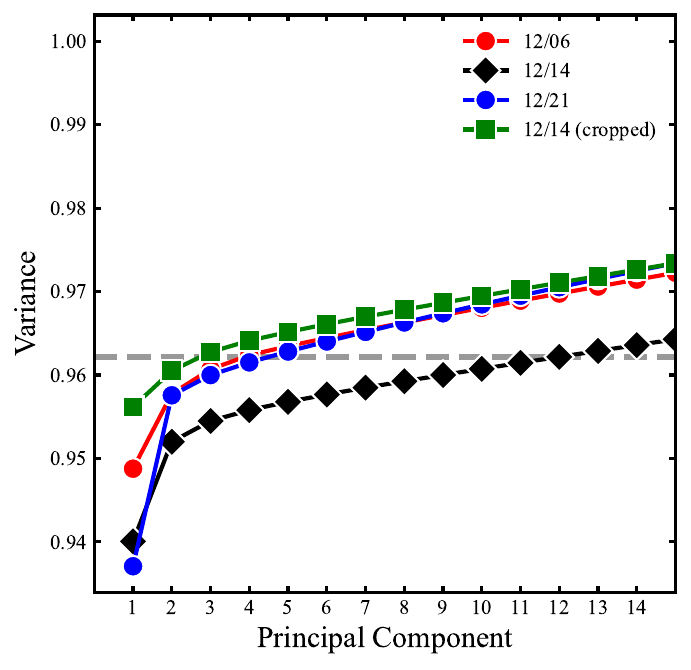}
    \caption{Cumulative variance projected onto each principal component for each night of IGRINS data. Because the number of principal components is determined by the rank of a matrix, sequences with fewer frames have fewer principal components, and therefore a higher fraction of variance is projected onto a given component compared to a shorted sequence.}
    \label{fig:variance}
\end{figure}

Investigating the fraction of total variance accounted for by each principal component, we can see that on average, about 1\% more variance is projected onto a given principal component for the post-eclipse nights compared to the same component from the pre-eclipse night (Figure \ref{fig:variance}). This is not surprising as the number of principal components a matrix has is equal to its rank, which in this case is the number of frames in each sequence, of which both of the post-eclipse nights had fewer. Therefore, a higher fraction of variance per component is necessary. Indeed, if we crop the pre-eclipse sequence to the same amount of frames, the variance contained per component matches that of the post-eclipse nights (green in Figure \ref{fig:variance}). 

About 0.67\% more total variance in contained in the first 4 principal components of the post-eclipse nights than the first 4 of the pre-eclipse night. It takes the first 12 pre-eclipse principal components to account for the same amount of variance. Based on injections of the IGRINS best-fit model, the contributed variance of the planet signal itself is less than this difference, and it is possible more of the planet signal was removed in the post-eclipse SVD process due to this difference. To determine whether the increased fraction of variance removed significantly affected the planet signal, we performed a retrieval on both the cropped pre-eclipse sequence with the first 4 principal components removed and the full sequence with the first 12 principal components removed. For the cropped sequence, both the constraints on the composition and P-T profile were largely unchanged. The confidence intervals on these quantities were slightly larger compared to the original retrieval but still smaller than those from the post-eclipse retrievals. The composition constraints from the 12-PC-removed case were also similar, and we can conclude that the higher fraction of variance removed per principal component is not an issue.

For the 12-PC-removed case, the median retrieved P-T profile was more isothermal, similar to the median post-eclipse profile (Figure \ref{fig:post exploration}, top row). This suggests the possibility that in the 4 PC case, the poorly constrained gas abundances and isothermal P-T profile are separate issues affecting the post-eclipse data. Some of the P-T parameters, especially $\alpha_1$, are correlated with the multiplicative scale factor $a$, which is constrained worse in the pre-eclipse 12 PC retrieval. Compared to the 4 PC retrieval, the 1$\sigma$ confidence intervals on log$_{10}a$ are about 2 times larger, and the marginalized posterior distribution is not approximately Gaussian as it was before. It is likely that after the removal of the first 12 principal components, we have less information about the overall line contrasts, which is why $a$ and by extension the P-T profile are more poorly constrained, but information about \textit{relative} line contrasts is preserved, allowing the absolute gas abundances to be estimated to similar precision as the 4 PC retrieval.

While the 3 PC retrieval on the post-eclipse data yields significant improvement over the 4 PC case and places bounded constraints on the H$_2$O and CO gas abundances consistent with the pre-eclipse constraints and only a factor of $\sim$2 worse precision (Figure \ref{fig:post exploration}), the P-T profile is still near-isothermal. The similarity to the 12 PC pre-eclipse retrieval (bounded gas constraints but poorly constrained P-T profile) while more variance still being removed compared to the pre-eclipse SVD process suggests that information about the absolute line contrasts and therefore the P-T profile could be the ``first to go" in the SVD process. However, this may also be a result of our 1D modeling framework failing to capture 3D atmospheric effects, such as a changing avgerage P-T profile with visible longitudes (discussed further in the next subsection). Telluric artifacts are prevalent in all orders of the the post-SVD matrices when only the first 2 PCs are removed, so we did not attempt a retrieval for a 2 PC removed case.

\begin{figure*}
    \centering
    \includegraphics[width=\textwidth]{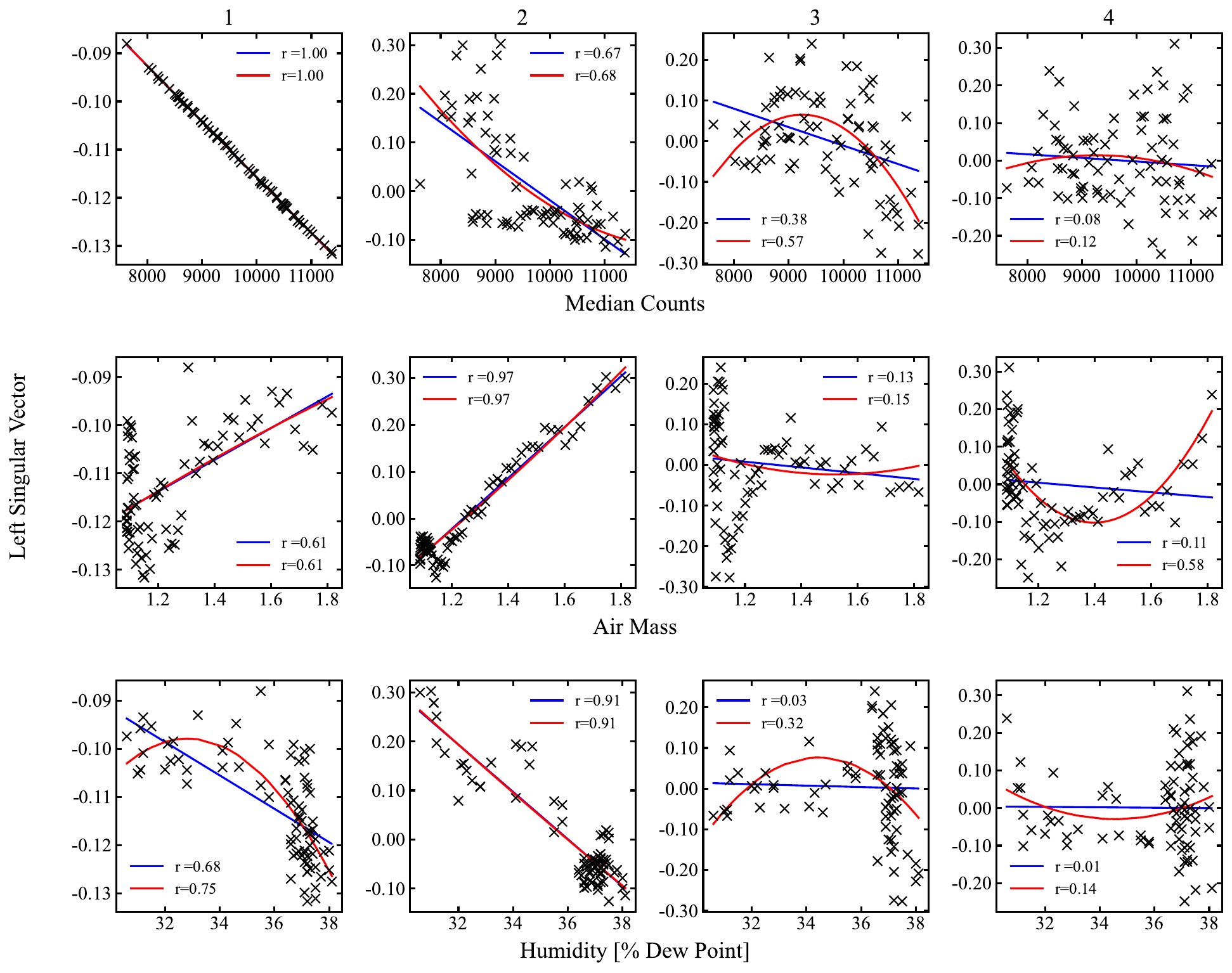}
    \caption{The first four left singular vectors for order 26 (1.69-1.71 micron) from the sequence taken on 12/21 compared to the continuum level, air mass, and humidity. Each column corresponds to the $n$th singular value, indicated at the top. For each left singular vector, we search for both linear (blue) and quadratic (red) correlations with these. The correlation coefficients for linear and quadratic fits are shown, with a  value $\geq$0.5 being considered a significant correlation. In this particular order, it appears that effects from humidity are captured in the first two singular vectors, whereas the continuum takes the first three and there still remains a correlation with air mass in the fourth.}
    \label{fig:LSVs}
\end{figure*}

Next, we investigate the observing conditions of each night and how well the components of the SVD capture them. Similar to \cite{dekok2013}, we can identify correlations with recorded observing conditions, such as the air mass and humidity, with the left singular values (LSVs) in each order. For each sequence and each order, we identify linear and quadratic correlations between the median count value (essentially the continuum level), air mass, and humidity with the first 3 or 4 LSVs (Figure \ref{fig:LSVs}), indicating that the removal of these first few singular values is indeed removing these from the data. However, the specific singular vectors that correlate to these quantities changes order-by-order. This implies that each order may have a unique number of singular values necessary to remove in the detrending process.

\begin{figure*}
    \centering
    \includegraphics[width=\textwidth]{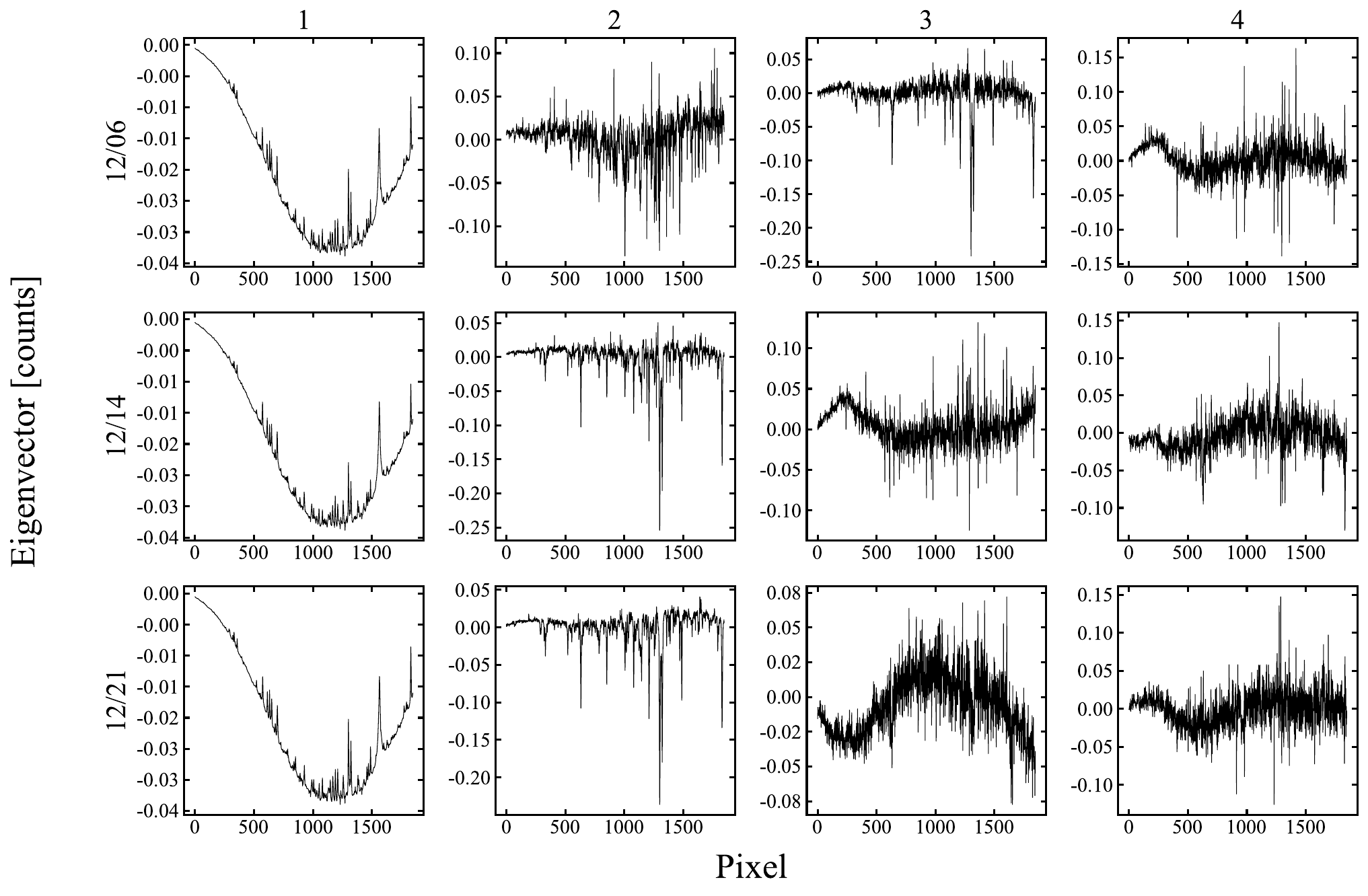}
    \caption{The first four eigenvectors (columns) from the SVD on order 26 for each night (rows). In this particular echelle order, all three nights have a common first eigenvector, while 12/06 appears to have an ``extra" eigenvector associated with the second singular value. The third eigenvector on 12/06 resembles the second eigenvectors on 12/14 and 12/21.}
    \label{fig:eigenvectors}
\end{figure*}

For most orders, the first one or two eigenvectors (right singular vectors) appear consistent across all three nights. However, in many orders, there appears to be either an ``extra" eigenvector on 12/06 that does not exist for 12/14 and 12/21, or 12/06 has similar eigenvectors but in a different order (Figure \ref{fig:eigenvectors}). The extra eigenvectors displace what would have been otherwise common eigenvectors with the other nights to lower singular values. For example, for the order shown in Figure \ref{fig:eigenvectors}, the third eigenvector on 12/06 closely matches the second on both 12/14 and 12/21 but has been ``pushed back" by that night's own unique second eigenvector.

Likewise, in orders heavily impacted by telluric features, the eigenvectors on 12/06 and 12/14 closely match and it is 12/21 that can have an ``extra" eigenvector associated with the second or third singular value. Again, the next eigenvector of 12/21 will match closely with eigenvectors of 12/06 and 12/14 but ranked differently by the singular values. We believe these ``extra" eigenvectors are the cause of the post-eclipse data's sensitivity to number of principal components removed. It appears that whenever an order has one of these extra eigenvectors, the additional information removed contains a large part of the planet signal. 

The trade off between which night and which order this extra vector occurs in appears to be related to the humidity. The humidity on 12/14 was relatively stable for the majority of the sequence and hence does not have this issue, while on 12/06 the humidity was both higher and more variable by 40\%. On 12/21 the humidity was the lowest of the three nights but it was also more variable than 12/14 by 25\%. It appears that with high humidity, 12/06 ``needs" the extra eigenvector and the first three singular values in orders with few telluric lines to remove the effects of humidity, while in orders with many telluric lines it only needs the first two components and eigenvectors to do so. The opposite is true on 12/21, where in the case of low humidity, the third and extra eigenvector is needed to remove the effects of humidity in orders with heavy telluric contamination while for the other orders only the first two are needed.

Because the extra eigenvector switches between orders and between nights, selecting the number of principal components to remove both by-night and by-order may be warranted in the future. Additionally, to obtain a more stable set of eigenvectors to remove, one could perform SVD on all three sequences concatenated together or some other large matrix of IGRINS data and then remove them via multilinear regression such as in \cite{lafarga2023}. This approach might be more successful and robust in future analyses. 

For such a more finely tuned approach to selecting the number of principal components to remove, it would be appropriate to develop a quantitative metric to indicate how many is ``enough." This is not trivial -- CCF signals can be spuriously over-optimized to particular models or velocities (see e.g., \citealp{cabot2019, cheverall2023}), and searching for correlations between the LSVs and physical components of the data requires a comprehensive record of such components. Air mass and humidity appear to be prevalent quantities captured by the SVD in this data, but information about e.g., the seeing or dew point were not available yet these may still impact the data.

\subsubsection{Salvaging the Post-eclipse Data By Combining with NIRSpec}
\label{subsubsec:salvage}

\begin{figure*}
    \centering
    \includegraphics[width=0.7\textwidth]{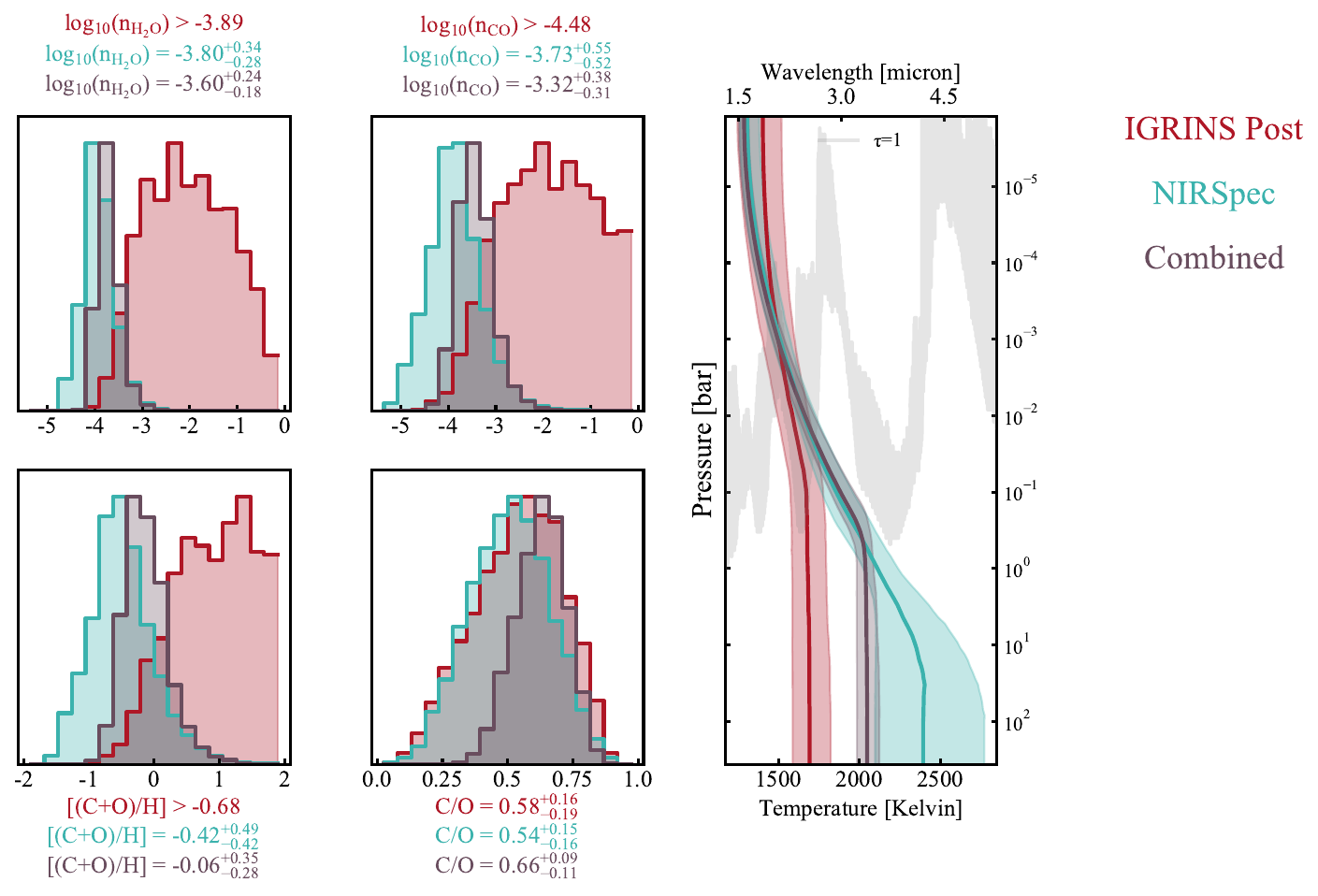}
    \caption{Marginalized posterior distributions of relevant parameters from retrievals on the post-eclipse data with 4 principal components removed (red), NIRSpec (blue), and the two combined (purple). Only lower limits would be placed on the abundances of H$_2$O and CO from the post-eclipse data alone. By combining it with the NIRSpec data, we are able to use what gas abundance information there was in that data and gain more precise estimates for these values than afforded by NIRSpec alone.}
    \label{fig:post and nirspec}
\end{figure*}

The retrieval on the post-eclipse data with 4 principal components removed struggled to produce informative results in large part due to the inability to break the degeneracy between the gas abundances and certain parameters for the analytic P-T profile. In principal, if information about the temperature in the deep atmosphere was regained and broke this degeneracy, informative inferences about the composition could be made from the post-eclipse data in the hypothetical scenario in which removing less principal components did not improve the signal. The NIRSpec data contains such information, so we combined it with the 4 PC-removed post-eclipse data in a single retrieval similar to the combination with the pre-eclipse data in Section \ref{subsec:igrins+nirspec}. This retrieval is able to place bounded constraints on H$_2$O and CO to precision slightly better than achieved with the NIRSpec data alone but not quite at the precision the pre-eclipse data provided (this exercise is summarized in Figure \ref{fig:post and nirspec}). This test confirms that once the abundance and P-T degeneracy was broken, we could extract gas abundance information from the post-eclipse data that improved upon the constraints possible with NIRSpec alone. Therefore, even if a high resolution data set seems fruitless, combining it with low-resolution data can unlock previously inaccessible information.

\subsection{Evidence for Thermal Inhomogeneity}
\label{subsec:TCGs}

\begin{figure*}
    \centering
    \includegraphics[width=0.7\textwidth]{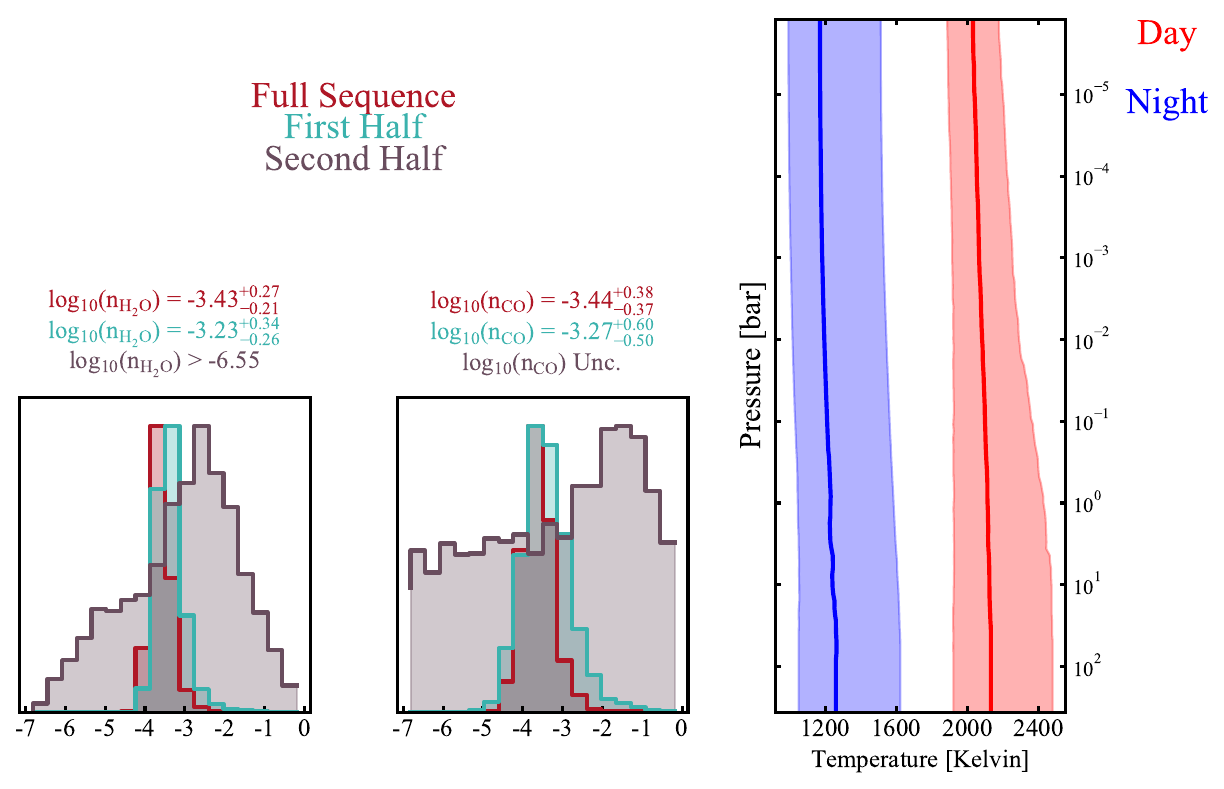}
    \caption{\textit{Left}: Marginalized posterior distributions of the H$_2$O and CO gas abundances from retrievals on the full post-eclipse sequence (red), the first half only (blue), and the second half only (purple). \textit{Right}: Median retrieved day- and night-side P-T profiles from the 2 P-T retrieval in Section \ref{subsec:TCGs}.}
    \label{fig:post 2 PT}
\end{figure*}

As can be seen in the CCF orbit trail in Section \ref{sec:cross correlation}, the post-eclipse signal disappears around phase 0.57. On both post-eclipse nights, the air mass increased over the course of the night, and the subsequent decrease in total flux of the planet may be reflected in the orbit trail. However, air mass is similarly increasing with time during the pre-eclipse night, yet the planet signal appears to be robust for the entire pre-eclipse sequence, including phases further away from eclipse (i.e., $\sim$100\% dayside visibility) than those covered in the post-eclispe data. Therefore, the strength of the CCF trail with time may instead be reflecting a physical change in the planet signal itself.

The morning terminator of the planet is rotating into view during the post-eclipse sequences. If WASP-77A b has efficient heat redistribution, which the retrieved P-T profile is consistent with \citep{line2021}, the hotspot should be offset to the east (evening) and the morning (western) terminator should be cooler than the evening terminator, which is visible in pre-eclipse phases. There is precedence for hotspot offsets affecting HRCCS data, such as \cite{herman2022} or \cite{vansluijs2023}, both of which inferred thermal inhomogeneities via a phase dependence of line contrasts.

The decrease in CCF signal may simply be from the hotspot rotating out of view, or it may also be due to patchy cloud coverage. While the day sides of hot Jupiters are too hot for clouds to form, it is expected that clouds will form on their night sides before blowing over and vaporizing again on the day side \citep{parmentier2016}. WASP-77A b's night side would be around the same temperature as the L/T-dwarf transition, which is marked by the appearance of cloud species such as forsterite and enstatite. WASP-77A b's UV transit depth is also consistent with silicate clouds \citep{turner2016}. Therefore, it is possible that clouds form on the night side and heterogeneously cover the morning limb and other parts of the western hemisphere and are affecting the IGRINS data.

As a first order test of whether we are sensitive to longitudinal thermal variations, we split the post-eclipse sequence into two halves demarcated at phase 0.57 and performed retrievals on each half individually. The constraints from the first half are virtually identical to those from the entire sequence, while for the second half only a lower limit can be placed on the abundance of H$_2$O and CO is entirely unconstrained (Figure \ref{fig:post 2 PT}). We can conclude that the full sequence constraints were driven almost entirely by the first half alone. This suggests we are indeed probing thermal gradients in the post-eclipse sequence and a longitudinally varying treatment of the atmosphere is warranted.

To test both (1) whether the increasing fraction of the cooler morning terminator and night side as we see different longitudes is ``diluting" the observed thermal emission spectrum and (2) whether clouds on these cooler parts of the planet are further obscuring the deep atmosphere, we repeat the retrieval analysis with a simplified 2 P-T atmosphere (such as in e.g., \citealp{feng2016}). The total outgoing spectrum is a phase dependent linear combination of a day and night side spectrum:

\begin{equation}
    F(t) = k(t) F_\mathrm{day} + (1 - k(t)) F_\mathrm{night}
\end{equation}
where $k$ is the fraction of the day side that is visible at a given phase angle $\alpha$:

\begin{equation}
    k(t) = \frac{1}{2} (1 + \cos\alpha(t)); \ \alpha(t) = 2 \pi \varphi(t) + \pi
    \label{eqn:k}
\end{equation}

Both the day and night side have an individual free P-T profile and a gray cloud opacity while we enforce that the global bulk composition of H$_2$O and CO remains constant and therefore are each single free parameters for the entire planet (see e.g., \citealp{cooper2006} for a discussion on the global chemical homogenization on hot Jupiters). With this framework, we are unable to measure a nightside cloud opacity, but we are able to measure two separate P-T profiles, with the night side profile cooler than the dayside by several standard deviations. The inclusion of the 2 P-T profiles is moderately favored over the baseline 1 P-T profile retrieval with a Bayes factor of 16.54 (2.88$\sigma$). Further analysis of the three dimensional nature of WASP-77A b's atmosphere is beyond the scope of this paper, but our 1D retrieval results are likely not significantly biased by the night side because the multiplicative scale factor $a$ can adequately account for dilution by thermal inhomogeneities \citep{taylor2020}. Nonetheless, our 1D retrieval models failing to capture underlying 3D effects might be one of several factors (along with lower SNR and the issues with detrending) contributing to the slightly less precise inferences offered by the post-eclipse data than compared to the pre-eclipse sequence.

\subsection{Conflicting Evidence for the Presence of $^{13}$CO?}

The measurement of the $^{13}$CO\new{/}$^{12}$CO isotopologue ratio first presented in \cite{line2021}'s analysis of the pre-eclipse IGRINS data was an intriguing result. This was the first such measurement in a transiting exoplanet atmosphere, and isotope ratios can potentially provide another avenue for shedding light on a planet's formation history \citep{pontoppidan2014}. We are able to recreate this measurement in the pre-eclipse data, and the inclusion of $^{13}$CO is favored by 4.2$\sigma$, but we are unable to measure the isotope ratio from the post-eclipse data alone and inclusion of $^{13}$CO is neither statistically favored or disfavored.

To test the sensitivity of the post-eclipse data to the isotopologue, we divide out of the data the best fit model from the 3-night IGRINS retrieval, then inject the same model but with the pre-eclipse isotope ratio back into the data at an offset $V_\mathrm{sys}$, and perform a retrieval again. We are still unable to place a bounded measurement on the isotopologue ratio, indicating that the quality of the post-eclipse data is preventing us from confirming the pre-eclipse measurement.

Similarly, from the NIRSpec data we are unable to measure the isotopologue ratio and instead place an upper limit consistent with the pre-eclipse IGRINS data at \new{1/1.48}. To test NIRSpec's sensitivity to the presence of $^{13}$CO, we perform a retrieval on the best-fit model from the three-night IGRINS retrieval. We post-process the model spectrum to the NIRSpec wavelength range, bin onto the data wavelength grid, then add a noise instance based on the error bars. The input value for $^{13}$CO/$^{12}$CO was \new{1/22}, and we again can only place an upper limit at \new{1/5.88}, similar to the retrieval on the real NIRSpec data. Therefore, we conclude that the NIRSpec data is also not sensitive enough to definitively rule out the presence of $^{13}$CO.

\section{Conclusions}
\label{sec:conclusions}

We have presented the first combined exoplanet retrieval analysis using both ground-based and JWST data as well as two new nights of IGRINS data covering the post-eclipse phases \new{of WASP-77A b}. Our findings are as follows:

\begin{itemize}
    \item In addition to the IGRINS pre-eclipse day side thermal emission data first presented in \cite{line2021}, we present here two additional nights capturing the planet's day side emission in the post-eclipse phases. Using traditional cross-correlation methods, we detect H$_2$O and CO absorption features in these additional data.
    \item We found no signatures of atmospheric dynamics and WASP-77A b's orbit is effectively circular. However, propagated ephemeris error manifested as measurements of both an effective eccentricity and an implausibly strong westward equatorial jet. Using updated midtransit times decreases the chances of false positive detections of dynamics.
    \item The additional post-eclipse IGRINS data is highly sensitive to the number of principal components removed in the SVD detrending process. Removing 4, as was done on the pre-eclipse data, removed too much of the planet signal to make informative inferences about WASP-77A b's atmosphere. Only by removing 3 could informative inferences on its composition be made. Investigation of the individual singular vectors produced by the SVD indicates that high and variable humidity played a large role in what information was projected onto a given principal component in a given order. Future observations would benefit from low or less variable humidity or a by-order choice of number of principal components to remove.
    \item We find moderate evidence for thermal inhomogeneity in the post-eclipse data and are able to retrieve both a hot dayside and cooler nightside P-T profile.
    \item The IGRINS and NIRSpec data are in excellent agreement with each other, lending much more confidence in the analysis of each. \new{We reproduce the sub-solar metallicity as measured by \cite{line2021} and \cite{august2023} and the solar C/O ratio as measured by \cite{line2021}. Recent studies suggest this combination may indicate formation beyond the CO$_2$ ice line, but formation within the H$_2$O ice line is also possible within different formation modeling frameworks.}
    \item Combining the pre-eclipse IGRINS data with the NIRSpec/G395H data allowed us to more stringently constrain both the composition and vertical thermal structure of WASP-77A b than possible with either data set alone. IGRINS is more sensitive to absolute gas abundances, while NIRSpec is more sensitive to the P-T profile.
    \item Our inferences from the IGRINS data alone largely predict the NIRSpec data well. Wavelengths in which IGRINS struggled to predict the NIRSpec data coincided with pressures both instruments probe, indicating that NIRSpec is able to estimate the P-T profile more accurately due to the preservation of continuum information in low resolution data.
    \item Neither the post-eclipse or NIRSpec data are sensitive enough to the $^{13}$CO\new{/}$^{12}$CO isotopologue ratio to either refute or confirm the measurement made with the pre-eclipse data.
    
\end{itemize}

As shown by its increased sensitivity to absolute gas abundances and ability to search for atmospheric dynamics, high resolution instruments like IGRINS are still relevant and necessary for exoplanet atmospheric science even with the advent of JWST. Combinations with low resolution transit spectroscopy can provide more powerful probes of exoplanetary atmospheres than either method can provide alone thanks to the complementary nature of the physical quantities each method is sensitive to. High resolution observations in emission are essentially partial spectroscopic phase curves, and as shown in this paper, there is potential to uncover three dimensional information about the thermochemical structure of hot Jupiters. However, thermochemical gradients of hot and ultra-hot Jupiters are largely unexplored in high resolution thermal emission studies. Dynamics, which are measurable with HRCCS, and thermal structure, measurable with low-resolution instruments, are intrinsically and physically linked, but the utility of combining high- and low-resolution data for probing thermochemical gradients remains to be seen.

\facilities{Gemini South (IGRINS), JWST (NIRSpec)}

\software{maptlotlib \citep{hunter2007}, numpy \citep{vanderWalt2011}, pymultinest \citep{buchner2016}, scipy \citep{virtanen2019}}

\section*{Acknowledgements}
M.R.L.\ and J.L.B. acknowledge support for this work from NSF grant AST-2307177. MM acknowledges support provided by NASA through the NASA Hubble Fellowship grant HST-HF2-51485.001-A awarded by the Space Telescope Science Institute, which is operated by AURA, Inc., for NASA, under contract NAS5-26555. Based on observations obtained at the international Gemini Observatory, a program of NSF’s NOIRLab, which is managed by the Association of Universities for Research in Astronomy (AURA) under a cooperative agreement with the National Science Foundation on behalf of the Gemini Observatory partnership: the National Science Foundation (United States), National Research Council (Canada), Agencia Nacional de Investigaci\'{o}n y Desarrollo (Chile), Ministerio de Ciencia, Tecnolog\'{i}a e Innovaci\'{o}n (Argentina), Minist\'{e}rio da Ci\^{e}ncia, Tecnologia, Inova\c{c}\~{o}es e Comunica\c{c}\~{o}es (Brazil), and Korea Astronomy and Space Science Institute (Republic of Korea). This work used the Immersion Grating Infrared Spectrometer (IGRINS) that was developed under a collaboration between the University of Texas at Austin and the Korea Astronomy and Space Science Institute (KASI) with the financial support of the US National Science Foundation 27 under grants AST-1229522 and AST-1702267, of the University of Texas at Austin, and of the Korean GMT Project of KASI. This publication makes use of data products from Exoplanet Watch, a citizen science project managed by NASA's Jet Propulsion Laboratory on behalf of NASA's Universe of Learning. This work is supported by NASA under award number NNX16AC65A to the Space Telescope Science Institute, in partnership with Caltech/IPAC, Center for Astrophysics|Harvard \& Smithsonian, and NASA Jet Propulsion Laboratory.

\bibliography{main.bib}{}

\begin{thebibliography}{}
\expandafter\ifx\csname natexlab\endcsname\relax\def\natexlab#1{#1}\fi
\providecommand{\url}[1]{\href{#1}{#1}}
\providecommand{\dodoi}[1]{doi:~\href{http://doi.org/#1}{\nolinkurl{#1}}}
\providecommand{\doeprint}[1]{\href{http://ascl.net/#1}{\nolinkurl{http://ascl.net/#1}}}
\providecommand{\doarXiv}[1]{\href{https://arxiv.org/abs/#1}{\nolinkurl{https://arxiv.org/abs/#1}}}

\bibitem[{{Arcangeli} {et~al.}(2018){Arcangeli}, {D{\'e}sert}, {Line}, {Bean}, {Parmentier}, {Stevenson}, {Kreidberg}, {Fortney}, {Mansfield}, \& {Showman}}]{arcangeli2018}
{Arcangeli}, J., {D{\'e}sert}, J.-M., {Line}, M.~R., {et~al.} 2018, \apjl, 855, L30, \dodoi{10.3847/2041-8213/aab272}

\bibitem[{{Asplund} {et~al.}(2009){Asplund}, {Grevesse}, {Sauval}, \& {Scott}}]{asplund2009}
{Asplund}, M., {Grevesse}, N., {Sauval}, A.~J., \& {Scott}, P. 2009, \araa, 47, 481, \dodoi{10.1146/annurev.astro.46.060407.145222}

\bibitem[{{Atreya} {et~al.}(2016){Atreya}, {Crida}, {Guillot}, {Lunine}, {Madhusudhan}, \& {Mousis}}]{atreya2016}
{Atreya}, S.~K., {Crida}, A., {Guillot}, T., {et~al.} 2016, arXiv e-prints, arXiv:1606.04510.
\newblock \doarXiv{1606.04510}

\bibitem[{{August} {et~al.}(2023){August}, {Bean}, {Zhang}, {Lunine}, {Xue}, {Line}, \& {Smith}}]{august2023}
{August}, P.~C., {Bean}, J.~L., {Zhang}, M., {et~al.} 2023, \apjl, 953, L24, \dodoi{10.3847/2041-8213/ace828}

\bibitem[{{Azzam} {et~al.}(2016){Azzam}, {Tennyson}, {Yurchenko}, \& {Naumenko}}]{azzam2016}
{Azzam}, A. A.~A., {Tennyson}, J., {Yurchenko}, S.~N., \& {Naumenko}, O.~V. 2016, \mnras, 460, 4063, \dodoi{10.1093/mnras/stw1133}

\bibitem[{{Barber} {et~al.}(2014){Barber}, {Strange}, {Hill}, {Polyansky}, {Mellau}, {Yurchenko}, \& {Tennyson}}]{barber2014}
{Barber}, R.~J., {Strange}, J.~K., {Hill}, C., {et~al.} 2014, \mnras, 437, 1828, \dodoi{10.1093/mnras/stt2011}

\bibitem[{{Baxter} {et~al.}(2020){Baxter}, {D{\'e}sert}, {Parmentier}, {Line}, {Fortney}, {Arcangeli}, {Bean}, {Todorov}, \& {Mansfield}}]{baxter2020}
{Baxter}, C., {D{\'e}sert}, J.-M., {Parmentier}, V., {et~al.} 2020, \aap, 639, A36, \dodoi{10.1051/0004-6361/201937394}

\bibitem[{{Bean} {et~al.}(2023){Bean}, {Xue}, {August}, {Lunine}, {Zhang}, {Thorngren}, {Tsai}, {Stassun}, {Schlawin}, {Ahrer}, {Ih}, \& {Mansfield}}]{bean2023}
{Bean}, J.~L., {Xue}, Q., {August}, P.~C., {et~al.} 2023, \nat, 618, 43, \dodoi{10.1038/s41586-023-05984-y}

\bibitem[{{Beltz} {et~al.}(2021){Beltz}, {Rauscher}, {Brogi}, \& {Kempton}}]{beltz2021}
{Beltz}, H., {Rauscher}, E., {Brogi}, M., \& {Kempton}, E. M.~R. 2021, \aj, 161, 1, \dodoi{10.3847/1538-3881/abb67b}

\bibitem[{{Birkby}(2018)}]{birkby2018}
{Birkby}, J.~L. 2018, arXiv e-prints, arXiv:1806.04617.
\newblock \doarXiv{1806.04617}

\bibitem[{{Birkby} {et~al.}(2013){Birkby}, {de Kok}, {Brogi}, {de Mooij}, {Schwarz}, {Albrecht}, \& {Snellen}}]{birkby2013}
{Birkby}, J.~L., {de Kok}, R.~J., {Brogi}, M., {et~al.} 2013, \mnras, 436, L35, \dodoi{10.1093/mnrasl/slt107}

\bibitem[{{Bitsch} {et~al.}(2022){Bitsch}, {Schneider}, \& {Kreidberg}}]{bitsch2022}
{Bitsch}, B., {Schneider}, A.~D., \& {Kreidberg}, L. 2022, \aap, 665, A138, \dodoi{10.1051/0004-6361/202243345}

\bibitem[{{Bonomo} {et~al.}(2017){Bonomo}, {Desidera}, {Benatti}, {Borsa}, {Crespi}, {Damasso}, {Lanza}, {Sozzetti}, {Lodato}, {Marzari}, {Boccato}, {Claudi}, {Cosentino}, {Covino}, {Gratton}, {Maggio}, {Micela}, {Molinari}, {Pagano}, {Piotto}, {Poretti}, {Smareglia}, {Affer}, {Biazzo}, {Bignamini}, {Esposito}, {Giacobbe}, {H{\'e}brard}, {Malavolta}, {Maldonado}, {Mancini}, {Martinez Fiorenzano}, {Masiero}, {Nascimbeni}, {Pedani}, {Rainer}, \& {Scandariato}}]{bonomo2017}
{Bonomo}, A.~S., {Desidera}, S., {Benatti}, S., {et~al.} 2017, \aap, 602, A107, \dodoi{10.1051/0004-6361/201629882}

\bibitem[{{Boucher} {et~al.}(2023){Boucher}, {Lafreni{\'e}re}, {Pelletier}, {Darveau-Bernier}, {Radica}, {Allart}, {Artigau}, {Cook}, {Debras}, {Doyon}, {Gaidos}, {Benneke}, {Cadieux}, {Carmona}, {Cloutier}, {Cort{\'e}s-Zuleta}, {Cowan}, {Delfosse}, {Donati}, {Fouqu{\'e}}, {Forveille}, {Grankin}, {H{\'e}brard}, {Martins}, {Martioli}, {Masson}, \& {Vinatier}}]{boucher2023}
{Boucher}, A., {Lafreni{\'e}re}, D., {Pelletier}, S., {et~al.} 2023, \mnras, 522, 5062, \dodoi{10.1093/mnras/stad1247}

\bibitem[{{Brogi} {et~al.}(2016){Brogi}, {de Kok}, {Albrecht}, {Snellen}, {Birkby}, \& {Schwarz}}]{brogi2016}
{Brogi}, M., {de Kok}, R.~J., {Albrecht}, S., {et~al.} 2016, \apj, 817, 106, \dodoi{10.3847/0004-637X/817/2/106}

\bibitem[{{Brogi} \& {Line}(2019)}]{brogi2019}
{Brogi}, M., \& {Line}, M.~R. 2019, \aj, 157, 114, \dodoi{10.3847/1538-3881/aaffd3}

\bibitem[{{Brogi} {et~al.}(2012){Brogi}, {Snellen}, {de Kok}, {Albrecht}, {Birkby}, \& {de Mooij}}]{brogi2012}
{Brogi}, M., {Snellen}, I. A.~G., {de Kok}, R.~J., {et~al.} 2012, \nat, 486, 502, \dodoi{10.1038/nature11161}

\bibitem[{{Brogi} {et~al.}(2023){Brogi}, {Emeka-Okafor}, {Line}, {Gandhi}, {Pino}, {Kempton}, {Rauscher}, {Parmentier}, {Bean}, {Mace}, {Cowan}, {Shkolnik}, {Wardenier}, {Mansfield}, {Welbanks}, {Smith}, {Fortney}, {Birkby}, {Zalesky}, {Dang}, {Patience}, \& {D{\'e}sert}}]{brogi2023}
{Brogi}, M., {Emeka-Okafor}, V., {Line}, M.~R., {et~al.} 2023, \aj, 165, 91, \dodoi{10.3847/1538-3881/acaf5c}

\bibitem[{{Buchner}(2016)}]{buchner2016}
{Buchner}, J. 2016, {PyMultiNest: Python interface for MultiNest}.
\newblock \doeprint{1606.005}

\bibitem[{{Burrows} \& {Sharp}(1999)}]{burrows1999}
{Burrows}, A., \& {Sharp}, C.~M. 1999, \apj, 512, 843, \dodoi{10.1086/306811}

\bibitem[{{Cabot} {et~al.}(2019){Cabot}, {Madhusudhan}, {Hawker}, \& {Gandhi}}]{cabot2019}
{Cabot}, S. H.~C., {Madhusudhan}, N., {Hawker}, G.~A., \& {Gandhi}, S. 2019, \mnras, 482, 4422, \dodoi{10.1093/mnras/sty2994}

\bibitem[{{Cheverall} {et~al.}(2023){Cheverall}, {Madhusudhan}, \& {Holmberg}}]{cheverall2023}
{Cheverall}, C.~J., {Madhusudhan}, N., \& {Holmberg}, M. 2023, \mnras, 522, 661, \dodoi{10.1093/mnras/stad648}

\bibitem[{{Coles} {et~al.}(2019){Coles}, {Yurchenko}, \& {Tennyson}}]{coles2019}
{Coles}, P.~A., {Yurchenko}, S.~N., \& {Tennyson}, J. 2019, \mnras, 490, 4638, \dodoi{10.1093/mnras/stz2778}

\bibitem[{{Cooper} \& {Showman}(2006)}]{cooper2006}
{Cooper}, C.~S., \& {Showman}, A.~P. 2006, \apj, 649, 1048, \dodoi{10.1086/506312}

\bibitem[{{Cort{\'e}s-Zuleta} {et~al.}(2020){Cort{\'e}s-Zuleta}, {Rojo}, {Wang}, {Hinse}, {Hoyer}, {Sanhueza}, {Correa-Amaro}, \& {Albornoz}}]{cortes}
{Cort{\'e}s-Zuleta}, P., {Rojo}, P., {Wang}, S., {et~al.} 2020, \aap, 636, A98, \dodoi{10.1051/0004-6361/201936279}

\bibitem[{{Coulombe} {et~al.}(2023){Coulombe}, {Benneke}, {Challener}, {Piette}, {Wiser}, {Mansfield}, {MacDonald}, {Beltz}, {Feinstein}, {Radica}, {Savel}, {Dos Santos}, {Bean}, {Parmentier}, {Wong}, {Rauscher}, {Komacek}, {Kempton}, {Tan}, {Hammond}, {Lewis}, {Line}, {Lee}, {Shivkumar}, {Crossfield}, {Nixon}, {Rackham}, {Wakeford}, {Welbanks}, {Zhang}, {Batalha}, {Berta-Thompson}, {Changeat}, {D{\'e}sert}, {Espinoza}, {Goyal}, {Harrington}, {Knutson}, {Kreidberg}, {L{\'o}pez-Morales}, {Shporer}, {Sing}, {Stevenson}, {Aggarwal}, {Ahrer}, {Alam}, {Bell}, {Blecic}, {Caceres}, {Carter}, {Casewell}, {Crouzet}, {Cubillos}, {Decin}, {Fortney}, {Gibson}, {Heng}, {Henning}, {Iro}, {Kendrew}, {Lagage}, {Leconte}, {Lendl}, {Lothringer}, {Mancini}, {Mikal-Evans}, {Molaverdikhani}, {Nikolov}, {Ohno}, {Palle}, {Piaulet}, {Redfield}, {Roy}, {Tsai}, {Venot}, \& {Wheatley}}]{coulombe2023}
{Coulombe}, L.-P., {Benneke}, B., {Challener}, R., {et~al.} 2023, \nat, 620, 292, \dodoi{10.1038/s41586-023-06230-1}

\bibitem[{{de Kok} {et~al.}(2013){de Kok}, {Brogi}, {Snellen}, {Birkby}, {Albrecht}, \& {de Mooij}}]{dekok2013}
{de Kok}, R.~J., {Brogi}, M., {Snellen}, I.~A.~G., {et~al.} 2013, \aap, 554, A82, \dodoi{10.1051/0004-6361/201321381}

\bibitem[{{Eastman} {et~al.}(2013){Eastman}, {Gaudi}, \& {Agol}}]{eastman2013}
{Eastman}, J., {Gaudi}, B.~S., \& {Agol}, E. 2013, \pasp, 125, 83, \dodoi{10.1086/669497}

\bibitem[{{Ehrenreich} {et~al.}(2020){Ehrenreich}, {Lovis}, {Allart}, {Zapatero Osorio}, {Pepe}, {Cristiani}, {Rebolo}, {Santos}, {Borsa}, {Demangeon}, {Dumusque}, {Gonz{\'a}lez Hern{\'a}ndez}, {Casasayas-Barris}, {S{\'e}gransan}, {Sousa}, {Abreu}, {Adibekyan}, {Affolter}, {Allende Prieto}, {Alibert}, {Aliverti}, {Alves}, {Amate}, {Avila}, {Baldini}, {Bandy}, {Benz}, {Bianco}, {Bolmont}, {Bouchy}, {Bourrier}, {Broeg}, {Cabral}, {Calderone}, {Pall{\'e}}, {Cegla}, {Cirami}, {Coelho}, {Conconi}, {Coretti}, {Cumani}, {Cupani}, {Dekker}, {Delabre}, {Deiries}, {D'Odorico}, {Di Marcantonio}, {Figueira}, {Fragoso}, {Genolet}, {Genoni}, {G{\'e}nova Santos}, {Hara}, {Hughes}, {Iwert}, {Kerber}, {Knudstrup}, {Landoni}, {Lavie}, {Lizon}, {Lendl}, {Lo Curto}, {Maire}, {Manescau}, {Martins}, {M{\'e}gevand}, {Mehner}, {Micela}, {Modigliani}, {Molaro}, {Monteiro}, {Monteiro}, {Moschetti}, {M{\"u}ller}, {Nunes}, {Oggioni}, {Oliveira}, {Pariani}, {Pasquini}, {Poretti}, {Rasilla}, {Redaelli}, {Riva}, {Santana Tschudi},
  {Santin}, {Santos}, {Segovia Milla}, {Seidel}, {Sosnowska}, {Sozzetti}, {Span{\`o}}, {Su{\'a}rez Mascare{\~n}o}, {Tabernero}, {Tenegi}, {Udry}, {Zanutta}, \& {Zerbi}}]{ehrenreich2020}
{Ehrenreich}, D., {Lovis}, C., {Allart}, R., {et~al.} 2020, \nat, 580, 597, \dodoi{10.1038/s41586-020-2107-1}

\bibitem[{{Feng} {et~al.}(2016){Feng}, {Line}, {Fortney}, {Stevenson}, {Bean}, {Kreidberg}, \& {Parmentier}}]{feng2016}
{Feng}, Y.~K., {Line}, M.~R., {Fortney}, J.~J., {et~al.} 2016, \apj, 829, 52, \dodoi{10.3847/0004-637X/829/1/52}

\bibitem[{{Feroz} {et~al.}(2009){Feroz}, {Hobson}, \& {Bridges}}]{feroz2009}
{Feroz}, F., {Hobson}, M.~P., \& {Bridges}, M. 2009, \mnras, 398, 1601, \dodoi{10.1111/j.1365-2966.2009.14548.x}

\bibitem[{{Flowers} {et~al.}(2019){Flowers}, {Brogi}, {Rauscher}, {Kempton}, \& {Chiavassa}}]{flowers2019}
{Flowers}, E., {Brogi}, M., {Rauscher}, E., {Kempton}, E. M.~R., \& {Chiavassa}, A. 2019, \aj, 157, 209, \dodoi{10.3847/1538-3881/ab164c}

\bibitem[{{Fortney} {et~al.}(2008){Fortney}, {Lodders}, {Marley}, \& {Freedman}}]{fortney2008}
{Fortney}, J.~J., {Lodders}, K., {Marley}, M.~S., \& {Freedman}, R.~S. 2008, \apj, 678, 1419, \dodoi{10.1086/528370}

\bibitem[{{Fulton} {et~al.}(2018){Fulton}, {Petigura}, {Blunt}, \& {Sinukoff}}]{fulton2018}
{Fulton}, B.~J., {Petigura}, E.~A., {Blunt}, S., \& {Sinukoff}, E. 2018, \pasp, 130, 044504, \dodoi{10.1088/1538-3873/aaaaa8}

\bibitem[{{Gandhi} {et~al.}(2022){Gandhi}, {Kesseli}, {Snellen}, {Brogi}, {Wardenier}, {Parmentier}, {Welbanks}, \& {Savel}}]{gandhi2022}
{Gandhi}, S., {Kesseli}, A., {Snellen}, I., {et~al.} 2022, \mnras, 515, 749, \dodoi{10.1093/mnras/stac1744}

\bibitem[{{Gandhi} {et~al.}(2019){Gandhi}, {Madhusudhan}, {Hawker}, \& {Piette}}]{gandhi2019}
{Gandhi}, S., {Madhusudhan}, N., {Hawker}, G., \& {Piette}, A. 2019, \aj, 158, 228, \dodoi{10.3847/1538-3881/ab4efc}

\bibitem[{{Gandhi} {et~al.}(2023){Gandhi}, {Kesseli}, {Zhang}, {Louca}, {Snellen}, {Brogi}, {Miguel}, {Casasayas-Barris}, {Pelletier}, {Landman}, {Maguire}, \& {Gibson}}]{gandhi2023}
{Gandhi}, S., {Kesseli}, A., {Zhang}, Y., {et~al.} 2023, \aj, 165, 242, \dodoi{10.3847/1538-3881/accd65}

\bibitem[{{Gharib-Nezhad} {et~al.}(2021){Gharib-Nezhad}, {Iyer}, {Line}, {Freedman}, {Marley}, \& {Batalha}}]{gharib-nezhad2021}
{Gharib-Nezhad}, E., {Iyer}, A.~R., {Line}, M.~R., {et~al.} 2021, \apjs, 254, 34, \dodoi{10.3847/1538-4365/abf504}

\bibitem[{{Gibson} {et~al.}(2020){Gibson}, {Merritt}, {Nugroho}, {Cubillos}, {de Mooij}, {Mikal-Evans}, {Fossati}, {Lothringer}, {Nikolov}, {Sing}, {Spake}, {Watson}, \& {Wilson}}]{gibson2020}
{Gibson}, N.~P., {Merritt}, S., {Nugroho}, S.~K., {et~al.} 2020, \mnras, 493, 2215, \dodoi{10.1093/mnras/staa228}

\bibitem[{{Hargreaves} {et~al.}(2020){Hargreaves}, {Gordon}, {Rey}, {Nikitin}, {Tyuterev}, {Kochanov}, \& {Rothman}}]{hargreaves2020}
{Hargreaves}, R.~J., {Gordon}, I.~E., {Rey}, M., {et~al.} 2020, \apjs, 247, 55, \dodoi{10.3847/1538-4365/ab7a1a}

\bibitem[{{Hawker} {et~al.}(2018){Hawker}, {Madhusudhan}, {Cabot}, \& {Gandhi}}]{hawker2018}
{Hawker}, G.~A., {Madhusudhan}, N., {Cabot}, S. H.~C., \& {Gandhi}, S. 2018, \apjl, 863, L11, \dodoi{10.3847/2041-8213/aac49d}

\bibitem[{{Haynes} {et~al.}(2015){Haynes}, {Mandell}, {Madhusudhan}, {Deming}, \& {Knutson}}]{haynes2015}
{Haynes}, K., {Mandell}, A.~M., {Madhusudhan}, N., {Deming}, D., \& {Knutson}, H. 2015, \apj, 806, 146, \dodoi{10.1088/0004-637X/806/2/146}

\bibitem[{{Herman} {et~al.}(2022){Herman}, {de Mooij}, {Nugroho}, {Gibson}, \& {Jayawardhana}}]{herman2022}
{Herman}, M.~K., {de Mooij}, E. J.~W., {Nugroho}, S.~K., {Gibson}, N.~P., \& {Jayawardhana}, R. 2022, \aj, 163, 248, \dodoi{10.3847/1538-3881/ac5f4d}

\bibitem[{{Hoeijmakers} {et~al.}(2019){Hoeijmakers}, {Ehrenreich}, {Kitzmann}, {Allart}, {Grimm}, {Seidel}, {Wyttenbach}, {Pino}, {Nielsen}, {Fisher}, {Rimmer}, {Bourrier}, {Cegla}, {Lavie}, {Lovis}, {Patzer}, {Stock}, {Pepe}, \& {Heng}}]{hoeijmakers2019}
{Hoeijmakers}, H.~J., {Ehrenreich}, D., {Kitzmann}, D., {et~al.} 2019, \aap, 627, A165, \dodoi{10.1051/0004-6361/201935089}

\bibitem[{{Householder} \& {Weiss}(2022)}]{householder2022}
{Householder}, A., \& {Weiss}, L. 2022, arXiv e-prints, arXiv:2212.06966, \dodoi{10.48550/arXiv.2212.06966}

\bibitem[{Hunter(2007)}]{hunter2007}
Hunter, J.~D. 2007, Computing In Science \& Engineering, 9, 90, \dodoi{10.1109/MCSE.2007.55}

\bibitem[{{Husser} {et~al.}(2013){Husser}, {Wende-von Berg}, {Dreizler}, {Homeier}, {Reiners}, {Barman}, \& {Hauschildt}}]{husser2013}
{Husser}, T.~O., {Wende-von Berg}, S., {Dreizler}, S., {et~al.} 2013, A\&A, 553, A6, \dodoi{10.1051/0004-6361/201219058}

\bibitem[{{JWST Transiting Exoplanet Community Early Release Science Team} {et~al.}(2023){JWST Transiting Exoplanet Community Early Release Science Team}, {Ahrer}, {Alderson}, {Batalha}, {Batalha}, {Bean}, {Beatty}, {Bell}, {Benneke}, {Berta-Thompson}, {Carter}, {Crossfield}, {Espinoza}, {Feinstein}, {Fortney}, {Gibson}, {Goyal}, {Kempton}, {Kirk}, {Kreidberg}, {L{\'o}pez-Morales}, {Line}, {Lothringer}, {Moran}, {Mukherjee}, {Ohno}, {Parmentier}, {Piaulet}, {Rustamkulov}, {Schlawin}, {Sing}, {Stevenson}, {Wakeford}, {Allen}, {Birkmann}, {Brande}, {Crouzet}, {Cubillos}, {Damiano}, {D{\'e}sert}, {Gao}, {Harrington}, {Hu}, {Kendrew}, {Knutson}, {Lagage}, {Leconte}, {Lendl}, {MacDonald}, {May}, {Miguel}, {Molaverdikhani}, {Moses}, {Murray}, {Nehring}, {Nikolov}, {Petit dit de la Roche}, {Radica}, {Roy}, {Stassun}, {Taylor}, {Waalkes}, {Wachiraphan}, {Welbanks}, {Wheatley}, {Aggarwal}, {Alam}, {Banerjee}, {Barstow}, {Blecic}, {Casewell}, {Changeat}, {Chubb}, {Col{\'o}n}, {Coulombe}, {Daylan}, {de Val-Borro},
  {Decin}, {Dos Santos}, {Flagg}, {France}, {Fu}, {Garc{\'\i}a Mu{\~n}oz}, {Gizis}, {Glidden}, {Grant}, {Heng}, {Henning}, {Hong}, {Inglis}, {Iro}, {Kataria}, {Komacek}, {Krick}, {Lee}, {Lewis}, {Lillo-Box}, {Lustig-Yaeger}, {Mancini}, {Mandell}, {Mansfield}, {Marley}, {Mikal-Evans}, {Morello}, {Nixon}, {Ortiz Ceballos}, {Piette}, {Powell}, {Rackham}, {Ramos-Rosado}, {Rauscher}, {Redfield}, {Rogers}, {Roman}, {Roudier}, {Scarsdale}, {Shkolnik}, {Southworth}, {Spake}, {Steinrueck}, {Tan}, {Teske}, {Tremblin}, {Tsai}, {Tucker}, {Turner}, {Valenti}, {Venot}, {Waldmann}, {Wallack}, {Zhang}, \& {Zieba}}]{ers2023}
{JWST Transiting Exoplanet Community Early Release Science Team}, {Ahrer}, E.-M., {Alderson}, L., {et~al.} 2023, \nat, 614, 649, \dodoi{10.1038/s41586-022-05269-w}

\bibitem[{{Karman} {et~al.}(2019){Karman}, {Gordon}, {van der Avoird}, {Baranov}, {Boulet}, {Drouin}, {Groenenboom}, {Gustafsson}, {Hartmann}, {Kurucz}, {Rothman}, {Sun}, {Sung}, {Thalman}, {Tran}, {Wishnow}, {Wordsworth}, {Vigasin}, {Volkamer}, \& {van der Zande}}]{karman2019}
{Karman}, T., {Gordon}, I.~E., {van der Avoird}, A., {et~al.} 2019, \icarus, 328, 160, \dodoi{10.1016/j.icarus.2019.02.034}

\bibitem[{{Kasper} {et~al.}(2021){Kasper}, {Bean}, {Line}, {Seifahrt}, {St{\"u}rmer}, {Pino}, {D{\'e}sert}, \& {Brogi}}]{kasper2021}
{Kasper}, D., {Bean}, J.~L., {Line}, M.~R., {et~al.} 2021, \apjl, 921, L18, \dodoi{10.3847/2041-8213/ac30e1}

\bibitem[{{Kasper} {et~al.}(2023){Kasper}, {Bean}, {Line}, {Seifahrt}, {Brady}, {Lothringer}, {Pino}, {Fu}, {Pelletier}, {St{\"u}rmer}, {Benneke}, {Brogi}, \& {D{\'e}sert}}]{kasper2023}
---. 2023, \aj, 165, 7, \dodoi{10.3847/1538-3881/ac9f40}

\bibitem[{{Khorshid} {et~al.}(2023){Khorshid}, {Min}, \& {D{\'e}sert}}]{khorshid2023}
{Khorshid}, N., {Min}, M., \& {D{\'e}sert}, J.~M. 2023, \aap, 675, A95, \dodoi{10.1051/0004-6361/202245469}

\bibitem[{{Khorshid} {et~al.}(2021){Khorshid}, {Min}, {D{\'e}sert}, {Woitke}, \& {Dominik}}]{khorshid2021}
{Khorshid}, N., {Min}, M., {D{\'e}sert}, J.~M., {Woitke}, P., \& {Dominik}, C. 2021, arXiv e-prints, arXiv:2111.00279.
\newblock \doarXiv{2111.00279}

\bibitem[{{Kokori} {et~al.}(2022){Kokori}, {Tsiaras}, {Edwards}, {Rocchetto}, {Tinetti}, {W{\"u}nsche}, {Paschalis}, {Agnihotri}, {Bachschmidt}, {Bretton}, {Caines}, {Cal{\'o}}, {Casali}, {Crow}, {Dawes}, {Deldem}, {Deligeorgopoulos}, {Dymock}, {Evans}, {Falco}, {Ferratfiat}, {Fowler}, {Futcher}, {Guerra}, {Hurter}, {Jones}, {Kang}, {Kim}, {Lee}, {Lopresti}, {Marino}, {Mallonn}, {Mortari}, {Morvan}, {Mugnai}, {Nastasi}, {Perroud}, {Pereira}, {Phillips}, {Pintr}, {Raetz}, {Regembal}, {Savage}, {Sedita}, {Sioulas}, {Strikis}, {Thurston}, {Tomacelli}, \& {Tomatis}}]{kokori2022}
{Kokori}, A., {Tsiaras}, A., {Edwards}, B., {et~al.} 2022, Experimental Astronomy, 53, 547, \dodoi{10.1007/s10686-020-09696-3}

\bibitem[{{Kolecki} \& {Wang}(2022)}]{kolecki2022}
{Kolecki}, J.~R., \& {Wang}, J. 2022, \aj, 164, 87, \dodoi{10.3847/1538-3881/ac7de3}

\bibitem[{{Kreidberg} {et~al.}(2014){Kreidberg}, {Bean}, {D{\'e}sert}, {Line}, {Fortney}, {Madhusudhan}, {Stevenson}, {Showman}, {Charbonneau}, {McCullough}, {Seager}, {Burrows}, {Henry}, {Williamson}, {Kataria}, \& {Homeier}}]{kreidberg2014}
{Kreidberg}, L., {Bean}, J.~L., {D{\'e}sert}, J.-M., {et~al.} 2014, \apjl, 793, L27, \dodoi{10.1088/2041-8205/793/2/L27}

\bibitem[{{Lafarga} {et~al.}(2023){Lafarga}, {Brogi}, {Gandhi}, {Cegla}, {Seidel}, {Doyle}, {Allart}, {Buchschacher}, {Lendl}, {Lovis}, \& {Sosnowska}}]{lafarga2023}
{Lafarga}, M., {Brogi}, M., {Gandhi}, S., {et~al.} 2023, \mnras, 521, 1233, \dodoi{10.1093/mnras/stad480}

\bibitem[{{Lee} \& {Gullikson}(2016)}]{lee1016}
{Lee}, J.-J., \& {Gullikson}, K. 2016, {Plp: V2.1 Alpha 3}, v2.1-alpha.3,  Zenodo, \dodoi{10.5281/zenodo.56067}

\bibitem[{{Li} {et~al.}(2015){Li}, {Gordon}, {Rothman}, {Tan}, {Hu}, {Kassi}, {Campargue}, \& {Medvedev}}]{li2015}
{Li}, G., {Gordon}, I.~E., {Rothman}, L.~S., {et~al.} 2015, \apjs, 216, 15, \dodoi{10.1088/0067-0049/216/1/15}

\bibitem[{{Line} {et~al.}(2013){Line}, {Wolf}, {Zhang}, {Knutson}, {Kammer}, {Ellison}, {Deroo}, {Crisp}, \& {Yung}}]{line2013}
{Line}, M.~R., {Wolf}, A.~S., {Zhang}, X., {et~al.} 2013, \apj, 775, 137, \dodoi{10.1088/0004-637X/775/2/137}

\bibitem[{{Line} {et~al.}(2021){Line}, {Brogi}, {Bean}, {Gandhi}, {Zalesky}, {Parmentier}, {Smith}, {Mace}, {Mansfield}, {Kempton}, {Fortney}, {Shkolnik}, {Patience}, {Rauscher}, {D{\'e}sert}, \& {Wardenier}}]{line2021}
{Line}, M.~R., {Brogi}, M., {Bean}, J.~L., {et~al.} 2021, \nat, 598, 580, \dodoi{10.1038/s41586-021-03912-6}

\bibitem[{{Lodders} \& {Fegley}(2002)}]{lodders2002}
{Lodders}, K., \& {Fegley}, B. 2002, \icarus, 155, 393, \dodoi{10.1006/icar.2001.6740}

\bibitem[{{Mace} {et~al.}(2018){Mace}, {Sokal}, {Lee}, {Oh}, {Park}, {Lee}, {Good}, {MacQueen}, {Oh}, {Kaplan}, {Kidder}, {Chun}, {Yuk}, {Jeong}, {Pak}, {Kim}, {Nah}, {Lee}, {Yu}, {Hwang}, {Park}, {Kim}, {Chinn}, {Peck}, {Diaz}, {Rutten}, {Prato}, {Jacoby}, {Cornelius}, {Hardesty}, {DeGroff}, {Dunham}, {Levine}, {Nofi}, {Lopez-Valdivia}, {Weinberger}, \& {Jaffe}}]{mace2016}
{Mace}, G., {Sokal}, K., {Lee}, J.-J., {et~al.} 2018, in Society of Photo-Optical Instrumentation Engineers (SPIE) Conference Series, Vol. 10702, Ground-based and Airborne Instrumentation for Astronomy VII, ed. C.~J. {Evans}, L.~{Simard}, \& H.~{Takami}, 107020Q, \dodoi{10.1117/12.2312345}

\bibitem[{{Madhusudhan} {et~al.}(2014){Madhusudhan}, {Amin}, \& {Kennedy}}]{madhu2014}
{Madhusudhan}, N., {Amin}, M.~A., \& {Kennedy}, G.~M. 2014, \apjl, 794, L12, \dodoi{10.1088/2041-8205/794/1/L12}

\bibitem[{{Madhusudhan} {et~al.}(2017){Madhusudhan}, {Bitsch}, {Johansen}, \& {Eriksson}}]{madhu2017}
{Madhusudhan}, N., {Bitsch}, B., {Johansen}, A., \& {Eriksson}, L. 2017, \mnras, 469, 4102, \dodoi{10.1093/mnras/stx1139}

\bibitem[{{Madhusudhan} \& {Seager}(2009)}]{madhu2009}
{Madhusudhan}, N., \& {Seager}, S. 2009, \apj, 707, 24, \dodoi{10.1088/0004-637X/707/1/24}

\bibitem[{{Mansfield} {et~al.}(2021){Mansfield}, {Line}, {Bean}, {Fortney}, {Parmentier}, {Wiser}, {Kempton}, {Gharib-Nezhad}, {Sing}, {L{\'o}pez-Morales}, {Baxter}, {D{\'e}sert}, {Swain}, \& {Roudier}}]{mansfield2021}
{Mansfield}, M., {Line}, M.~R., {Bean}, J.~L., {et~al.} 2021, Nature Astronomy, 5, 1224, \dodoi{10.1038/s41550-021-01455-4}

\bibitem[{{Mansfield} {et~al.}(2022){Mansfield}, {Wiser}, {Stevenson}, {Smith}, {Line}, {Bean}, {Fortney}, {Parmentier}, {Kempton}, {Arcangeli}, {D{\'e}sert}, {Kilpatrick}, {Kreidberg}, \& {Malik}}]{mansfield2022}
{Mansfield}, M., {Wiser}, L., {Stevenson}, K.~B., {et~al.} 2022, \aj, 163, 261, \dodoi{10.3847/1538-3881/ac658f}

\bibitem[{{Maxted} {et~al.}(2013){Maxted}, {Anderson}, {Collier Cameron}, {Doyle}, {Fumel}, {Gillon}, {Hellier}, {Jehin}, {Lendl}, {Pepe}, {Pollacco}, {Queloz}, {S{\'e}gransan}, {Smalley}, {Southworth}, {Smith}, {Triaud}, {Udry}, \& {West}}]{maxted2013}
{Maxted}, P.~F.~L., {Anderson}, D.~R., {Collier Cameron}, A., {et~al.} 2013, \pasp, 125, 48, \dodoi{10.1086/669231}

\bibitem[{{Meibom} {et~al.}(2007){Meibom}, {Krot}, {Robert}, {Mostefaoui}, {Russell}, {Petaev}, \& {Gounelle}}]{meibom2007}
{Meibom}, A., {Krot}, A.~N., {Robert}, F., {et~al.} 2007, \apjl, 656, L33, \dodoi{10.1086/512052}

\bibitem[{{Mordasini} {et~al.}(2016){Mordasini}, {van Boekel}, {Molli{\`e}re}, {Henning}, \& {Benneke}}]{mordasini2016}
{Mordasini}, C., {van Boekel}, R., {Molli{\`e}re}, P., {Henning}, T., \& {Benneke}, B. 2016, \apj, 832, 41, \dodoi{10.3847/0004-637X/832/1/41}

\bibitem[{{Moses} {et~al.}(2013){Moses}, {Line}, {Visscher}, {Richardson}, {Nettelmann}, {Fortney}, {Barman}, {Stevenson}, \& {Madhusudhan}}]{moses2013}
{Moses}, J.~I., {Line}, M.~R., {Visscher}, C., {et~al.} 2013, \apj, 777, 34, \dodoi{10.1088/0004-637X/777/1/34}

\bibitem[{{Mousis} {et~al.}(2019){Mousis}, {Ronnet}, \& {Lunine}}]{mousis2019}
{Mousis}, O., {Ronnet}, T., \& {Lunine}, J.~I. 2019, \apj, 875, 9, \dodoi{10.3847/1538-4357/ab0a72}

\bibitem[{{{\"O}berg} \& {Bergin}(2016)}]{oberg2016}
{{\"O}berg}, K.~I., \& {Bergin}, E.~A. 2016, \apjl, 831, L19, \dodoi{10.3847/2041-8205/831/2/L19}

\bibitem[{{{\"O}berg} {et~al.}(2011){{\"O}berg}, {Murray-Clay}, \& {Bergin}}]{oberg2011}
{{\"O}berg}, K.~I., {Murray-Clay}, R., \& {Bergin}, E.~A. 2011, \apjl, 743, L16, \dodoi{10.1088/2041-8205/743/1/L16}

\bibitem[{{Park} {et~al.}(2014){Park}, {Jaffe}, {Yuk}, {Chun}, {Pak}, {Kim}, {Pavel}, {Lee}, {Oh}, {Jeong}, {Sim}, {Lee}, {Nguyen Le}, {Strubhar}, {Gully-Santiago}, {Oh}, {Cha}, {Moon}, {Park}, {Brooks}, {Ko}, {Han}, {Nah}, {Hill}, {Lee}, {Barnes}, {Yu}, {Kaplan}, {Mace}, {Kim}, {Lee}, {Hwang}, \& {Park}}]{park2014}
{Park}, C., {Jaffe}, D.~T., {Yuk}, I.-S., {et~al.} 2014, in Society of Photo-Optical Instrumentation Engineers (SPIE) Conference Series, Vol. 9147, Ground-based and Airborne Instrumentation for Astronomy V, ed. S.~K. {Ramsay}, I.~S. {McLean}, \& H.~{Takami}, 91471D, \dodoi{10.1117/12.2056431}

\bibitem[{{Parmentier} {et~al.}(2016){Parmentier}, {Fortney}, {Showman}, {Morley}, \& {Marley}}]{parmentier2016}
{Parmentier}, V., {Fortney}, J.~J., {Showman}, A.~P., {Morley}, C., \& {Marley}, M.~S. 2016, \apj, 828, 22, \dodoi{10.3847/0004-637X/828/1/22}

\bibitem[{{Parmentier} {et~al.}(2018){Parmentier}, {Line}, {Bean}, {Mansfield}, {Kreidberg}, {Lupu}, {Visscher}, {D{\'e}sert}, {Fortney}, {Deleuil}, {Arcangeli}, {Showman}, \& {Marley}}]{parmentier2018}
{Parmentier}, V., {Line}, M.~R., {Bean}, J.~L., {et~al.} 2018, \aap, 617, A110, \dodoi{10.1051/0004-6361/201833059}

\bibitem[{{Pelletier} {et~al.}(2021){Pelletier}, {Benneke}, {Darveau-Bernier}, {Boucher}, {Cook}, {Piaulet}, {Coulombe}, {Artigau}, {Lafreni{\`e}re}, {Delisle}, {Allart}, {Doyon}, {Donati}, {Fouqu{\'e}}, {Moutou}, {Cadieux}, {Delfosse}, {H{\'e}brard}, {Martins}, {Martioli}, \& {Vandal}}]{pelletier2021}
{Pelletier}, S., {Benneke}, B., {Darveau-Bernier}, A., {et~al.} 2021, \aj, 162, 73, \dodoi{10.3847/1538-3881/ac0428}

\bibitem[{{Pelletier} {et~al.}(2023){Pelletier}, {Benneke}, {Ali-Dib}, {Prinoth}, {Kasper}, {Seifahrt}, {Bean}, {Debras}, {Klein}, {Bazinet}, {Hoeijmakers}, {Kesseli}, {Lim}, {Carmona}, {Pino}, {Casasayas-Barris}, {Hood}, \& {St{\"u}rmer}}]{pelletier2023}
{Pelletier}, S., {Benneke}, B., {Ali-Dib}, M., {et~al.} 2023, arXiv e-prints, arXiv:2306.08739, \dodoi{10.48550/arXiv.2306.08739}

\bibitem[{{Pino} {et~al.}(2022){Pino}, {Brogi}, {D{\'e}sert}, {Nascimbeni}, {Bonomo}, {Rauscher}, {Basilicata}, {Biazzo}, {Bignamini}, {Borsa}, {Claudi}, {Covino}, {Di Mauro}, {Guilluy}, {Maggio}, {Malavolta}, {Micela}, {Molinari}, {Molinaro}, {Montalto}, {Nardiello}, {Pedani}, {Piotto}, {Poretti}, {Rainer}, {Scandariato}, {Sicilia}, \& {Sozzetti}}]{pino2022}
{Pino}, L., {Brogi}, M., {D{\'e}sert}, J.~M., {et~al.} 2022, \aap, 668, A176, \dodoi{10.1051/0004-6361/202244593}

\bibitem[{{Piskorz} {et~al.}(2018){Piskorz}, {Buzard}, {Line}, {Knutson}, {Benneke}, {Crockett}, {Lockwood}, {Blake}, {Barman}, {Bender}, {Deming}, \& {Johnson}}]{piskorz2018}
{Piskorz}, D., {Buzard}, C., {Line}, M.~R., {et~al.} 2018, \aj, 156, 133, \dodoi{10.3847/1538-3881/aad781}

\bibitem[{{Polanski} {et~al.}(2022){Polanski}, {Crossfield}, {Howard}, {Isaacson}, \& {Rice}}]{polanski2022}
{Polanski}, A.~S., {Crossfield}, I. J.~M., {Howard}, A.~W., {Isaacson}, H., \& {Rice}, M. 2022, Research Notes of the American Astronomical Society, 6, 155, \dodoi{10.3847/2515-5172/ac8676}

\bibitem[{{Polyansky} {et~al.}(2018){Polyansky}, {Kyuberis}, {Zobov}, {Tennyson}, {Yurchenko}, \& {Lodi}}]{polyansky2018}
{Polyansky}, O.~L., {Kyuberis}, A.~A., {Zobov}, N.~F., {et~al.} 2018, \mnras, 480, 2597, \dodoi{10.1093/mnras/sty1877}

\bibitem[{{Pontoppidan} {et~al.}(2014){Pontoppidan}, {Salyk}, {Bergin}, {Brittain}, {Marty}, {Mousis}, \& {{\"O}berg}}]{pontoppidan2014}
{Pontoppidan}, K.~M., {Salyk}, C., {Bergin}, E.~A., {et~al.} 2014, in Protostars and Planets VI, ed. H.~{Beuther}, R.~S. {Klessen}, C.~P. {Dullemond}, \& T.~{Henning}, 363--385, \dodoi{10.2458/azu_uapress_9780816531240-ch016}

\bibitem[{{Reggiani} {et~al.}(2022){Reggiani}, {Schlaufman}, {Healy}, {Lothringer}, \& {Sing}}]{reggiani2022}
{Reggiani}, H., {Schlaufman}, K.~C., {Healy}, B.~F., {Lothringer}, J.~D., \& {Sing}, D.~K. 2022, \aj, 163, 159, \dodoi{10.3847/1538-3881/ac4d9f}

\bibitem[{{Rothman} {et~al.}(2010){Rothman}, {Gordon}, {Barber}, {Dothe}, {Gamache}, {Goldman}, {Perevalov}, {Tashkun}, \& {Tennyson}}]{rothman2010}
{Rothman}, L.~S., {Gordon}, I.~E., {Barber}, R.~J., {et~al.} 2010, \jqsrt, 111, 2139, \dodoi{10.1016/j.jqsrt.2010.05.001}

\bibitem[{{Schneider} \& {Bitsch}(2021)}]{schneider2021}
{Schneider}, A.~D., \& {Bitsch}, B. 2021, \aap, 654, A71, \dodoi{10.1051/0004-6361/202039640}

\bibitem[{{Sing} {et~al.}(2016){Sing}, {Fortney}, {Nikolov}, {Wakeford}, {Kataria}, {Evans}, {Aigrain}, {Ballester}, {Burrows}, {Deming}, {D{\'e}sert}, {Gibson}, {Henry}, {Huitson}, {Knutson}, {Lecavelier Des Etangs}, {Pont}, {Showman}, {Vidal-Madjar}, {Williamson}, \& {Wilson}}]{sing2016}
{Sing}, D.~K., {Fortney}, J.~J., {Nikolov}, N., {et~al.} 2016, \nat, 529, 59, \dodoi{10.1038/nature16068}

\bibitem[{{Snellen} {et~al.}(2010){Snellen}, {de Kok}, {de Mooij}, \& {Albrecht}}]{snellen2010}
{Snellen}, I. A.~G., {de Kok}, R.~J., {de Mooij}, E. J.~W., \& {Albrecht}, S. 2010, \nat, 465, 1049, \dodoi{10.1038/nature09111}

\bibitem[{{Southworth}(2011)}]{southworth2011}
{Southworth}, J. 2011, \mnras, 417, 2166, \dodoi{10.1111/j.1365-2966.2011.19399.x}

\bibitem[{{Taylor} {et~al.}(2020){Taylor}, {Parmentier}, {Irwin}, {Aigrain}, {Lee}, \& {Krissansen-Totton}}]{taylor2020}
{Taylor}, J., {Parmentier}, V., {Irwin}, P. G.~J., {et~al.} 2020, \mnras, 493, 4342, \dodoi{10.1093/mnras/staa552}

\bibitem[{{Taylor} {et~al.}(2023){Taylor}, {Radica}, {Welbanks}, {MacDonald}, {Blecic}, {Zamyatina}, {Roth}, {Bean}, {Parmentier}, {Coulombe}, {Feinstein}, {Espinoza}, {Benneke}, {Lafreni{\`e}re}, {Doyon}, \& {Ahrer}}]{taylor2023}
{Taylor}, J., {Radica}, M., {Welbanks}, L., {et~al.} 2023, \mnras, 524, 817, \dodoi{10.1093/mnras/stad1547}

\bibitem[{{Tennyson} {et~al.}(2020){Tennyson}, {Yurchenko}, {Al-Refaie}, {Clark}, {Chubb}, {Conway}, {Dewan}, {Gorman}, {Hill}, {Lynas-Gray}, {Mellor}, {McKemmish}, {Owens}, {Polyansky}, {Semenov}, {Somogyi}, {Tinetti}, {Upadhyay}, {Waldmann}, {Wang}, {Wright}, \& {Yurchenko}}]{tennyson2020}
{Tennyson}, J., {Yurchenko}, S.~N., {Al-Refaie}, A.~F., {et~al.} 2020, \jqsrt, 255, 107228, \dodoi{10.1016/j.jqsrt.2020.107228}

\bibitem[{{Turner} {et~al.}(2016){Turner}, {Pearson}, {Biddle}, {Smart}, {Zellem}, {Teske}, {Hardegree-Ullman}, {Griffith}, {Leiter}, {Cates}, {Nieberding}, {Smith}, {Thompson}, {Hofmann}, {Berube}, {Nguyen}, {Small}, {Guvenen}, {Richardson}, {McGraw}, {Raphael}, {Crawford}, {Robertson}, {Tombleson}, {Carleton}, {Towner}, {Walker-LaFollette}, {Hume}, {Watson}, {Jones}, {Lichtenberger}, {Hoglund}, {Cook}, {Crossen}, {Jorgensen}, {Romine}, {Thompson}, {Villegas}, {Wilson}, {Sanford}, {Taylor}, \& {Henz}}]{turner2016}
{Turner}, J.~D., {Pearson}, K.~A., {Biddle}, L.~I., {et~al.} 2016, \mnras, 459, 789, \dodoi{10.1093/mnras/stw574}

\bibitem[{{van der Walt} {et~al.}(2011){van der Walt}, {Colbert}, \& {Varoquaux}}]{vanderWalt2011}
{van der Walt}, S., {Colbert}, S.~C., \& {Varoquaux}, G. 2011, Computing in Science and Engineering, 13, 22, \dodoi{10.1109/MCSE.2011.37}

\bibitem[{{van Sluijs} {et~al.}(2023){van Sluijs}, {Birkby}, {Lothringer}, {Lee}, {Crossfield}, {Parmentier}, {Brogi}, {Kulesa}, {McCarthy}, \& {Charbonneau}}]{vansluijs2023}
{van Sluijs}, L., {Birkby}, J.~L., {Lothringer}, J., {et~al.} 2023, \mnras, 522, 2145, \dodoi{10.1093/mnras/stad1103}

\bibitem[{{Virtanen} {et~al.}(2019){Virtanen}, {Gommers}, {Oliphant}, {Haberland}, {Reddy}, {Cournapeau}, {Burovski}, {Peterson}, {Weckesser}, {Bright}, {van der Walt}, {Brett}, {Wilson}, {Jarrod Millman}, {Mayorov}, {Nelson}, {Jones}, {Kern}, {Larson}, {Carey}, {Polat}, {Feng}, {Moore}, {Vand erPlas}, {Laxalde}, {Perktold}, {Cimrman}, {Henriksen}, {Quintero}, {Harris}, {Archibald}, {Ribeiro}, {Pedregosa}, {van Mulbregt}, \& {Contributors}}]{virtanen2019}
{Virtanen}, P., {Gommers}, R., {Oliphant}, T.~E., {et~al.} 2019, arXiv e-prints, arXiv:1907.10121.
\newblock \doarXiv{1907.10121}

\bibitem[{{Welbanks} {et~al.}(2019){Welbanks}, {Madhusudhan}, {Allard}, {Hubeny}, {Spiegelman}, \& {Leininger}}]{welbanks2019}
{Welbanks}, L., {Madhusudhan}, N., {Allard}, N.~F., {et~al.} 2019, \apjl, 887, L20, \dodoi{10.3847/2041-8213/ab5a89}

\bibitem[{{Welbanks} {et~al.}(2023){Welbanks}, {McGill}, {Line}, \& {Madhusudhan}}]{welbanks2023}
{Welbanks}, L., {McGill}, P., {Line}, M., \& {Madhusudhan}, N. 2023, \aj, 165, 112, \dodoi{10.3847/1538-3881/acab67}

\bibitem[{{Wilson} \& {Rood}(1994)}]{wilson1994}
{Wilson}, T.~L., \& {Rood}, R. 1994, \araa, 32, 191, \dodoi{10.1146/annurev.aa.32.090194.001203}

\bibitem[{{Wong} {et~al.}(2020){Wong}, {Shporer}, {Daylan}, {Benneke}, {Fetherolf}, {Kane}, {Ricker}, {Vanderspek}, {Latham}, {Winn}, {Jenkins}, {Boyd}, {Glidden}, {Goeke}, {Sha}, {Ting}, \& {Yahalomi}}]{wong2020}
{Wong}, I., {Shporer}, A., {Daylan}, T., {et~al.} 2020, \aj, 160, 155, \dodoi{10.3847/1538-3881/ababad}

\bibitem[{{Zahnle} {et~al.}(2009){Zahnle}, {Marley}, {Freedman}, {Lodders}, \& {Fortney}}]{zahnle2009}
{Zahnle}, K., {Marley}, M.~S., {Freedman}, R.~S., {Lodders}, K., \& {Fortney}, J.~J. 2009, \apjl, 701, L20, \dodoi{10.1088/0004-637X/701/1/L20}

\bibitem[{{Zellem} {et~al.}(2020){Zellem}, {Pearson}, {Blaser}, {Fowler}, {Ciardi}, {Biferno}, {Massey}, {Marchis}, {Baer}, {Ball}, {Chasin}, {Conley}, {Dixon}, {Fletcher}, {Hernandez}, {Nair}, {Perian}, {Sienkiewicz}, {Tock}, {Vijayakumar}, {Swain}, {Roudier}, {Bryden}, {Conti}, {Hill}, {Hergenrother}, {Dussault}, {Kane}, {Fitzgerald}, {Boyce}, {Peticolas}, {Gee}, {Cominsky}, {Zimmerman-Brachman}, {Smith}, {Creech-Eakman}, {Engelke}, {Iturralde}, {Dragomir}, {Jovanovic}, {Lawton}, {Arbouch}, {Kuchner}, \& {Malvache}}]{zellum2020}
{Zellem}, R.~T., {Pearson}, K.~A., {Blaser}, E., {et~al.} 2020, \pasp, 132, 054401, \dodoi{10.1088/1538-3873/ab7ee7}

\bibitem[{{Zhang} {et~al.}(2021){Zhang}, {Snellen}, {Bohn}, {Molli{\`e}re}, {Ginski}, {Hoeijmakers}, {Kenworthy}, {Mamajek}, {Meshkat}, {Reggiani}, \& {Snik}}]{zhang2021}
{Zhang}, Y., {Snellen}, I. A.~G., {Bohn}, A.~J., {et~al.} 2021, \nat, 595, 370, \dodoi{10.1038/s41586-021-03616-x}

\end{thebibliography}
\bibliographystyle{aasjournal}

\appendix

\section{Eccentricity and Wind Models}
\label{appendix:eccentricty model}

In Section \ref{sec:velocities} we introduced two alternate versions of the planet velocity, $V_P(t)$ to use in Eqn. \ref{eqn:velocity}. The first is an eccentric orbit, for which $V_P(t)$ becomes:

\begin{equation}
    V_P(t) = K_P [\cos(\nu(t) + \omega_P) + e\cos(\omega_P)]
    \label{eqn:eccentric RV}
\end{equation}
where $\nu$(t) is the true anomaly, $\omega_P$ is the argument of periastron of the planet, and $e$ is the eccentricity. $\nu(t)$ is defined as

\begin{equation}
    \nu(t) = E(t) + 2 \arctan \frac{\beta \sin E(t)}{1 - \beta\cos E(t)}; \ \beta = \frac{e}{1 + \sqrt{1-e^2}}
\end{equation}
where $E(t)$ is the eccentric anomaly, which we use the Newton-Raphson method to solve for from the mean anomaly $M(t)$:

\begin{equation}
    M(t) = \frac{2 \pi}{P} (t - T_P) = E(t) - e \sin E(t)
\end{equation}
where $T_P$ is the time of periastron.

$\omega_P$ should not be confused with $\omega_\star$, the argument of periastron of the star\footnote{The many RV codes in the literature are inconsistent between which $\omega$ is used and the sign of line-of-sight velocity in Eqn. \ref{eqn:eccentric RV}. For example, \cite{cortes} use \texttt{ExoFast} \citep{eastman2013} and report $\omega_\star$, so we add 180$^\circ$ to this value to get $\omega_P$. This subtlety caused the authors many headaches. See \cite{householder2022} for a more detailed overview of this problem.} -- the two are different by 180$^\circ$. To be consistent with \cite{cortes}, we use $\omega_\star$ as the input value for our RV code and convert to $\omega_P$ for Eqn. \ref{eqn:eccentric RV}. It is common in the literature to not fit directly for $e$ and $\omega_\star$ but instead for the two quantities $A = \sqrt{e}\cos\omega_\star$ and $B = \sqrt{e}\sin\omega_\star$ (see e.g., \citealp{eastman2013} and \citealp{fulton2018} for discussion on choice of RV parameterization). $e$ and $\omega_\star$ are obtained from the two via:

\begin{equation}
    e = A^2 + B^2; \ \omega_\star = \arctan\frac{B}{A}
    \label{eqn:A and B}
\end{equation}

For both of these we set a uniform prior from -1 to 1.

For the jet model, we find the disk averaged line-of-sight velocity from a function of both longitude $\theta$ and latitude $\phi$ on the visible disk:

\begin{equation}
    v_\mathrm{LOS}(\theta, \phi) = u \sin\theta + v_\mathrm{rot}\sin\theta \cos \phi
\end{equation}
where $u$ is the jet speed and $v_\mathrm{rot}$ is the equatorial rotation velocity, which we set to 4.52 km s$^{-1}$ assuming tidal locking. At a given orbital phase, we calculate $v_\mathrm{LOS}$ on a grid of of disk longitudes and latitudes overlapping with the visible dayside then take the weighted average where the weights are

\begin{equation}
    \mu(\theta, \phi) = \cos\theta \cos\phi
\end{equation}

and the average line-of-sight velocity is then:

\begin{equation}
    \langle v_\mathrm{LOS} \rangle = \frac{\sum_{i,j} \mu(\theta_i, \phi_j) v_\mathrm{LOS}(\theta_i, \phi_j)}{\sum_{i,j} \mu(\theta_i, \phi_j)} 
\end{equation}

This is added to the velocity from a circular orbit:

\begin{equation}
    V_P(t) = K_P \sin[2 \pi \varphi(t)] + \langle v_\mathrm{LOS}(\varphi(t)) \rangle
\end{equation}

\section{Corner Plots}
\label{appendix:corner}

The associated corner plots for all of the velocity inferences and retrieval analyses are available in \href{https://zenodo.org/records/10382053?token=eyJhbGciOiJIUzUxMiJ9.eyJpZCI6ImUwYzQ2OWY2LWUxMWItNGNmYy05MWNmLWNhN2YyYzc4YjEwMiIsImRhdGEiOnt9LCJyYW5kb20iOiI3MWIzOWZhNTI2MzAyOWIxODQ2NWEyOWFlNzQ0ZjVkMyJ9.2zZfrcXislbWzgsZg78Ra18eZzeOfALwPiUsSe4GQyApRu_IVNv34fe5cAhJmC26bkP93jzsK-kMZUOAmfA-JA}{a public Zenodo repository linked here}. Note: by default our plotting routine lists the marginalized posterior medians and 1$\sigma$ confidence intervals, even if the posterior distribution is against a prior bound. The utility of corner plots is to qualitatively inspect the correlation between model parameters. For quantitative estimates, refer to those listed in Table \ref{tab:results} or in the text.

\end{document}